\documentclass[twocolumn,twocolappendix]{aastex7}

\definecolorset{HTML}{CL}{}{Rosewater,DC8A78;Flamingo,DD7878;Pink,EA76CB;Mauve,8839EF;Red,D20F39;Maroon,E64553;Peach,FE640B;Yellow,DF8E1D;Green,40A02B;Teal,179299;Sky,04A5E5;Sapphire,209FB5;Blue,1E66F5;Lavender,7287FD;Text,4C4F69;Subtext1,5C5F77;Subtext0,6C6F85;Overlay2,7C7F93;Overlay1,8C8FA1;Overlay0,9CA0B0;Surface2,ACB0BE;Surface1,BCC0CC;Surface0,CCD0DA;Base,EFF1F5;Mantle,E6E9EF;Crust,DCE0E8}
\hypersetup{linkcolor=CLRed,citecolor=CLBlue,filecolor=CLSky,urlcolor=CLMauve}

\usepackage[T1]{fontenc}
\usepackage{amsmath}  
\usepackage{bm}  
\usepackage{newunicodechar}

\usepackage{xfrac}\usepackage{anyfontsize}  
\usepackage{booktabs}   
\usepackage[capitalise,noabbrev]{cleveref}  
\usepackage{enumitem}
\usepackage{etoolbox}
\usepackage{fancyvrb}\VerbatimFootnotes{}  

\makeatletter\pretocmd\@sect{\def\@currentcounter{#1}}{}{\fail}\makeatother

\graphicspath{{./}{}}

\usepackage{mathtools}

\usepackage{siunitx}  
\DeclareSIUnit{\hubble}{\mathnormal{h}}  
\DeclareSIUnit{\parsec}{pc}
\DeclareSIUnit{\angstrom}{\Angstrom}

\usepackage[acronym]{glossaries-extra}
\setabbreviationstyle[acronym]{long-short}
\glsdisablehyper{}  
\newacronym{2lpt}{2LPT}{second-order Lagrangian perturbation theory}
\newacronym{2pcf}{2PCF}{two-point correlation function}
\newacronym[
longplural={baryon acoustic oscillations}
shortplural={BAO}
]{bao}{BAO}{baryon acoustic oscillations}
\newacronym{cdm}{CDM}{cold dark matter}
\newacronym{cic}{CIC}{cloud-in-cell}
\newacronym{de}{DE}{dark energy}
\newacronym{dm}{DM}{dark matter}
\newacronym{dtfe}{DTFE}{Delaunay tesselation field estimator}
\newacronym{fof}{FOF}{friends-of-friends}
\newacronym{ic}{IC}{initial condition}
\newacronym{lpt}{LPT}{Lagrangian perturbation theory}
\newacronym{lss}{LSS}{large-scale structure}
\newacronym{msc}{MSC}{Morse-Smale complex}
\newacronym{mse}{MSE}{Morse-Smale complex extraction}
\newacronym{ph}{PH}{persistent homology}
\newacronym{pm}{PM}{particle-mesh}
\newacronym{tda}{TDA}{topological data analysis}
\newacronym{tpm}{TreePM}{tree-particle-mesh}
\newacronym{za}{ZA}{Zel'dovich approximation}
\newacronym{nh}{NH}{normal hierarchy}
\newacronym{ih}{IH}{inverted hierarchy} 
\newacronym{cmb}{CMB}{cosmic microwave background}
\newacronym{dr1}{DR1}{Data Release 1}
\newacronym{dr2}{DR2}{Data Release 2}
\newacronym{dr3}{DR3}{Data Release 3}

\received{2026 February 23}
\submitjournal{\apjs}

\begin{document}

\shorttitle{Persistent Homology, Cosmic Filaments, and Massive Neutrinos}
\shortauthors{G. Rossi et al. (2026)}
\title{The Cosmic Web and Its Filaments: Neutrino Mass from Topology and Persistent Homology}

\author[sname=Rossi,gname=Graziano]{Graziano Rossi}  
\affiliation{Department of Physics and Astronomy\\
  Sejong University, Seoul, 143--747, Republic of Korea}
\email[show]{graziano@sejong.ac.kr}

\correspondingauthor{Graziano Rossi}

\author[sname=Yu,gname=Hogyun]{Hogyun Yu}
\affiliation{Department of Physics and Astronomy\\
  Sejong University, Seoul, 143--747, Republic of Korea}
\email{...@...}  

\author[sname=Michaux,gname=Michaël]{Micha{\"e}l Michaux}  
\affiliation{Department of Physics and Astronomy\\
  Sejong University, Seoul, 143--747, Republic of Korea}
\email{...@...}


\begin{abstract}

We apply discrete Morse theory, global topology, and persistent homology to characterize the impact of massive neutrinos on the \textit{multiscale} 
cosmic web, focusing on filaments. The topology of the cosmic web is sensitive to neutrino imprints, and persistence diagrams provide more information 
than commonly used summary statistics by quantifying the longevity of topological features across densities. 
This scale-adaptive, parameter-free formalism is powerful, as massive neutrinos affect halos, walls, filaments, and voids in distinct ways. Within this framework, 
we \textit{simultaneously} assess their impact on tracers and skeleton structures and capture their \textit{multiscale} signals across cosmic time. 
Discrete Morse theory is also well suited for particle-based neutrino implementations, often affected by Poisson shot noise, as it preserves 
the salient features of the underlying smooth field. Using two independent sets of $N$-body simulations, we present filament statistics and persistence 
diagrams in massive-neutrino cosmologies. Our results show that neutrinos leave distinct imprints on filaments and skeleton 
connectivity, producing mass-dependent signatures most pronounced at high redshift ($z \simeq 2$) and detectable at the few-percent 
level for masses as small as $M_\nu \sim 0.1$ eV. Filaments thus provide an ideal environment for isolating neutrino effects. We also compare 
two implementations of massive neutrinos to assess systematics. Our study establishes a promising avenue for leveraging cosmic web topology, 
persistent homology, and environment-based statistics to constrain or directly detect neutrino 
mass and infer the mass hierarchy---a long-standing challenge in particle physics and a major 
objective of ongoing and upcoming galaxy redshift surveys (e.g., DES, DESI, Euclid, Rubin-LSST).

\end{abstract}




\section{Introduction}       \label{sec_introduction}

Neutrino cosmology provides a unique connection between particle physics and astrophysics, linking 
the properties of fundamental particles to the formation of structure on the largest observable scales. 
Since oscillation experiments established that neutrinos are massive (Nobel Prize in Physics 2015), 
determining their absolute mass scale has become a major objective in both particle physics and 
cosmology, prompting intense and renewed activity in both fields. 
Oscillation measurements constrain only neutrino mass-squared differences and mixing parameters 
\citep[e.g.,][]{Esteban2020,Esteban2024,Navas2024},
leaving the absolute mass 
scale and hierarchy undetermined; cosmological observations therefore provide one of the most powerful complementary probes.
In this context, current strategic roadmaps for next-generation surveys 
\citep[e.g., the 2023 Particle Physics Project Prioritization Panel or P5;][]{Asai2024}
have reaffirmed the measurement of the summed neutrino mass $M_{\nu}$, 
together with the nature of dark matter (DM) and the evolution of dark energy (DE), as key scientific goals for 
Stage-IV and next-generation Stage-V precision experiments.

Although neutrinos interact only weakly, their cosmic abundance allows them to influence structure formation through gravity. 
Their large thermal velocities produce significant free-streaming lengths that suppress clustering below characteristic scales, 
leaving distinct signatures in cosmological observables that depend on their total mass. 
At linear order, the impact of massive neutrinos on cosmological perturbations is well understood---see \cite{LesgourguesPastor2006} 
and \cite{Rossi2017}, Section 3, for detailed discussions.
Their homogeneous density and pressure modify the expansion history, 
while their perturbations affect metric fluctuations and the growth of structure. 
These effects alter both the matter power spectrum and angular anisotropy 
spectra on large scales. However, linear theory cannot capture several key aspects of structure formation, particularly in 
the nonlinear regime and on small scales where baryonic processes and mode coupling become important.
In this nonlinear regime, the influence of massive neutrinos is more complex and comparatively less explored. 
High-resolution numerical simulations that incorporate massive 
neutrino components currently provide the most reliable approach for quantifying these effects. 
To this end, \cite{Rossi2017} has shown that neutrinos produce scale-dependent 
distortions in the nonlinear matter and flux power spectra, including the characteristic \textit{spoon-like} 
suppression whose amplitude and scale evolve with redshift and depend on neutrino mass. 
These features suggest the existence of characteristic 
nonlinear scales that may serve as diagnostics of neutrino properties.

Motivated by these developments, there has been increasing interest 
in exploiting the cosmic web as a complementary cosmological laboratory for neutrino physics, 
offering new avenues for constraining massive neutrinos with large-scale surveys.
In this context, substantial  progress in neutrino cosmology has been driven by state-of-the-art 
observations of the cosmic microwave background (CMB) 
and large-scale structure (LSS), enabled by major collaborations such as Planck \citep{Planck2016,Planck2020cosmo}, 
Atacama Cosmology Telescope \citep[ACT;][]{ACT2025}, South Pole Telescope  \citep[SPT;][]{SPT2025}, 
Sloan Digital Sky Survey  \citep[SDSS;][]{York2000}---particularly its fourth phase \citep[SDSS-IV;][]{Blanton2017}, including the
extended Baryon Oscillation Spectroscopic Survey \citep[eBOSS;][]{Dawson2016}---as well as
Dark Energy Survey  \citep[DES;][]{DES2005}
and Dark Energy Spectroscopic Instrument  \citep[DESI;][]{DESICollaboration2016a}.

Joint analyses combining CMB and LSS measurements have progressively tightened upper bounds on the summed neutrino mass. 
For example, results based on eBOSS Data Release 16 (DR16) together with  Planck reported $M_\nu < 0.115~\mathrm{eV}$ \citep{eBOSS2021}, 
while more recent studies incorporating baryon acoustic oscillations (BAO) and full-shape clustering from DESI Data Release 1 (DR1) 
with CMB datasets (Planck+ACT+SPT) obtained $M_\nu < 0.072~\mathrm{eV}$ under standard priors \citep{DESI-DR1-Cosmo} 
within the $\Lambda$CDM framework, the concordance spatially flat model dominated by cold dark matter (CDM) 
and a cosmological constant ($\Lambda$) DE component.
The latest analyses using DESI Data Release 2 (DR2) along with a combination of CMB data further tighten this bound to $M_\nu < 0.0642~\mathrm{eV}$ 
under the same baseline cosmological assumptions \citep{DESI-DR2-Cosmo}.
These limits, however, depend sensitively on the adopted dataset combinations, assumed cosmological model, treatment of neutrino mass states, and are prior-dependent.  
In fact, cosmological constraints on $M_\nu$ are always derived within a specified model framework, and posterior distributions can be 
strongly influenced by prior choices, with some analyses showing evidence of `prior-weight-dominated'  behavior. Consequently, 
present bounds on the neutrino mass should be interpreted as model-dependent inferences rather than direct measurements.

Interestingly, the latest neutrino mass constraints from DESI DR2 \citep{DESI-DR2-Cosmo} present a puzzling situation. 
Neutrino oscillation experiments measure only mass-squared differences, not the absolute mass scale; consequently, 
the minimal summed mass is $M_\nu \simeq 0.057~\mathrm{eV}$ for the normal hierarchy (NH) and $M_\nu \simeq 0.097~\mathrm{eV}$ for the inverted hierarchy (IH). 
In this context, the DESI DR2 bounds not only rule out the IH scenario but, in some aggressive analyses where the prior on $M_\nu$ is left unconstrained, 
even approach---or formally fall below---the minimal value allowed for NH. 
This has prompted renewed discussion in the literature, including speculative interpretations such as the controversial notion of `negative neutrino mass' 
\citep[e.g.,][]{Naredo-Tuero2024,GreenMeyers2025}.
Such results arise because relaxing positivity priors can yield formally unphysical regions of parameter space within a $\Lambda$CDM inference framework.
These findings may reflect residual data systematics, parameter degeneracies, or limitations of the assumed concordance cosmological model used to interpret the measurements, 
potentially pointing to the need for extensions beyond $\Lambda$CDM. 
Indeed, repeating the same DESI DR2 analysis within an evolving 
DE framework ($w_0w_a$CDM) 
yields a substantially weaker constraint, $M_\nu < 0.163~\mathrm{eV}$ (95\%), 
which alleviates the apparent tension but underscores the strong model dependence of current limits \citep{DESI-DR2-Cosmo}.
A similarly shifted preferred value, $M_\nu < 0.149~\mathrm{eV}$ (95\%), is obtained when adopting an alternative-to-$\Lambda$CDM scenario  
mediated by stellar collapse to cosmologically coupled black holes \citep[CCBH;][]{Ahlen2025}.

Taken together, these results highlight the need to reassess how cosmological neutrino constraints are extracted,  
to develop more robust probes of neutrino effects in structure formation, and to deepen our understanding of their impact in the nonlinear regime. 
In particular, these tensions motivate the exploration of alternative statistical approaches beyond traditional two-point power spectrum
and correlation function analyses, especially when combining multiple cosmological datasets and background assumptions.

A rich body of work has investigated the effects of massive neutrinos using a wide range of tracers and observables spanning 
different physical scales. A highly incomplete list includes studies based on the matter power spectrum, bispectrum, velocity and marked statistics, 
three-point functions, weak lensing, Sunyaev-Zel'dovich measurements, magnification bias, peculiar velocities, void statistics, and Lyman-$\alpha$ forest analyses 
\citep[e.g.,][]{Ajani2020,Bolliet2020,Hahn2020,Kuruvilla2020,Rossi2017,Rossi2020,Zhang2020,Bose2021,Massara2021,Whitford2022,
Verza2023,Yankelevich2023,Cueli2024,Thiele2024,Labate2025,Luchina2025,Maggiore2025}.
These approaches exploit diverse probes of LSS, including galaxy clustering, weak lensing, Lyman-$\alpha$ 
forest measurements, cluster abundances, and combinations thereof. 
Despite their differences, most existing methodologies share a common feature: 
they rely primarily on a single observable and/or on information extracted from a single or limited range of scales.

The considerations above suggest that further progress will require not only improved data but also new analysis strategies 
capable of extracting complementary information from LSS. In particular, standard summary statistics used to derive 
tight neutrino upper bounds---such as the two-point correlation function and power spectrum---while highly successful, 
compress the density field and therefore discard higher-order and structural 
information that may encode additional sensitivity to neutrino physics. 
Massive neutrinos affect structure formation in ways that are 
intrinsically geometric and \textit{multiscale}, altering connectivity, 
morphology, and the hierarchical organization of matter.
This motivates the development of alternative probes that directly characterize the 
topology and global structure of the cosmic web, 
which can capture signatures inaccessible to conventional statistics and may provide 
more robust and physically interpretable constraints on neutrino properties. 
Such a possibility is offered by topological methods, and in particular by persistent homology, 
which provides a unified \textit{multiscale} and 
\textit{multiprobe} description of structure. By retaining information that standard statistics compress or discard, these approaches may reveal 
additional neutrino signatures and thus offer a more complete view of their cosmological effects. This is precisely the objective of the present work. 

Topological and homological methods are emerging as powerful tools for characterizing the LSS of the Universe. 
Early applications relied on global descriptors such as genus statistics and Minkowski functionals, whereas more recent developments have 
introduced computational topology techniques that enable a \textit{multiscale}, noise-robust characterization of structure directly from discrete tracer distributions.
In Section~\ref{sec_theory}, we provide a concise pedagogical review of key concepts in \textit{discrete} Morse theory, 
persistent homology, and global topology in a cosmological context. 
Building on advances in computational topology, modern analyses of LSS increasingly employ  discrete Morse theory 
and persistent homology \citep{Forman1998,Forman2002,Edelsbrunner2002,Robins2000}. These frameworks extend 
classical Morse theory \citep{Milnor1963}---originally formulated for smooth, noise-free functions---to realistic datasets that are discrete, irregular, and affected by sampling noise. 
In this setting, discrete Morse complexes and their Morse-Smale counterparts provide a rigorous way to decompose space into regions defined by gradient-flow connectivity, while 
persistence quantifies the significance and stability of topological features across scales and enables the removal of noise-induced structures \citep{Sousbie2011a}. 
Applied to cosmology, this formalism establishes a direct correspondence between critical points of the density field and components of the cosmic web 
(voids, walls, filaments, and clusters), allowing a \textit{multiscale} and hierarchical characterization 
of structure that is robust to sampling effects and sensitive to underlying physical differences.
For recent developments, see, e.g., \cite{Feldbrugge2019}, 
\cite{Pranav2019}, \cite{Pranav2021}, \cite{ Wilding2021}, \cite{Biagetti2022}, \cite{Heydenreich2022}, \cite{Calles2025}.

Within this framework, a key question is whether topological descriptors can capture the subtle imprints of additional physical 
components that affect structure formation, such as massive neutrinos. 
Massive neutrinos introduce scale-dependent growth suppression, modify the connectivity of the density 
field, and alter the hierarchical organization of cosmic structures,
producing signatures that are inherently \textit{multiscale} and nonlocal and therefore particularly well suited to investigation with topological diagnostics. 
The feasibility of this approach was first demonstrated by \citet{Rossi2022}, who presented the first application of persistent homology 
to cosmologies with massive neutrinos, and was further developed in \cite{MoonRossiYu2023}, which introduced the first analysis of critical points in such models 
and quantified their clustering statistics, cross-correlations, and sensitivity to neutrino mass through large-scale features such as
BAO peak amplitudes and inflection scales.
Despite these initial advances, the use of topological methods in cosmologies with massive neutrinos remains comparatively unexplored, 
with only a small number of subsequent studies beginning to investigate related directions \citep{Jalali2024,Yip2024,Prat2026}.

Motivated by this gap, and building on the first application of persistent homology to massive-neutrino cosmologies introduced in \citet{Rossi2022}, 
we employ here discrete Morse theory, global topology, and persistent homology to characterize how neutrinos 
simultaneously affect multiple cosmic environments---halos, walls, filaments, and voids---across the
\textit{multiscale} cosmic web, and to assess their potential as probes of neutrino physics beyond conventional statistics. 
Using two independent $N$-body simulation suites with different neutrino implementations and normalization conventions, 
we quantify these effects within a parameter-free, scale-adaptive framework that is robust to Poisson shot noise, operates directly on particle distributions, 
and requires no smoothing or pre-processing of the density field.

Our methodology and key algorithms, described in Section~\ref{sec_methods}, enable the construction of filament-based statistics and 
persistence diagrams in massive-neutrino cosmologies. 
Persistence diagrams, which encode substantially more information than commonly used topological summary statistics, provide simultaneous 
access to \textit{multiscale} information across tracers and epochs, enabling a more comprehensive view of how massive neutrinos shape cosmic structures 
by exploiting their \textit{combined} imprints rather than isolated effects. In particular, they offer a richer statistical description by quantifying the longevity and 
stability of topological features across all density thresholds. We also compare two independent neutrino implementations to assess potential systematic effects.

Among our main results, presented in Section~\ref{sec_homtop_results}, we show that massive neutrinos leave distinct and measurable imprints on 
cosmic filaments and skeleton structures, directly modifying the connectivity of the cosmic web. In particular, we identify coherent, mass-dependent signatures 
in the persistent filamentary network that become especially pronounced at high redshift ($z \simeq 2$) and remain detectable at the few-percent level 
even for neutrino masses as small as $M_\nu \sim 0.1~\mathrm{eV}$. These findings point to cosmic filaments as particularly 
sensitive environments for isolating neutrino-induced effects on structure formation.
We further find that high-persistence apex points trace sharp connectivity transitions and percolation processes in the web, 
encoding dominant neutrino signatures in the topology of the density field. 
The trends we observe are consistent across independent simulation suites and distinct neutrino implementations.

Taken together, our results demonstrate that topological observables can reveal physically interpretable signatures
inaccessible to conventional summary statistics. 
They thus establish cosmic web topology, persistent homology, and environment-based statistics as complementary components of a physically motivated 
and intrinsically \textit{multiscale} framework for probing the effects of massive neutrinos beyond standard two-point clustering analyses. 
The present work therefore provides both methodological foundations and practical tools for a broader program aimed at extracting neutrino information from the topology and persistent homology of LSS, 
as well as from environment-dependent statistics. This direction is particularly timely given ongoing
surveys such as DES, DESI, and Euclid \citep{Laureijs2011}, 
as well as forthcoming experiments including the Rubin Observatory Legacy Survey of Space and Time \citep[LSST;][]{LSST2019}, the Nancy Grace Roman Space Telescope \citep[Roman;][]{Spergel2015}, 
and the Prime Focus Spectrograph \citep[PFS;][]{Takada2014}, whose statistical precision and redshift coverage will enable these techniques to be applied directly to observational data, 
especially in the high redshift regime. More broadly, these results open a pathway toward constraining neutrino mass---along with additional cosmological parameters---using topological, 
homological, and environment-sensitive observables as complementary probes for next-generation precision cosmology.

The paper is organized as follows. 
Section \ref{sec_simulations} briefly describes the two simulation suites used in this study, which differ in their massive-neutrino 
implementations as well as in mass resolution, box size, and fiducial cosmology. 
Section \ref{sec_theory} summarizes the theoretical framework, reviewing \textit{discrete} Morse theory, persistent homology, and global topology in a cosmological setting, 
and outlining how these formalisms extend to scenarios with massive neutrinos. 
Section \ref{sec_methods} details our methodology, key algorithms, and analysis pipeline, with additional technical material provided in Appendix \ref{sec_appendix_A}.
Section \ref{sec_homtop_results} presents the main results of our analysis, namely the application of topological and homological methods to massive-neutrino cosmologies.
We first examine critical points, skeleton structures, and the impact of massive neutrinos; we then analyze filamentary statistics of the cosmic web and their dependence on neutrino mass.
Persistence diagrams are subsequently used to quantify neutrino-induced topological differences, followed by an investigation of redshift evolution 
and possible systematics arising from variations in simulation setup and neutrino implementation.
Finally, Section \ref{sec_conclusion} summarizes the main results and outlines directions for future work.
 

\section{Simulations}     \label{sec_simulations}

In this section, we briefly describe the simulations used in our study.
For this work, we focus specifically on DM-only $N$-body realizations
and consider two different implementations of massive neutrinos. 
In the first simulation suite, neutrinos are incorporated through 
an analytic linear-response approximation, 
while in the second, they are represented as an additional particle 
component alongside the DM species.
In forthcoming related publications, we will extend our methodologies 
to dedicated hydrodynamical runs 
\citep[e.g., \texttt{Sejong Suite};][]{Rossi2020}
and address the complexity of baryonic physics at small scales and high redshift.   

\subsection{\texttt{MassiveNuS} Simulation Suite}

The Cosmological Massive Neutrino Simulations
\citep[\texttt{MassiveNuS};][]{Liu2018}
are a suite of $N$-body runs produced using a modified version of \texttt{Gadget-2}
\citep{Springel2005}, supplemented with a neutrino patch \citep{Ali-Haimoud2013, Bird2018}.
Initial conditions are generated with \texttt{S-GenIC}\footnote{\url{https://github.com/sbird/S-GenIC}}
at an initial redshift of $z_{\rm in} = 99$, and initial power spectra are obtained with the Boltzmann code \texttt{CAMB} \citep{Lewis2000}.
The average transfer functions of CDM and baryons, weighted by their respective densities, are used.
A regular grid of collisionless CDM particles is then perturbed,
with displacements computed via the Zel'dovich approximation \citep{Zeldovich1970} at sufficiently high redshift,
ensuring that second-order Lagrangian perturbation effects remain negligible.
All simulations share the same phase, as an identical seed was used to generate the initial conditions.
No relativistic corrections are implemented, except for the background density of neutrinos,
which includes a relativistic correction.
The baseline cosmology follows a flat $\Lambda$CDM model,
with a Hubble parameter of $h = 0.7$, a primordial scalar spectral index of $n_{\rm s} = 0.97$,
a baryon density of $\Omega_{\rm b} = 0.046$, and a DE equation of state $w = -1$.
Three cosmological parameters are varied: the total neutrino mass
$M_{\nu}$ in the range of $0.0 - 0.6~{\rm eV}$, the total matter density $\Omega_{\rm m}$
(which includes massive neutrinos when present), and the
primordial power spectrum amplitude $A_{\rm s}$ at the pivot scale $k_0 = 0.05 {\rm Mpc}^{-1}$.
Notably, under these assumptions, the linear theory rms matter fluctuation in $8h^{-1} {\rm Mpc}$ spheres today (i.e., $\sigma_8$)
is a derived parameter for each model. 
 The flatness condition
$\Omega_{\Lambda} + \Omega_{\rm m} = 1$ is imposed,
where $\Omega_{\Lambda}$ is the DE density parameter.
The simulations are performed in periodic boxes with a side length of
$512$ $h^{-1}$Mpc, at a resolution of
$N_{\rm CDM} = 1024^3$ CDM particles, corresponding to a
mass resolution of $\sim10^{10} h^{-1}{\rm M}_\odot$.
Snapshots are output every $126 h^{-1}{\rm Mpc}$, from $z = 45$
down to $z = 0$.

\begin{table}
\centering
\caption{Details of the simulations used in this study.}
\doublerulesep2.0pt
\renewcommand\arraystretch{1.5}
\begin{tabular}{ccc} 
\hline \hline  
  &  \texttt{MassiveNuS} & \texttt{QUIJOTE} \\
\hline
{\centering \it Relevant Parameters} & \\
\hline
$\Omega_{\rm m}$ &                 0.3 & 0.3175\\
$\Omega_{\rm \Lambda}$   &   0.7 &          0.6825    \\
$\Omega_{\rm b}$ &      0.046 &         0.049   \\
$n_{\rm s}$  &                  0.97 &    0.9624\\
$h$             &             0.7 & 0.6711\\
$w$             &             -1 & -1 \\
$M_{\nu}$ [eV]              &      0.0, 0.1, 0.6 & 0.0, 0.1, 0.4\\
$A_{\rm S}$ [baseline]   &             2.1 $\times$ $10^{-9}$  & 2.13 $\times$ $10^{-9}$  \\  
$\sigma_8 (z=0) $ [baseline]   &  0.8523  &  0.834 \\
$k_0$   [Mpc$^{-1}$]& 0.05 & 0.05 \\  
\hline
{\centering \it Simulation Specs} & \\
\hline
Box [$h^{-1}$Mpc] &   512   & 1000 \\
$N_{\rm CDM}$ &   $1024^3$ & $512^3$  \\
$N_{\nu}$ &   -- & $512^3$    \\
$z_{\rm in}$ &   99 & 127  \\
ICs Type &   Zel'dovich &   Zel'dovich \\
$z$ used & 0,1,2 &  0,1,2 \\
\hline
\hline
\label{table_sims_details}
\end{tabular}
\end{table}

\begin{figure*}
\centering
\includegraphics[angle=0,width=1.00\textwidth]{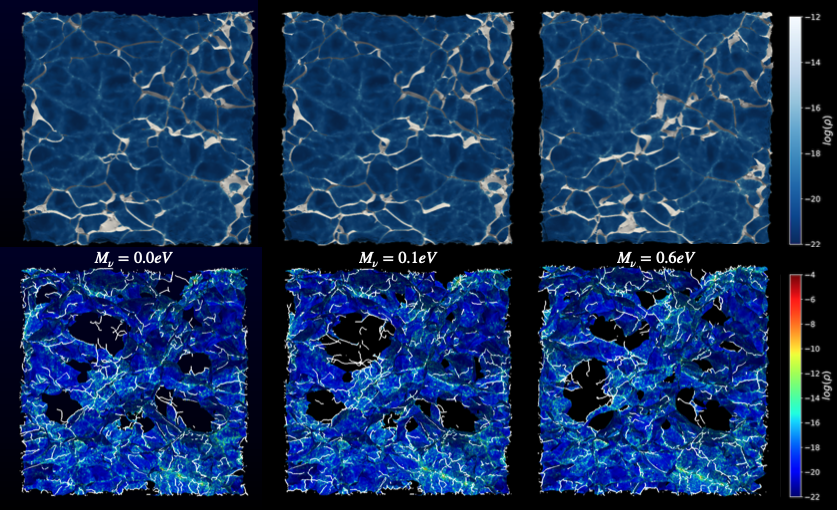} 
\caption{Visualizations of $100 \times 100 \times 50h^{-1}{\rm Mpc}$ slices at $z=0$ with identical spatial locations, 
from the fiducial \texttt{MassiveNuS} simulations. From left to right, the neutrino mass values are 
$M_{\nu} = 0.0~{\rm eV}$ (left), $M_{\nu} = 0.1~{\rm eV}$ (center), and $M_{\nu} = 0.6~{\rm eV}$ (right). 
The top panels display voids (depicted in blue-to-white colors, varying with the corresponding density as indicated by the side color bar), 
overlaid with wall structures (represented in a uniform grey scale). The bottom panels show walls (colored from dark blue to dark red, varying with the corresponding density
as reported in the side color bar), with superimposed filaments shown in a uniform white palette. All structures are identified using the \texttt{DisPerSE} code, as detailed in Section \ref{sec_methods}.}
\label{fig_visualization_1b}
\end{figure*}

In our analysis, we used selected realizations from \texttt{MassiveNuS}.
Specifically, we adopted the three fiducial runs characterized by
$A_{\rm s}^{\rm fid} = 2.1 \times 10^9$, $\Omega_{\rm m}^{\rm fid} = 0.3$,
and either $M_{\nu} = 0.0~{\rm eV}$, $M_{\nu} = 0.1~{\rm eV}$,
or $M_{\nu} = 0.6~{\rm eV}$.
In the massless neutrino cosmology, the derived $\sigma_{8}$
parameter at $z = 0$ is $0.8523$.
We first focused on simulation snapshots (i.e., DM particles) at $z = 0$,
then extended our analysis to snapshots at $z = 1$ and $z = 2$
to characterize redshift evolution effects.\footnote{Unfortunately,
the fiducial simulation with $M_{\nu} = 0.6~{\rm eV}$ was only available at $z = 0$.}
The main characteristics of these simulations are summarized
in the middle column of Table \ref{table_sims_details},
for ease of comparison with the second suite considered in this study.

The selection of the simulation subset is primarily driven by theoretical considerations, as detailed in \cite{Rossi2017,Rossi2020}.
As originally pointed out by \cite{LesgourguesPastor2006}, the most suitable normalization for quantifying the impact of massive 
neutrinos on key cosmological observables---relative to a reference neutrinoless cosmology---is achieved 
by keeping $A_{\rm s}$ and $\Omega_{\rm m}$ fixed, allowing only $\sigma_8$ to vary at $z = 0$.
In our previous works, we referred to this choice as the `UN' or `UNNORM' convention.
In \cite{Rossi2017,Rossi2020}, we also introduced a second normalization, termed `NORM', specifically for grid simulations.
In this approach, all realizations were tuned to match the `best-guess' central value of $\sigma_8$ 
at $z = 0$, corresponding to the fiducial neutrinoless cosmology.
This convention is observationally motivated, as $\sigma_8$ at $z = 0$ is directly constrained by observations.
The second simulation suite considered in this work follows this normalization.
In contrast, the normalization used in other \texttt{MassiveNuS} 
simulations---which varies $M_{\nu}$, $\Omega_{\rm m}$, and $A_{\rm s}$ 
simultaneously---poses challenges for theoretical interpretations.
For this reason, we focus exclusively on the fiducial models in this study.

\begin{figure*}
\centering
\includegraphics[angle=0,width=0.98\textwidth]{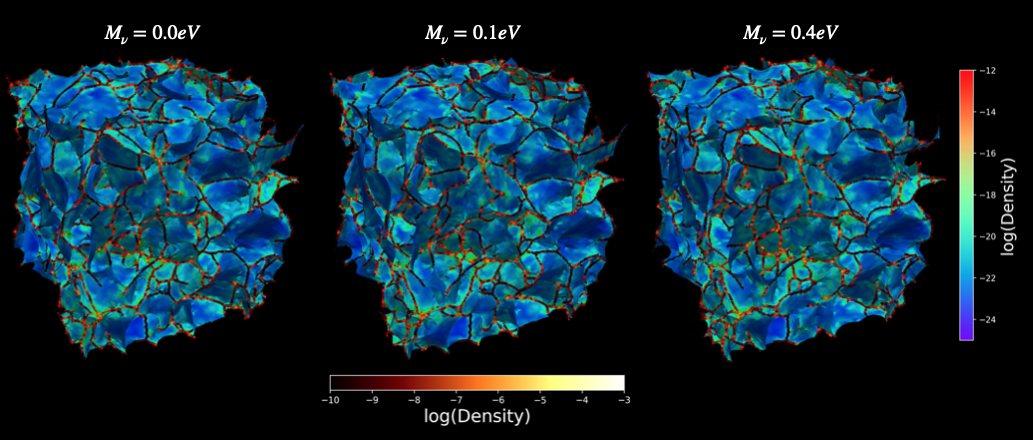} 
\caption{
Three cubic slices at $z=0$ with identical spatial locations and $100 h^{-1}{\rm Mpc}$ side, 
obtained from the \texttt{QUIJOTE} simulations with $M_{\nu} = 0.0~{\rm eV}$ (left, fiducial model), 
$M_{\nu} = 0.1~{\rm eV}$ (center), and $M_{\nu} = 0.4~{\rm eV}$ (right), respectively. 
In all of the panels, wall structures are shown using a blue-to-red palette, with the corresponding density indicated by the right-side color bar. 
Superimposed, the filamentary skeleton structures are displayed, rendered with a dark-red-to-yellow density scale (bottom color bar). 
All of the structures are identified using the entire snapshot particles contained in the selected cube, 
with a persistence threshold $\sigma = 3$.}
\label{fig_visualization_1c}
\end{figure*}

Regarding the implementation of massive neutrinos, 
in the  \texttt{MassiveNuS} simulations their effect is 
included using an analytic linear-response approximation
\citep{Ali-Haimoud2013}. 
A normal hierarchy is assumed, with $m_1 < m_2 < m_3$, where the total neutrino mass is given by $M_{\nu} = m_1 + m_2 + m_3$.
Neutrinos are evolved perturbatively, but their clustering responds to the nonlinear CDM potential.
The simulations account for the effects of radiation on the background expansion, 
as well as the clustering of neutrinos in response to the nonlinear evolution of DM.
The core idea behind this implementation is that, while the CDM density field becomes highly nonlinear, 
neutrino perturbations remain suppressed due to their free-streaming.
As a result, neutrinos can still be described using linear perturbation theory, 
provided their clustering is sourced by the fully nonlinear CDM density.
This approach eliminates the need to treat neutrinos as a separate particle species, significantly reducing computational 
and memory costs---bringing them close to those of a massless neutrino run---while also avoiding shot noise.
However, the main drawback is that the neutrino power spectrum is not accurately captured on small scales,
which presents challenges, particularly for high redshift studies and modeling the Lyman-$\alpha$ forest.
For further details on the simulations, see \cite{Liu2018}.

As an example, Figure \ref{fig_visualization_1b} shows six visualizations at $z=0$ of $100 \times 100h^{-1}{\rm Mpc}$ 
slices with identical spatial location and a thickness of $50h^{-1}{\rm Mpc}$, 
from the fiducial \texttt{MassiveNuS} simulations having
$M_{\nu} =0.0~{\rm eV}$ (left),  $M_{\nu} =0.1~{\rm eV}$ (center),  and $M_{\nu} =0.6~{\rm eV}$ (right),        
respectively. 
The top panels display voids (colored from blue to white, with color intensity varying according to the corresponding density, 
as indicated by the color bar), along with wall structures (depicted in a uniform grey scale).
The bottom panels show walls (colored from dark blue to dark red, varying with the corresponding density, 
as indicated by the color bar), with superimposed filaments shown in a uniform white palette.
All structures are identified using the \texttt{DisPerSE} code \citep{Sousbie2011a}, based on $1\%$ 
of snapshot particles 
and a persistence level corresponding to $\sigma = 6.5$, 
as described in Section \ref{sec_methods}. 

\subsection{\texttt{QUIJOTE} Simulation Suite}

The \texttt{QUIJOTE} simulations \citep{VillaescusaNavarro2020} consist of a set of 44,100 publicly available $N$-body 
runs\footnote{\url{https://quijote-simulations.readthedocs.io/en/latest/}}  
covering over 7,000 cosmological models, produced using the Tree Particle Mesh (TreePM) code \texttt{GADGET-III} \citep{Springel2005}. 
All realizations are characterized by a periodic box size of 1$h^{-1}$Gpc, tracking the evolution of $256^3$, $512^3$, 
or $1024^3$ particles over a total volume of $44,100~(h^{-1}$Gpc$)^3$. 
The baseline fiducial cosmological parameters (see Table \ref{table_sims_details}, right column) are:
$\Omega_{\rm m} = 0.3175$, $\Omega_{\Lambda} = 0.6825$, 
$\Omega_{\rm b} = 0.049$, 
$n_{\rm s} = 0.9624$, 
$h = 0.6711$, 
and $w = -1$. These parameters are closer to those reported by the Planck Collaboration in 2016 \citep{Planck2016}. 
The gravitational evolution of $N_{\rm CDM}$ CDM and $N_{\nu}$ neutrino particles is followed from $z_{\rm in} = 127$ to $z = 0$, 
using either the Zel'dovich approximation or second-order Lagrangian perturbation theory (2LPT). 
All massive neutrino realizations assume degenerate masses. 
The input matter power spectra and transfer functions are obtained via \texttt{CAMB}. 
For models with massive neutrinos, the rescaling method developed in \cite{Zennaro2017} is used to determine the initial conditions. 
The gravitational softening is set to $1/40$ of the mean inter-particle distance for both CDM and neutrinos. Random seeds for the initial 
conditions are consistent across identical realizations in different models but vary between different realizations of the same model. 
Snapshots are saved at $z = 0$, $0.5$, $1$, $2$, and $3$. 

The neutrino runs, spanning a range of neutrino masses, are performed using a particle-based implementation, 
where neutrinos are modeled as a collisionless and pressureless fluid, similar to CDM. 
Additionally, the forces at small scales for the neutrino species are accurately computed. 
Despite challenges related to shot noise, the particle-based approach automatically captures the full nonlinear clustering of neutrinos. 
This method is particularly suitable for high-redshift studies because it allows for precise quantification of the response of the power spectrum to variations 
in individual parameters and varying neutrino masses. It also enables disentangling the $M_{\nu}-\sigma_8$ degeneracy and 
accounting for the effects of baryons at small scales, as detailed in \cite{Rossi2017, Rossi2020}.
Note that the variety of techniques used in the literature to implement neutrinos 
reflects the fact that neutrinos can be 
treated either as a fluid or as an ensemble of particles. Their evolution can be described using linear theory or a 
full nonlinear treatment. While all of these complementary methods may be faster in terms of CPU time and production runs, 
the particle-based method accurately reproduces the nonlinear evolution at small scales, which is crucial 
for obtaining reliable and robust cosmological constraints.

In this study, we consider a subset of the \texttt{QUIJOTE} suite. 
Specifically, we use 100 independent realizations for each of the following cosmologies: the fiducial framework without massive neutrinos, 
and two massive neutrino models characterized by $M_{\nu}=0.1~{\rm eV}$ and $0.4~{\rm eV}$, respectively. 
The baseline massless neutrino cosmology is characterized by $A_{\rm s} = 2.13 \times 10^{-9}$, so that $\sigma_8(z=0) = 0.834$. 
For the massive neutrino models, the normalization is such that $\sigma_8$ is fixed at $z=0$ as in the fiducial cosmology 
(a convention termed `NORM', as previously discussed), implying that $A_{\rm s}$ varies with different neutrino masses. We utilize full snapshots at $z = 0, 1, 2$.

In this regard, it is interesting (and instructive) to compare results from the two simulation suites examined in our study (\texttt{MassiveNuS} versus \texttt{QUIJOTE}), 
as this allows us to assess the impact of systematics on the overall results due to differences in resolution, simulation codes, neutrino implementations, and normalization conventions 
(see Section \ref{sec_z_evo_systematics}). Moreover, the \texttt{QUIJOTE} suite enables proper assessment of cosmic variance, since we have sets of 100 realizations 
sharing the same cosmology but with different initial random seeds. The main characteristics of the simulations are reported in the last column of 
Table \ref{table_sims_details}, and further technical details can be found in \cite{VillaescusaNavarro2020}.

Figure \ref{fig_visualization_1c} shows three small cubic slices of side 100 $h^{-1}{\rm Mpc}$ 
with identical spatial location (center of the simulation boxes) at $z=0$, 
obtained from the \texttt{QUIJOTE} simulations having $M_{\nu} = 0.0~{\rm eV}$ (left, fiducial model), 
$M_{\nu} = 0.1~{\rm eV}$ (center), and $M_{\nu} = 0.4~{\rm eV}$ (right), respectively. 
In all of the panels, wall structures identified with the \texttt{DisPerSE} code are shown using a 
blue-to-red palette (with the corresponding density indicated by the right-side color bar). 
Superimposed filamentary skeleton structures are displayed, rendered with a dark-red-to-yellow density scale (bottom color bar). 
All structures are classified using the entire snapshot particles contained in the selected cube, 
with a persistence level corresponding to $\sigma = 3$ (see again Section \ref{sec_methods} for extensive details).
 

\section{Theoretical Background}     \label{sec_theory}
 
In this section, we outline the key concepts of \textit{discrete} Morse theory, persistent homology, and global topology 
within the context of cosmology. This part is intended to be rather pedagogical, ensuring the work is self-consistent:
an experienced reader familiar with these topics may skip it without compromising their understanding of the main results. 
We then briefly discuss the extension of this framework to massive neutrino cosmologies, 
with a particular focus on the persistent cosmic web filamentary network.

\subsection{Discrete Morse Theory and Persistence Homology: Overview and Cosmological Applications}

Discrete Morse theory \citep[for pioneering studies, see][]{Forman1998,Forman2002} and persistent homology, 
first introduced by \cite{Edelsbrunner2000}, \cite{Edelsbrunner2002}, and \cite{Robins1998,Robins2000}, are powerful branches 
of computational topology. These methods effectively address the challenges of applying traditional, mathematically rigorous 
Morse theory \citep[e.g.,][]{Milnor1963}, which was developed for idealized, well-defined, and smooth functions, to real-world data. 
By contrast, discrete Morse theory and persistent homology make it possible to handle noisy, 
irregular, and discrete datasets, making them particularly useful in a variety of practical applications.

In its original formulation, Morse theory provides a powerful method for capturing the complex relationship between 
the geometric and topological properties of a generic smooth and well-behaved function or field, by relying on its gradient and flow.
In Morse theory terms, ascending and descending $k$-manifolds partition the space into $k$-domains, which are defined by the gradient 
of the field and its network. The branching of these manifolds corresponds to intersections, and the \textit{critical points} are those where the 
gradient is null and the Hessian has non-null eigenvalues. These manifolds are further classified according to the order of critical points at their origin or destination.
Thus, Morse theory offers a compelling way to study the topological structure of smooth manifolds by examining the (non-degenerate) 
critical points of a Morse function defined on them. In particular, topological properties capture how points are connected, 
and these properties are invariant under smooth continuous transformations.
Central to Morse theory is the concept of the Morse complex, i.e., the ensemble of $k$-manifolds, with an embedded notion of hierarchy and neighborhood. 
A refinement of this concept is the Morse-Smale complex, which represents the transverse-only intersection of ascending 
and descending $k$-manifolds of a Morse-Smale function. In this way, the space is tessellated into regions called $p$-cells (or Morse cells), 
where all the integral lines (curves tangent to the gradient field at every point) have the same origin and destination. 
Thus, a Morse-Smale complex subdivides a generic space into subspaces of similar `flow'.
Morse theory has been widely applied in various scientific fields, particularly for data-intensive analyses.

However, as pointed out by \cite{Sousbie2011a}, Morse theory faces two critical issues:
(1) the overly idealized requirement of a smooth, twice-differentiable continuous function; and
(2) the presence of severe Poisson noise due to sparse sampling.
Both challenges make direct application to realistic datasets unfeasible.
Nevertheless, these challenges are addressed by {\it discrete} Morse theory and 
persistent homology, as we briefly explain next.

As opposed to classical Morse theory, its \textit{discrete} counterpart applies to intrinsically \textit{discrete} 
functions defined over simplicial complexes---see the seminal 
works of \cite{Gyulassy2008} and \cite{Zomorodian2009} for extensive details. 
The building blocks of {\it discrete} Morse theory are $k$-simplices, the simplest 
geometric objects with $k+1$ vertices, denoted as $\alpha_{\rm k}$ in our notation.
In this framework, a \textit{discrete} Morse function $f$ associates a real value $f(\alpha_{\rm k})$ 
to each simplex $\alpha_{\rm k}$, which belongs 
to the simplicial complex $K$, i.e., the set of $k$-simplices such that if a $k$-simplex $\alpha_{\rm k}$ 
belongs to $K$, all of its faces also belong to $K$. 
The classical concept of gradient is replaced by its discrete counterpart, where integral lines that 
define ascending/descending $k$-manifolds are substituted 
by $V$-paths, and critical points of order $k$ become critical $k$-simplices. Moreover, tracing a $V$-path simply 
consists of following the direction indicated by gradient pairs.
The Hessian non-degeneracy condition in smooth Morse theory is translated 
into a condition on the values of the function in its \textit{discrete} counterpart. 
A key consequence of this is the appearance of an asymmetry (unique to \textit{discrete} theory), 
since the opposite of a \textit{discrete} Morse function $f$ may not itself be Morse. 
In three-dimensional space, this occurs because minima and maxima correspond to 
vertices and tetrahedrons, respectively, rather than simple points in space.
However, this challenge is addressed by defining $-f$ over the dual complex. 
The \textit{discrete} function $-f$, which assigns $-f(\alpha_{\rm k})$ 
to the dual cell of each simplex, is Morse. 
In this formalism, the notion of the {\it discrete} Morse complex plays a central role. 
This is further refined by the {\it discrete} Morse-Smale complex, which consists 
of the ensemble of \textit{discrete} {\it extended} ascending/descending $k$-manifolds, 
with the space divided into a cell-complex of \textit{discrete} $p$-cells. 
These $p$-cells define a natural partition induced by the gradient flow, 
where all of the $V$-paths have the same origin and destination.
Notably, $0$-cells are simply points, $1$-cells are \textit{arcs} (backbones of filaments), 
$2$-cells are called \textit{quads}, and $3$-cells are termed \textit{crystals}.

The challenge of Poisson noise that arises from sparse sampling---a particularly critical issue for massive neutrino 
implementations in $N$-body codes---is addressed by persistence homology. Originally described in the context of simplicial homology 
for functions defined over a simplicial complex (see the seminal work of \citeauthor{Edelsbrunner2002} \citeyear{Edelsbrunner2002} for details), 
persistence provides a methodology to quantify the importance and robustness of the topological components of a space in the presence of noise. 
It allows one to remove spurious topological features. This \textit{filtration} process, referred to as 
\textit{topological simplification}, ensures that persistence 
pairs represent genuine topological components. Features with high persistence are stable, long-lived, and prominent, 
while features with low persistence are short-lived or transient, often arising due to noise.
In this view, persistence represents a general concept independent of the smoothness or discreteness of a specific function. 

More formally, for smooth functions, persistence is based on the evolving properties of their sublevel sets (or equivalently, their excursion sets) 
as they vary with the threshold value. Persistence threshold levels are generally expressed in terms of the robustness of the topological features 
(i.e., $k$-cycles) to noise and are specified as the number of $\sigma$ relative to the noise level. 
Hence, persistence measures the lifetime (creation and destruction) of the corresponding topological features in the 
sublevel/excursion sets.\footnote{Topological features are referred to as \textit{$k$-cycles}. 
Specifically, \textit{$0$-cycles} correspond to individual features or isolated islands; \textit{$1$-cycles} are rings with holes in the middle; 
\textit{$2$-cycles} represent three-dimensional empty regions.}
In its discrete version, persistence measures the lifetime of $k$-cycles in a \textit{filtration} of a finite simplicial complex $K$ (or tessellation) 
induced by a discrete function. The `arrival time' of each critical simplex corresponds to the creation or destruction of a topological feature. 
Therefore, persistence describes how much a function must change to remove a specific $k$-cycle \citep[e.g.,][]{Sousbie2011a}.

A crucial aspect, of central importance to our study, is that persistence can always be used to recover the actual topology and Morse (or Morse-Smale) 
complex of an underlying function, even in the presence of severe noise. It allows for the local cancellation of non-persistence pairs. 
Furthermore, persistence enables the assessment of the significance of critical points and is directly related to 
critical events---pairs of critical points with vanishing persistence. We discuss this interesting dual aspect in a follow-up publication.
See also Section \ref{sec_homtop_z0_persistence}, where we present persistence diagrams that contain much richer and 
more profound information than typical summary statistics, as they show features across all densities. This is because persistence facilitates 
the assessment of the multiscale nature of topological features of distinct dimensions at different epochs, 
due to the fundamental non-local character of topological measurements.

The powerful mathematical framework described above, in its \textit{discrete} version, is readily applicable to various 
scientific fields requiring data-intensive analyses, and in particular to the study and characterization of the cosmic web in cosmology.
If the large-scale matter density field can be approximated as a Morse-Smale function, then each critical point of the 
density field corresponds to a topological feature of the cosmic web, with its geometry described by an ascending or descending manifold. 
The arcs define a hierarchical neighborhood relation among these features, a property 
known as the \textit{combinatorial property} (e.g., number of nodes, branches, cycles, etc.)---see again \cite{Sousbie2011a} for extensive details.
At the core of this approach lies the idea that all salient features of the cosmic web have a direct, 
mathematically well-defined equivalent in \textit{discrete} Morse theory, namely:
\begin{itemize}
\item Ascending 3-manifolds trace voids, corresponding to critical points of order 0 (minima).
\item Ascending 2-manifolds trace walls, corresponding to critical points of order 1 (wall-like saddles).
\item Ascending 1-manifolds trace filaments, corresponding to critical points of order 2 (filament-like saddles).
\item Ascending 0-manifolds (or equivalently, descending 3-manifolds, the dual of voids) trace peak patches, 
clusters, or halos, corresponding to critical points of order 3 (maxima).
\end{itemize}
The tessellation of space is achieved via the scale-adaptive and parameter-free \textit{Delaunay Tessellation Field Estimator} (DTFE), 
described in the next section. DTFE allows for the classification of points (0-simplices), lines (1-simplices), surfaces (2-simplices), 
and volumes (3-simplices), ultimately enabling the computation of the Morse-Smale complex.
In particular, filaments are composed of two arcs joining two maxima through a tube-like saddle point, 
while `void filaments' (or `anti-filaments') link two minima via a wall-like saddle. 
The upper skeleton traces the filaments. Additionally, the filamentary network, along with its associated voids, 
walls, and peaks, is simplified by canceling pairs of critical points based on a persistence criterion 
recast in terms of significance relative to shot noise.
In this framework, high-persistence points are particularly significant, as they trace the most prominent 
features of the cosmic web (clusters, filaments, and voids). Furthermore, a key advantage of this approach is its ability 
to capture the structural nested hierarchy of the cosmic web, allowing one to study how its components connect as the level set varies. 
In fact, the corresponding changes in topology provide a highly informative description of web connectivity.
Persistence, in this context, relates the birth (creation) of topological features such as holes in the mass distribution 
to their eventual annihilation (death) as the level set evolves. Thus, differences in the shape and morphology of persistence 
diagrams reflect deeper variations in the multiscale structure and hierarchical evolution of the structural components 
of the cosmic web. For further insights, see \cite{Cautun2014},  \cite{Feldbrugge2019},
\cite{Pranav2019}, \cite{Pranav2021}, \cite{Wilding2021}, \cite{Calles2025},
and our Section \ref{sec_homtop_z0_persistence}.

\subsection{Discrete Morse Theory and Persistence Homology in Massive Neutrino Cosmologies}

Massive neutrinos alter the main constituents of the cosmic web---halos, walls,  filaments, and voids---across various 
scales and dynamical regimes, 
impacting LSS observables in different ways.
Traditional approaches to constraining neutrino mass typically rely on individual tracers
(e.g., galaxies, clusters, voids) at specific scales, 
using summary statistics  
to quantify neutrino-induced modifications and indirectly infer their mass.
In contrast, our topological framework---particularly through persistence diagrams---provides \textit{simultaneous} 
access to multiscale information across all tracers and epochs. 
This enables a more comprehensive view of how massive neutrinos shape cosmic structures, 
leveraging their \textit{combined} imprints rather than isolated effects. 
Notably, recent work \citep{Rossi2022} demonstrated, for the first time, the utility of persistent homology 
in massive neutrino cosmologies.
To this end, in what follows we briefly 
review the impact of massive neutrinos on cosmic structures at both linear and 
nonlinear levels in relation to conventional statistical methods, and highlight selected 
key insights gained from \textit{discrete} Morse theory and topological 
analyses---topics that will be explored in detail in the following sections.

At the linear level, the effects of neutrino mass and the impact of their free-streaming on the 
evolution of cosmological perturbations are well 
understood \citep[i.e.,][]{LesgourguesPastor2006,Rossi2017}. In essence, massive neutrinos 
alter the evolution of other perturbations (baryons, CDM, and photons) through two main mechanisms: 
their homogeneous density and pressure, which impact the Friedmann equations, 
and their direct gravitational backreaction, which modifies the evolution of metric perturbations.
During matter- and $\Lambda$-domination   neutrino free-streaming suppresses the growth of small-scale 
matter perturbations by damping fluctuations and preventing gravitational clustering, while still affecting the homogeneous expansion. 
The net result is a delay in radiation-matter equality: 
the linear total matter power spectrum $P(k)$ 
is enhanced only at small wave numbers $k$ (i.e., large scales), while small-scale fluctuations are suppressed.
The impact on the linear angular power spectrum is less intuitive, as it primarily reflects the physical evolution before recombination, 
when neutrinos were ultrarelativistic. This influence manifests at the background level, where a delayed matter-radiation equality 
leads to an enhancement of higher CMB peaks, particularly the first and third ones. 
 
At the nonlinear level, the effects of massive neutrinos on cosmic structures are more intricate and remain less explored, 
particularly at small scales, where baryonic physics introduces additional complexities. 
In this regime, linear theory is insufficient, especially when baryonic processes become significant.
High-resolution numerical simulations that incorporate massive neutrinos---ideally coupled with full 
hydrodynamical treatments, such as those in the \texttt{Sejong Suite} \citep[][]{Rossi2020}---currently provide the most effective approach for 
quantifying neutrino mass signatures in this regime. For a detailed discussion of their impact on conventional statistics, 
including the small-scale nonlinear 3D matter power spectrum
 $P_{\rm t}(k,z)$ and  Lyman-$\alpha$ forest observables, see \cite{Rossi2017,Rossi2020}.
Notably, our findings indicate that massive neutrinos induce a nontrivial, scale-dependent distortion of the total matter power spectrum shape, 
accompanied by a combined evolution in the $P_{\rm t}(k,z)$ amplitude when compared to a baseline 
$\Lambda$CDM cosmology. This manifests as the characteristic \textit{spoon-like} effect, where the suppression 
of the power spectrum increases with redshift, revealing significant small-scale deviations from linear theory.
A similar trend is observed in the small-scale flux power spectrum, a key observable for probing structure formation and growth, 
as well as for constraining cosmological parameters and the neutrino mass through its full shape, amplitude, and tomographic evolution. 
Furthermore, we have provided a halo model interpretation of these effects and argued that the amplitude and position of the 
maximal suppression in the \textit{spoon-like} feature define a characteristic nonlinear scale. 
This scale has the potential to serve as a diagnostic tool for determining or constraining neutrino properties using LSS observables.

The previous results on neutrino effects---both linear and nonlinear---on conventional statistics
serve as valuable guidance throughout this paper, providing physical intuition within the novel framework we propose. 
However, our Morse-based topological approach offers a significantly broader and more insightful perspective, as it is not restricted to 
a single scale or individual tracers -- see also \cite{MoonRossiYu2023} for a related complementary analysis of critical points in massive neutrino cosmologies.
In particular, this framework allows us to systematically map and quantify the fingerprints of massive neutrinos on the filamentary network and cosmic web 
skeleton as structures form and evolve. In this study, our approach is primarily focused on numerical simulations and we do not attempt a first-principles 
theoretical modeling of \textit{discrete} Morse theory and persistent homology in the presence of massive neutrinos: 
a more theoretically oriented exploration of these effects will be pursued in future work.

Nevertheless, in Section \ref{sec_homtop_results}, we extract a number of key physical insights---especially from persistence diagrams---
that offer a deeper understanding of the \textit{multiscale} impact of massive neutrinos on individual cosmic web constituents. 
In particular, we anticipate that massive neutrinos alter the connectivity of the cosmic web across different redshifts and scales, 
leading to fundamental changes in global topology---changes that cannot be fully captured by power spectra or correlation functions alone.
Moreover, massive neutrinos break the notion of topological self-similarity, and this disruption could serve as a potential smoking gun for the direct detection of neutrino mass.
 

\section{Methodology}       \label{sec_methods}
 
In this section, we briefly outline our methods, key algorithms, and pipeline, based on 
the computation of the \textit{discrete} Morse-Smale complex combined with 
topological simplification via persistence.

\subsection{Methods and Key Algorithms}  

Our primary goal is to efficiently compute the discrete Morse-Smale complex (i.e., the network of critical points connected by $p$-cells) 
of the density field and use its properties to identify and characterize the structures of the cosmic web in both massless and massive neutrino cosmologies. 
The Morse-Smale complex captures the intricate relationship between the gradient (or gradient pairs) of the underlying density field, its topology,
and the topology of the manifold on which it is defined. It induces a tessellation of space into $p$-cells, where all integral lines have the same origin and destination, 
with critical points as extremities. The Morse-Smale complex is computed directly from the Delaunay tessellation, which is the dual graph of the Voronoi diagram. 
We exploit the Delaunay Tessellation Field Estimator (DTFE) for this task and follow the heuristic methodology and core algorithms introduced 
by \cite{Sousbie2011a} to obtain the discrete Morse-Smale complex.

We then use persistent homology to assess the significance of the topological structures and remove low-significance 
persistence pairs through a filtration/simplification process. The filtration process functions as a scale-adaptive filter, 
where the scale is determined solely by the local topology, allowing less robust topological features to be systematically removed. 
Importantly, there is no need to recompute the Morse-Smale complex from scratch; instead, we only determine how it is modified 
by canceling low-persistence pairs. A key aspect of this process is that the original resolution of the network is preserved, even after persistence-based simplification.

At the end of this process, we obtain the discrete Morse-Smale complex as a function of the persistence level, expressed in units of $\sigma$. 
Within a specific cosmological scenario, we can thus confidently identify real cosmic structures, i.e., voids, walls, filaments, and peak patches or halos. 
See Appendix \ref{sec_appendix_A} for more details, and the seminal work of \cite{Sousbie2011a} for extensive information.
We note that in a previous directly related work \citep{MoonRossiYu2023}, we instead adopted a complementary 
technique, referred to as the  `\textit{density-threshold-based}' approach in configuration space, to
construct density fields from the output of $N$-body simulations and extract and classify critical points---faithful tracers of
cosmological structures.

\begin{figure}
\centering
\includegraphics[angle=0,width=0.48\textwidth]{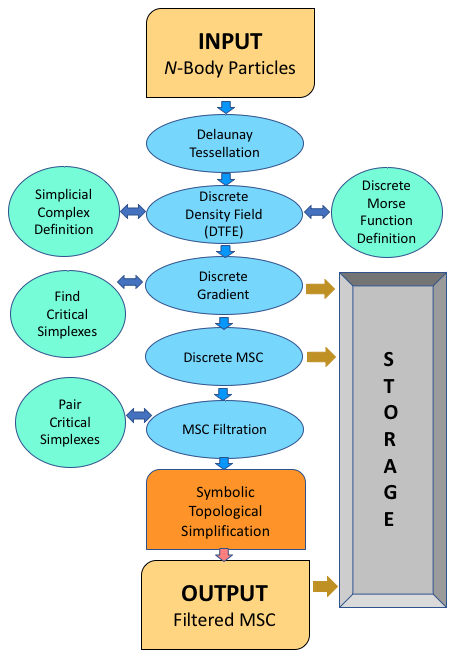} 
\caption{Schematic view of our pipeline for computing the filtered discrete Morse-Smale complex, 
starting from a subset of $N$-body particles. See the main text and Appendix \ref{sec_appendix_A} for more details.}
\label{fig_pipeline}
\end{figure}

\begin{figure*}
\centering
\includegraphics[angle=0,width=0.95\textwidth]{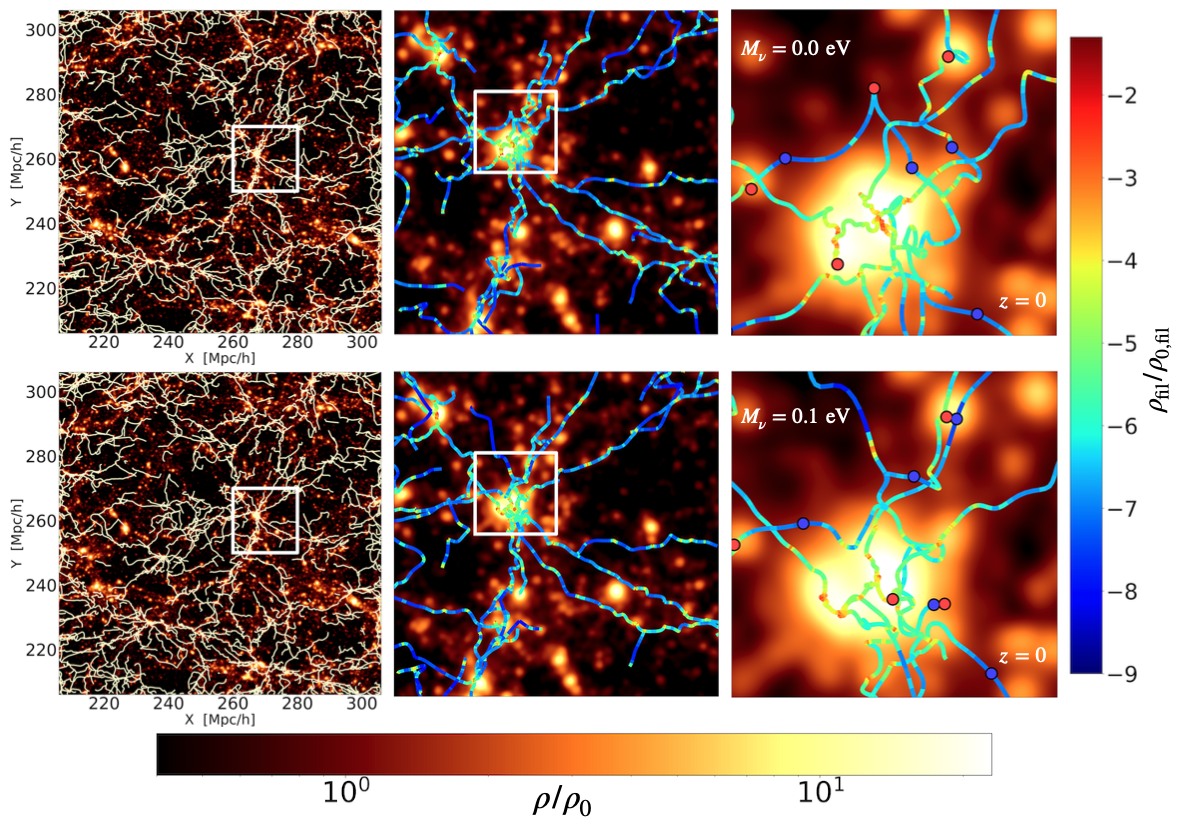}
\caption{Examples of upskeletons at $z=0$ from \texttt{DisPerSE} applied to selected \texttt{MassiveNuS} simulations. 
Top panels show a massless neutrino cosmology, while bottom panels depict a scenario with $M_{\nu}=0.1$ eV. 
Left panels display 2D density fields (in $100 \times 100h^{-1}{\rm Mpc}$ projections) with the density scale indicated in the bottom color bar, 
and the upskeletons traced in grey.
Insets show progressive zoom-ins, with upskeleton arcs now colored by density (right-side color bar). 
Critical points of order 3 (red; maxima) and 2 (blue; saddle-2) are also highlighted in the left panels. 
The figure illustrates the stark differences in filamentary structure and critical point distribution between massless and massive 
neutrino cosmologies, demonstrating the utility of filament-based statistics in probing tiny neutrino mass effects.}
\label{fig_visualization_1d}
\end{figure*}

\subsection{Pipeline: Details}  

Our pipeline is schematically shown in Figure \ref{fig_pipeline}
and can be summarized by the following steps:

\begin{enumerate}
\item Computation of the Delaunay tessellation from a subset of $N$-body simulation particles in a given cosmology.
\item Definition of the simplicial complex $K$ from the Delaunay tessellation.  
\item Computation of the discrete density field using the DTFE prescription. 
\item Heuristic definition of the discrete Morse function over $K$ via density assignment to its elements $\{\alpha_{\rm k}\}$.
\item Computation of the discrete gradient.
\item Determination of critical simplexes. 
\item Deduction of the discrete Morse-Smale complex. 
\item Filtration of the Morse-Smale complex according to the values of the discrete Morse function in terms of $\sigma$-levels. 
\item Pairing of critical simplexes into persistence pairs. 
\item Symbolic persistence-based topological simplification of the discrete Morse-Smale complex.
\item Output of the final \textit{filtered} discrete Morse-Smale complex, as a function of the persistence level. 
\end{enumerate} 
 
To perform most of the tasks above, we use the Discrete Persistent Structure Extractor \citep[\texttt{DisPerSE};][]{Sousbie2011a} code, 
optimized for our computing system and augmented by a number of customized Python scripts to interface \texttt{DisPerSE} with our 
computational architecture,
extract selected data, and analyze the various outputs.\footnote{We have efficiently carried out all of the 
calculations using our SJCOSMO cluster at Sejong University, which is composed of a 
Xeon Silver 4114 master node architecture and four Xeon Gold 6126 computing nodes.} 
A key asset of \texttt{DisPerSE} is that it keeps track of the relative connectivity, 
relying on just a single tunable parameter---namely the significance of the retained features, expressed in $\sigma$ units in terms of robustness with respect to noise.
Moreover, there is no requirement for any pre-treatment of the density field (i.e., smoothing, manipulation, etc.), nor a particular scale involved. 
In fact, \texttt{DisPerSE} preserves the geometry and connectivity of the cosmic web, even with a very sparse sample. 
Ultimately, the optimal resolution of the filamentary cosmic web is simply dictated by the initial particle distribution.
Since we are dealing with $N$-body simulations, in this work we always adopt periodic boundary conditions. 
Otherwise, special care would need to be taken with boundaries during the topological simplification process, 
as the persistence of critical pairs formed with a boundary simplex generally contains 
spurious persistence ratios due to the \texttt{DisPerSE} compactification procedure.
We also note that the discrete Morse-Smale complex is always computed at the sampling resolution limit of the 
simulation in question; therefore, there is no information loss, even after topological simplification.
A final note on smoothing, in relation to filament-based statistics that we present in the next section: in \texttt{DisPerSE}, smoothing is performed at the level of structure identification. 
Specifically, filaments are individually smoothed by fixing the critical points and averaging the position of 
each non-fixed segment's endpoint with the position of its closest neighboring endpoints a specified number of times.


\section{Topology and Homology in Massive Neutrino Cosmologies}     \label{sec_homtop_results}

This section presents the main results of our analysis. We begin by discussing 
critical points, skeleton structures, 
and the impact of massive neutrinos. We then examine the filamentary statistics of the 
cosmic web and their dependence on neutrino mass. 
Next, we analyze persistence diagrams in massive neutrino cosmologies to assess topological differences. 
Finally, we explore the effects of redshift evolution and potential systematics 
arising from distinct simulation setups and neutrino implementations.

\subsection{Critical Points and Skeleton Structures} 

Critical points, a subset of special points in position-smoothing space, trace cosmological structures and 
have recently gained increased attention for their peculiar role in representing the essence of the cosmic web 
\citep{Cadiou2020,Shim2021,Kraljic2022} and their remarkable potential for revealing subtle 
neutrino mass signatures \citep{Rossi2022, MoonRossiYu2023}. 
These points exhibit significant 
topological properties, offering a \textit{multiscale} perspective on the filamentary network, and can be 
associated with physical structures of the same type---with their size depending on the smoothing scale. 
Robust to systematic effects, 
critical points efficiently compress 3D density 
field information, highlighting its most salient features---crucial for cosmology and 
galaxy formation. Remarkably, the topology of the 
initial density field is encoded in the positions and heights of critical points at a fixed smoothing scale, 
allowing their tomographic characterization across redshift slices to enable predictions of cosmic web evolution. 
Moreover, it is precisely their drift with smoothing that defines the skeleton tree, capturing topological 
variations that evolve with increasing smoothing scale \citep{Sousbie2008,Pogosyan2009,Gay2012},
and their merging corresponds to the merging of halos, walls, filaments, and voids \citep{Cadiou2020}. 

Notably, in \cite{MoonRossiYu2023}, a study fully dedicated to critical points  in massive neutrino cosmologies,
we presented for the first time  
their clustering statistics as a function of neutrino mass, including cross-correlations, and quantified their imprints on the key web's constituents. 
In particular, we highlighted their potential to reveal tiny neutrino mass effects through their combined impact on BAO peak amplitudes 
in all critical point correlation functions above and below the rarity threshold, as well as the positions of their corresponding inflection points at large scales.

The characterization of critical points is central to the methodology adopted in this work, 
as they serve as the building blocks of discrete Morse theory and enable the construction of the discrete Morse-Smale 
complex (as discussed in Section \ref{sec_theory}), alongside the assessment of their persistence. 
In fact, according to Morse theory, the topology of a field is intrinsically linked to the presence, location, and nature of its singularities: 
namely, the topology of a manifold changes when a singularity is introduced or removed as the level set varies. 
Consequently, the existence and connectivity of topological features are \textit{entirely} determined by the location 
and nature of critical points in the density field.  
To illustrate this, Figure \ref{fig_visualization_1d} presents an example of the upskeleton 
 at $z=0$, obtained by running \texttt{DisPerSE} on our first set of 
$N$-body simulations (\texttt{MassiveNuS}). The top panels correspond to a massless neutrino cosmology, 
while the bottom panels show the scenario with $M_{\nu}=0.1$ eV. 
Specifically, the left panels display 2D density fields in a box of thickness 
of $100 \times 100h^{-1}{\rm Mpc}$, with the bottom color bar indicating the density field scale, rendered using a dark-red-to-yellow color palette. 
The corresponding upskeleton is shown in grey in both panels.
The white insets represent progressive zoom-ins into the same spatial location, shown in full in the middle panels. 
Similarly, white insets in the middle panels zoom into the same spatial location, shown in full in the right panels. 
In the central and right panels, the enlarged upskeletons are
colored according to the density in the arcs, 
using a blue-to-red palette (as shown in the right-side color bar).
The left panels also highlight critical points of order 3 (red; maxima) and order 2 (blue; saddle-2). 
This figure is generated using a 
4\% subsampling of the  $N$-body particles in the corresponding cosmologies. 
Note that filaments connect two maxima through saddle-2 types and are traced by skeleton structures. 
Even visually, one can clearly observe how the upskeleton arcs map the underlying density field accurately, and notice 
the stark changes in filamentary structure and critical point distribution between the massless neutrino cosmology (top panels) 
and the massive neutrino scenario with $M_{\nu}=0.1$ eV (bottom panels).
While the underlying density fields show no appreciable visual variations (i.e., background maps in all panels),
differences in terms of critical points, connectivity, skeleton structures, and filaments between the two distinct cosmological 
frameworks are remarkable. This highlights why filament-based statistics are ideal for characterizing neutrino mass effects, even in
the presence of very small neutrino masses (Sections \ref{sec_filament_stats} and \ref{sec_z_evo_systematics}).

\begin{figure}
\centering
\includegraphics[angle=0,width=0.45\textwidth]{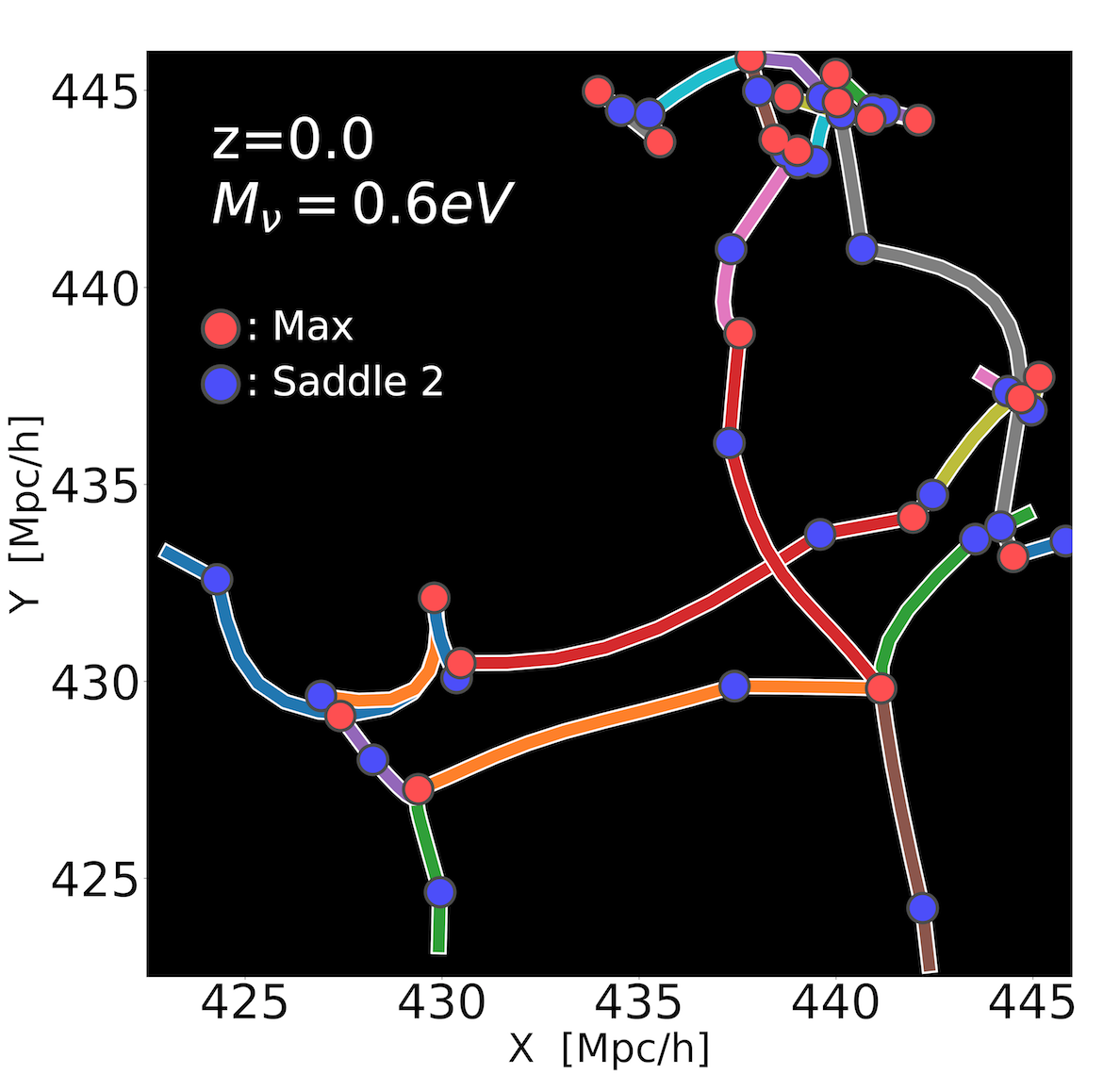}
\caption{A $25 \times 25h^{-1}{\rm Mpc}$ 2D slice at $z=0$ from a $M_{\nu}=0.6$ eV cosmology in the \texttt{MassiveNuS} suite, 
illustrating the definition of \texttt{DisPerSE} filaments. Filaments are derived from the upskeleton, where arcs connect saddle-2 points and maxima. 
Each filament consists of two arcs sharing a common saddle-2 point. Individual filaments are shown in distinct colors for clarity.}
\label{fig_visualization_1e}
\end{figure}

As a further example, Figure \ref{fig_visualization_1e} shows a  $25 \times 25h^{-1}{\rm Mpc}$ 2D slice at $z=0$
from a $M_{\nu}=0.6$ eV cosmology  in the \texttt{MassiveNuS} suite, illustrating how \texttt{DisPerSE} filaments are defined. 
Specifically, filaments are derived from the upskeleton, where arcs (a subset of the ascending 1-manifolds) connect saddle-2 points and maxima. 
Each filament consists of two arcs sharing a common saddle-2 point. In the figure, individual 
filaments are represented in distinct colors for clarity. Note that arcs can only intersect at critical points,
and they define how critical points are connected by integral lines (or V-paths).

Next, we focus on filament-based statistics in massive neutrino scenarios.

\begin{figure*}
\centering
\includegraphics[angle=0,width=0.85\textwidth]{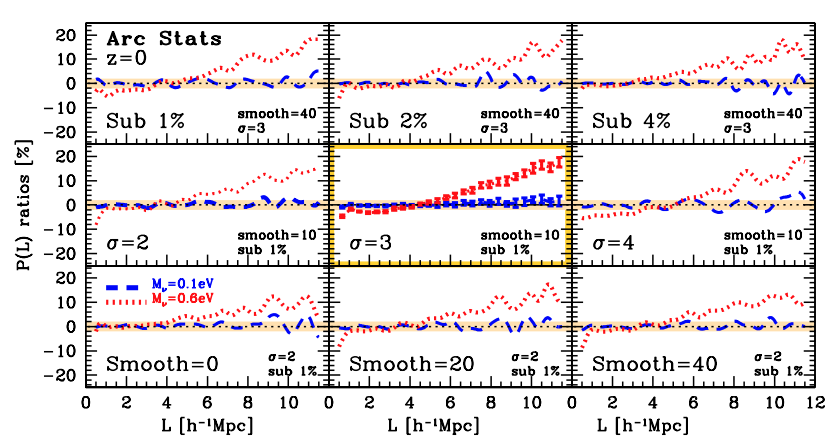}
\caption{
Summary of tests assessing the impact of subsampling, smoothing, and persistence thresholds on arc statistics in the \texttt{MassiveNuS} simulations at $z=0$. 
Each panel shows the ratio of arc lengths in $M_{\nu}=0.1$ eV (dashed blue lines) and $M_{\nu}=0.6$ eV (dotted red lines) cosmologies relative to the massless neutrino scenario, 
with the shaded light brown regions representing a $\pm  2\%$ variation. 
These tests reveal the influence of each parameter on arc statistics, guiding the optimal choices for our study. The central middle panel (highlighted in yellow) represents 
the reference setup for subsequent filament statistics and persistence diagram analyses at $z=0$, with error bars computed 
from subsampling uncertainty. The effect of nonzero $M_{\nu}$ on arc statistics is evident, 
including a notable `transition' scale around $4h^{-1}{\rm Mpc}$, which appears nearly independent of $M_{\nu}$. See the main text for further details.}
\label{fig_arc_length_stats_ratios}
\end{figure*}

\subsection{Filamentary Statistics of the Cosmic Web: Effects of Massive Neutrinos}       \label{sec_filament_stats}

We characterize filamentary statistics in massive neutrino cosmologies by first examining
the $z=0$ case to assess the impact of key analysis choices: 
(1) \textit{subsampling}, (2) \textit{smoothing}, and (3) \textit{persistence threshold}.
We focus here on the \texttt{MassiveNuS} simulation set,  
which offers the highest resolution, while the second series (\texttt{QUIJOTE}) is more appropriate for statistical analyses.
Redshift evolution and systematic effects---such as normalization conventions, massive neutrino implementations, 
and simulation specifics---are addressed in Section \ref{sec_z_evo_systematics}, where  
we consider both simulation suites.

Our topology-based technique relies on computing the Delaunay tessellation from 
a subset of $N$-body simulation particles in a given cosmology.
Since using the full particle set is computationally unfeasible, \textit{subsampling} is a 
necessary first step in our pipeline.
It is therefore crucial to assess its impact, ensuring that subsampling preserves the key characteristics 
of the cosmic web without significant information loss.
To this end, we performed a series of tests. 
Notably, \cite{Sousbie2011b} applied subsampling as severe as $0.2\%$ in terms of the total number of particles (see their Figure 4), 
demonstrating that \texttt{DisPerSE} can effectively capture the main features of the cosmic 
web even with an extremely sparse sample.
An alternative approach, explored in a companion study (Yu et al., in preparation) and tailored for applications towards
large-scale observational surveys, 
involves starting from halo catalogs rather than $N$-body particles.
In such case, the full dataset can be used without the need for subsampling.
 
Moreover, a key advantage of our topology-based technique is that it does not require any 
pre-processing of the density field, such as \textit{smoothing} or resampling, nor does it rely on a predefined scale.
In contrast, our complementary study \citep{MoonRossiYu2023} employed a `\textit{density-threshold-based}' approach, 
which necessitates smoothing at the density field level.
Here, smoothing is applied instead at the stage of structure identification.\footnote{In  \texttt{DisPerSE}, the \textit{smooth} 
option applies geometric smoothing to the extracted skeleton by iteratively averaging vertex positions along filaments.}
Specifically, in  \texttt{DisPerSE} the skeleton is smoothed $N$ times, with the critical points (nodes) fixed in place. 
Each filament is then smoothed by averaging the coordinates of each point along the filament with those of its neighboring points. 
After $N$ iterations, filaments are smoothed over $N$ sampling points, ensuring a gradual and controlled adjustment of the structure.
Hence, this process affects only the visual appearance of the skeleton and does not alter its underlying topology or persistence.

Regarding the selection of a \textit{persistence threshold}, this step is crucial as it helps eliminate persistence pairs that are likely to be induced by noise. 
The elimination is based on a confidence level threshold, expressed in units of $\sigma$, similar to the Gaussian case: 
persistence levels of 1, 2, 3, and 4 $\sigma$ correspond to 
probabilities of 0.68, 0.95, 0.997, and 0.99937, respectively, and so on.

\begin{figure*}
\centering
\includegraphics[angle=0,width=0.40\textwidth]{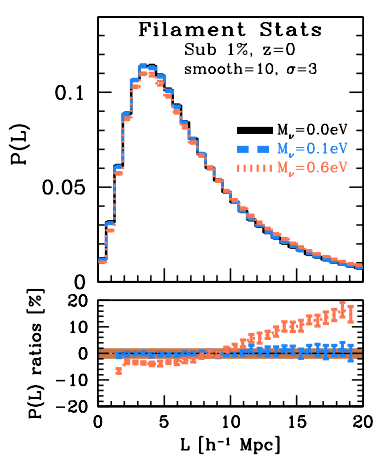}        
\includegraphics[angle=0,width=0.50\textwidth]{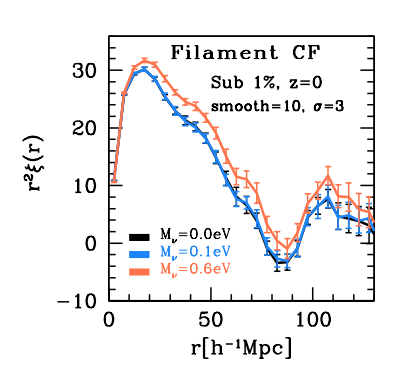}
\caption{[Top Left] Filament length distribution $P(L)$ at $z=0$
for massless ($M_{\nu}=0$ eV, black solid line) and massive neutrino cosmologies ($M_{\nu}=0.1$ eV, dashed light blue; 
$M_{\nu}=0.6$ eV, dotted orange). 
[Bottom Left] Ratios relative to the massless scenario, with the shaded light brown region indicating a 
$\pm 2\%$ variation. A transition scale around $10h^{-1}{\rm Mpc}$, nearly independent of $M_{\nu}$,
is observed, alongside a neutrino-mass-dependent trend featuring an excess of longer filaments and a deficit of shorter ones 
(adopting the `UNNORM' convention). 
Error bars reflect subsampling uncertainties. Variations in filament lengths as a function of $M_{\nu}$ 
are comparable to those observed in the arc statistics (i.e., Figure \ref{fig_arc_length_stats_ratios}).
[Right] Corresponding filament auto-correlation functions at $z=0$ for the same cosmologies shown in the left panel, 
using an identical color scheme and a persistence threshold $\sigma=3$. Filament-like saddle points (i.e., critical points of order 2, tracing filaments) 
are identified with the \texttt{DisPerSE} algorithm. Error bars 
represent $1\sigma$ variations from subsampling. The BAO peak is distinctly visible, with 
higher amplitudes associated with larger neutrino masses, as first noted in \cite{MoonRossiYu2023}.
Given their reduced nonlinearity, saddle-point statistics are particularly promising for detecting subtle neutrino mass effects.}
\label{fig_filaments_stats}
\end{figure*}

Figure \ref{fig_arc_length_stats_ratios}
summarizes the tests  
conducted to assess the impact of subsampling, smoothing, and persistence threshold on arc statistics 
in the \texttt{MassiveNuS} simulations at $z=0$. 
Specifically, each panel shows the ratio of arc lengths (with lengths expressed in units of $h^{-1}{\rm Mpc}$) measured in a $M_{\nu}=0.1$ eV cosmology 
(dashed blue lines) and a $M_{\nu}=0.6$ eV framework (dotted red lines) 
relative to the corresponding massless neutrino scenario.
The shaded light brown region in each panel represents a $\pm 2\%$ variation.
We consider subsampling levels of $1\%$, $2\%$, and $4\%$, persistence thresholds of $\sigma=2$, $3$, and $4$, 
and smoothing values of $0$ (i.e., no smoothing), $20$, and $40$. 
From left to right, the top panels vary subsampling from $1\%$ to $4\%$, while keeping smoothing fixed at $40$ and the persistence threshold at $\sigma=3$. 
The central panels vary the persistence threshold from $\sigma=2$ to $\sigma=4$, while  maintaining a fixed smoothing of $10$ and a subsampling level of $1\%$. 
The bottom panels explore the effect of \texttt{DisPerSE} smoothing, increasing it from $0$ to $40$ with the persistence threshold set at $\sigma=2$ and subsampling at $1\%$. 
This setup allows for a direct comparison of the effects of each parameter on arc statistics.
 
Based on these tests, we infer that subsampling has a minor effect on arc statistics, 
hence a 1\% subsampling rate is sufficient for our $z=0$ investigations in the \texttt{MassiveNuS} simulations. 
However, the optimal choice depends on the specific simulation setup, and in general increased subsampling is preferable.  
In Section  \ref{sec_z_evo_systematics}, 
we adopt a higher 
subsampling rate of 2\% for \texttt{MassiveNuS} when evaluating redshift effects, and a 16\% 
subsampling rate for the \texttt{QUIJOTE} suite, given its lower resolution.
Regarding smoothing, its effect becomes significant only at larger values (e.g., 40 or more) due to the 
intrinsic nature of smoothing implemented in \texttt{DisPerSE}, as previously discussed (i.e., purely aesthetic or numerical in purpose, 
affecting only the appearance of the skeleton, not its topology). 
To this end, a smoothing value of $10$ is consistently applied in our study.
As for the persistence threshold, this is topologically relevant: a very low persistence is likely to include mostly noise, while a very 
high threshold may lead to the omission of key structures. 
This effect is already evident when comparing the middle panels, as we move from $\sigma=2$ to $\sigma=4$. 
For our study at $z=0$, we adopt a persistence threshold of $\sigma=3$. 
However, for redshift evolution tests (Section \ref{sec_z_evo_systematics}), we use a 
more conservative value of $\sigma=4$, as noise becomes more pronounced at higher redshifts, 
with structures still in the process of formation.

Besides providing insight into the impact of key analysis choices, 
Figure \ref{fig_arc_length_stats_ratios} highlights several important 
effects. First, regardless of the presence of massive neutrinos, higher persistence thresholds (e.g., $\sigma=4$, middle right panel) 
generally result in longer arcs. 
This trend is expected, as shorter arcs are typically attributed to noise, which is mitigated by increasing the persistence threshold.
Additionally, the effect of nonzero $M_{\nu}$ on arc statistics is already evident. 
This behavior is characteristic of massive neutrinos, which slow down perturbations and free-stream below a distinctive scale, determined by their total mass 
and the background expansion.
Consequently, one would expect an excess of longer arcs and a deficit of smaller arcs in massive neutrino cosmologies, 
provided the $\sigma_{8}$ normalization convention ensures that $A_{\rm s}$ remains fixed across all cosmologies, as in the \texttt{MassiveNuS} set. 
Under the `NORM' convention (as in the \texttt{QUIJOTE} suite), the opposite trend would be observed.
Interestingly, the `transition' scale or crossing point in the panels, around $4h^{-1}{\rm Mpc}$, appears almost independent of
the value of $M_{\nu}$, while the overall abundance of short versus long arcs varies with neutrino mass.  
Variations are within $2\%$ when $M_{\nu}=0.1$ eV,  but can reach up to $\sim 15\%$ when $M_{\nu}=0.6$ eV.
These effects become  more pronounced at higher redshift, as we show in Section  \ref{sec_z_evo_systematics}.

Given the previous considerations, the central middle panel of Figure \ref{fig_arc_length_stats_ratios}, highlighted in yellow, 
represents our reference optimal setup 
(i.e., smoothing = 10, subsampling = 1\%, persistence threshold $\sigma = 3$), 
which we use for our subsequent investigations at $z=0$ of 
filament statistics, as well as for the persistence diagrams presented in Section \ref{sec_homtop_z0_persistence}.
For this specific configuration, we also display error bars, which were computed by 
considering the uncertainty due to subsampling.

Arc statistics are fundamental to filament statistics because arcs serve as the building blocks of filament structures.
In fact, an arc is defined as a 1-cell or integral line connecting two critical points whose order difference is 1, 
following the gradient direction (see Section \ref{sec_theory}). 
Since filaments are formed by unions of arcs, they are intrinsically linked to how critical points are connected (i.e., Figure \ref{fig_visualization_1e}). 
This connectivity shapes the overall filamentary structure, making arc statistics an essential tool for studying filament properties.

We next proceed to characterize filament statistics at $z=0$.  
The upper left panel of Figure \ref{fig_filaments_stats}
shows the filament length distribution, $P(L)$, 
obtained using the optimal combination of smoothing, subsampling, 
and persistence threshold determined by the tests above.
In the plot, the black solid line represents the massless neutrino scenario, 
the dashed light blue line corresponds to the $M_{\nu}=0.1$ eV cosmology,
and the dotted orange line indicates the $M_{\nu}=0.6$ eV framework.
Using the same color scheme, the lower left panel displays the ratio of 
filament lengths in the two massive neutrino cosmologies 
relative to the baseline massless neutrino model, with the 
shaded light brown region representing a $\pm 2\%$ variation.
The associated error bars are derived by accounting for the uncertainty due to subsampling.
Similar to the arc statistics, we detect a trend in filament lengths induced by a nonzero neutrino mass---namely,
an excess of longer filaments and a deficit of smaller ones 
(note again the adopted `UNNORM' convention; see Section \ref{sec_simulations})---with the relative 
abundance varying according to the value of $M_{\nu}$.
Furthermore, there also appears to be a `transition' scale, but now around $10h^{-1}{\rm Mpc}$,
that is nearly independent of $M_{\nu}$.
The variations in filament lengths are comparable to those observed in the arc statistics: within 
$2\%$ for $M_{\nu}=0.1$ eV,
and up to approximately 
$\sim 15\%$ for $M_{\nu}=0.6$ eV.

The right panel of Figure \ref{fig_filaments_stats} shows the corresponding filament auto-correlation functions
(i.e., two-point clustering statistics of the critical points of order 2, or filament-like saddles, as obtained using the
 \texttt{DisPerSE} algorithm) for the same cosmologies and 
parameter choices as in the filament length distribution analysis at $z=0$, with an identical color scheme. 
The minimal-variance Landy-Szalay (LS) estimator \citep{LandySzalay1993} is used to compute these correlation functions, 
assuming periodic boundary conditions and random catalogs at least $20$ times larger than the considered data.
Error bars represent the corresponding $1\sigma$ variations, estimated from subsampling. 
The spatial separation, $r$, is expressed in units of $h^{-1}{\rm Mpc}$, adopting a bin size of $5h^{-1}{\rm Mpc}$.
As seen in the plot, the filament auto-correlation $\xi(r)$ for $M_{\nu} = 0.6$ eV clearly differs from that 
of the massless neutrino scenario, 
while $\xi(r)$ for $M_{\nu} = 0.1$ eV is more challenging to disentangle.
As pointed out in our closely related work \citep{MoonRossiYu2023}, critical points are less sensitive to systematic effects,
trace the BAO peak similarly to DM, halos, and galaxies, and faithfully represent their complementary structures.  
Moreover, the presence of massive neutrinos affects the BAO peak amplitudes of all critical point 
correlation functions above/below the rarity or persistence threshold, as well as the positions of their 
corresponding inflection points at large scales. Specifically,  a nonzero neutrino mass
generally corresponds to higher BAO amplitudes, regardless of the specific 
critical point type, and these amplitudes increase as $M_{\nu}$ is augmented---with
departures from the massless neutrino scenario reaching up to $\sim 7\%$ at $z=0$ when
$M_{\nu} = 0.1$ eV. This trend is clearly evident in the figure, especially when $M_{\nu} = 0.6$ eV.

\cite{MoonRossiYu2023} also highlighted that saddle-point statistics, such as those involving filamentary structures, are more advantageous 
for extracting cosmological information because their cosmic evolution is less nonlinear. In addition, the two-point autocorrelations of filaments provide 
valuable information on the characteristic sizes of halos, while the two-point autocorrelations of walls offer insights into the characteristic sizes of voids. 
Furthermore, autocorrelations involving less nonlinear critical points (e.g., filaments and walls) appear mostly 
insensitive to redshift evolution effects, making saddle-point statistics particularly promising for cosmological analyses and neutrino mass constraints.

Next, we extend our analysis beyond filaments by examining persistence diagrams at $z=0$
in massive neutrino cosmologies. 

\begin{figure*}
\centering
\includegraphics[angle=0,width=0.92\textwidth]{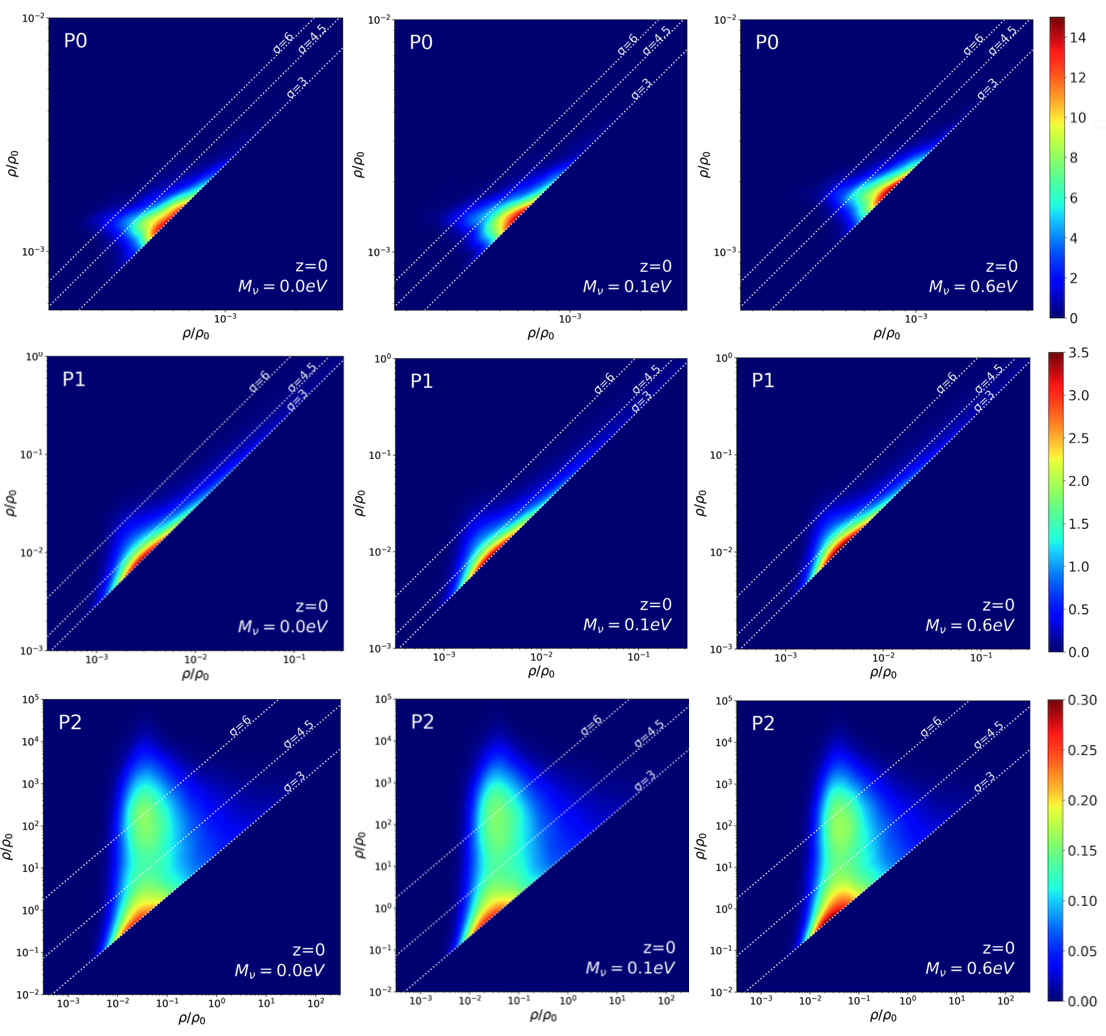}
\caption{Persistence diagrams of the \texttt{MassiveNuS} simulations at \(z=0\) for massless and massive neutrino cosmologies (\(M_\nu=0.1,\,0.6\,\mathrm{eV}\)). 
Rows show \(P0\), \(P1\), and \(P2\) persistence pairs, while columns correspond to increasing neutrino mass. White dotted lines indicate persistence thresholds at \(\sigma=3\), \(4.5\), and \(6\). 
High-persistence apexes trace key connectivity transitions of the cosmic web and exhibit systematic shifts with neutrino mass, 
reflecting neutrino-induced modifications to the hierarchical formation of voids, walls, and filaments.}
\label{fig_persistence_diagrams_3a}
\end{figure*}

\subsection{Persistence Diagrams and Massive Neutrinos}     \label{sec_homtop_z0_persistence}

Persistence diagrams characterize the stability and prominence of topological features across all density thresholds, 
providing a richer and more informative multiscale view of the cosmic web than traditional summary statistics. 
By tracking persistence pairs and quantifying feature lifetimes in excursion sets, these diagrams reveal the hierarchical assembly of filaments, walls, and voids, 
capturing non-linear structure formation beyond what Euler characteristics,
Betti numbers, or Betti curves can describe. Robust to sampling noise and independent of function smoothness, persistence 
diagrams are particularly effective at detecting subtle effects of massive neutrinos, making them a powerful and robust 
tool for cosmic web analysis in massive-neutrino cosmologies.

More specifically, persistence pairs describe the \textit{birth} and \textit{death} of topological features  as the parameter of a given filtration varies. 
In this work, we consider connected components, loops, and cavities of the density-field sublevel sets, parameterized by a density threshold. 
Each pair \(P_i = [p_i, q_{i+1}]\) consists of two critical points of order \(i\) and \(i+1\). As the density threshold increases, a new isolated sublevel set component appears at a 
local minimum, marking the \textit{birth} of a feature. When two isolated components merge at a saddle-1 critical point, the \textit{death} of a feature occurs, 
defining a persistence pair. The pair includes the saddle-1 and the minimum whose threshold difference with that saddle-1 is smallest: the ``youngest'' feature dies, while the ``oldest'' persists.
This construction extends to higher-dimensional features. Two components merging at a saddle-1 can give rise to a new loop, which subsequently dies at a saddle-2 point. 
Similarly, the collapse of a loop at a saddle-2 can create a cavity, which dies at a maximum. In all cases, persistence pairs are composed of two critical points whose orders differ by exactly one. 
We denote by \(P0\) the pairs connecting minima and saddle-1 points (wall-like saddles, order~1), by \(P1\) the saddle-1 to saddle-2 (filament-like saddles, order~2) pairs, and by \(P2\) the saddle-2 to maxima pairs. 
In this convention, filaments are primarily traced by \(P2\) pairs (with additional connectivity information present in  \(P1\) pairs), 
walls by \(P1\) pairs, and voids by \(P0\) pairs. High-persistence \(P2\) pairs indicate prominent filamentary structures.

Within this framework, a persistence diagram is a scatter plot in which each point corresponds to a persistence pair. 
Both axes represent the filtration parameter, designated as ``birth''  and ``death''. 
Each pair \(P_i \) is represented by the coordinates
\[ \left[\rho(p_i), \rho(q_{i+1})\right]/\rho_0 .\]
In our convention, the birth density threshold (the lower-density critical point) is shown on the horizontal axis, 
while the death density threshold (the higher-density critical point) 
is shown on the vertical axis. The diagonal corresponds to zero persistence. Since death always occurs at a density threshold greater than or equal to birth, all persistence pairs lie above the
\(\rho_{\mathrm{birth}} = \rho_{\mathrm{death}}\) line (i.e., the top-left region of the diagrams).

The persistence of a feature is proportional to its distance from this diagonal: the farther a point lies from the line, the more persistent the feature. In noisy density fields, numerous 
low-persistence pairs accumulate near the diagonal. A persistence threshold, defined in terms of a given number of \(\sigma\), is applied by drawing a line parallel to the diagonal; 
pairs below this line are removed during topological simplification. Clusters of points indicate recurring topological features across the simulation volume, while phase transitions 
appear as sharp changes in the number density of persistence pairs aligned with the birth or death axes. 
Spurious critical points arising from Poisson noise are typically filtered out enforcing a minimum threshold of \(3\sigma\).

\begin{figure*}
\centering 
\includegraphics[angle=0,width=0.82\textwidth]{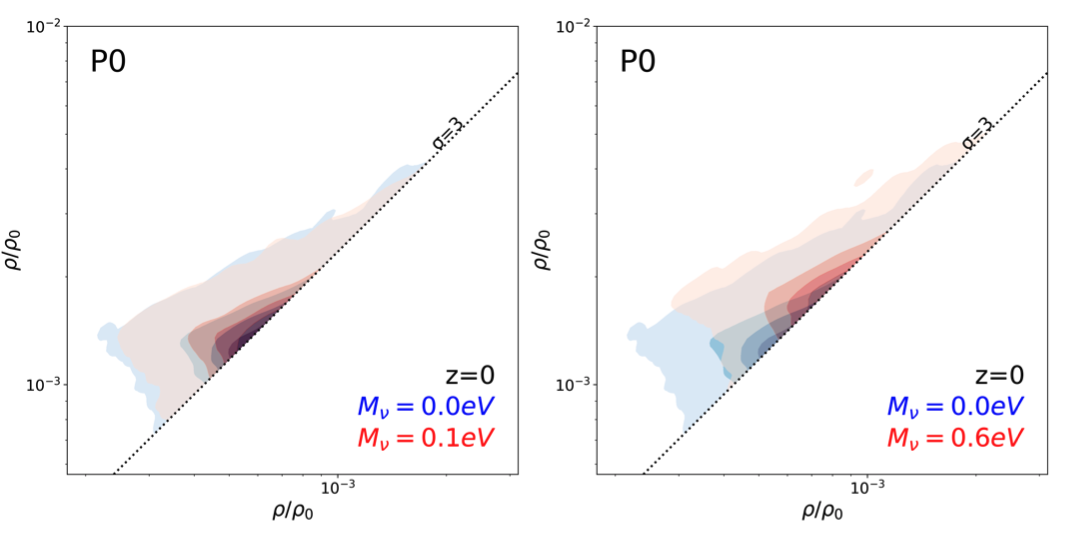}
\caption{Comparison of \(P0\) (minima to saddle-1) persistence diagrams for the fiducial, massless-neutrino cosmology (blue) and 
massive-neutrino models with \(M_{\nu} = 0.1\,\mathrm{eV}\) (left panel, red) and \(M_{\nu} = 0.6\,\mathrm{eV}\) (right panel, red). Filled contours indicate the number density of persistence pairs, 
with higher opacity corresponding to more pairs. Massive neutrinos shift the baseline toward higher density thresholds, reflecting a delayed evacuation of cosmic voids as neutrino mass increases.}
\label{fig_persistence_diagrams_3b}
\end{figure*}

To this end, Figure~\ref{fig_persistence_diagrams_3a} shows the persistence diagrams for the \texttt{MassiveNuS} simulations at \(z = 0\), using a \(1\,\%\) subsampling and a 
minimal persistence simplification threshold of \(\sigma = 3\). From left to right, the panels correspond to the massless-neutrino cosmology, a massive-neutrino cosmology with \(M_{\nu} = 0.1\,\mathrm{eV}\), 
and a massive-neutrino cosmology with \(M_{\nu} = 0.6\,\mathrm{eV}\). From top to bottom, the panels show the \(P0\), \(P1\), and \(P2\) persistence pairs. 
In each panel, the white dotted lines indicate 
persistence thresholds at \(\sigma = 3\), \(4.5\), and \(6\), from lowest to highest, respectively. Each persistence diagram is displayed as a two-dimensional histogram, 
with colors representing the probability density of persistence pairs in each bin. The three colorbars on the right-hand side of the figure are shared within each respective row.

\begin{figure*}
\centering 
\includegraphics[angle=0,width=0.85\textwidth]{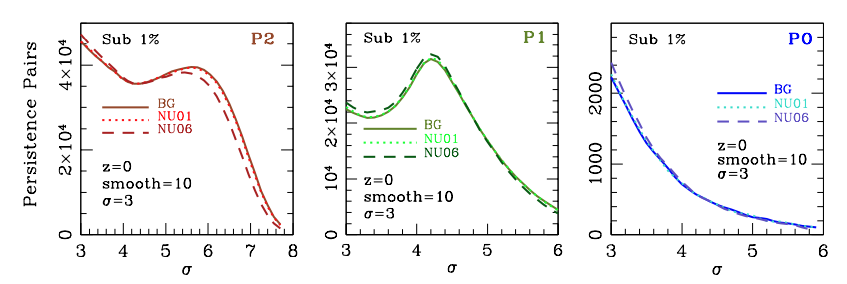}
\caption{Number of persistence pairs as a function of significance threshold (\(\sigma\)) for \(P2\) (left), \(P1\) (center), and \(P0\) (right) pairs from the \texttt{MassiveNuS} simulations at \(z=0\). 
The massless-neutrino model is shown as a solid line (``BG''), \(M_{\nu}=0.1\,\mathrm{eV}\) as a dotted line (``NU01''), and \(M_{\nu}=0.6\,\mathrm{eV}\) as a dashed line (``NU06''). 
The \(\sigma \sim 6\) bump in \(P2\) indicates high-persistence halos and filaments, which decrease with increasing neutrino mass, while lower-persistence pairs increase. 
Similar but smaller shifts are observed in \(P1\), and \(P0\) pairs show a monotonic decrease with \(\sigma\), with only minor changes due to neutrino mass at low persistence.}
\label{fig_persistence_diagrams_3c}
\end{figure*}

As pointed out by \citet{Wilding2021}, persistence diagrams reveal the presence of \textit{apexes}--sharp, high-persistence features that mark key 
connectivity transitions and act as topological signatures of the dynamical evolution of the cosmic web. High-persistence points are of particular interest because they
trace the most prominent structures (i.e., clusters, filaments, and voids). In our framework, we define apexes as persistence pairs with maximal persistence values.
For the \(P0\) persistence diagram, the characteristic triangular shape allows us to identify a typical birth density below which few features are created and a typical death density 
above which few features survive. The intersection of these regimes defines the apex of the triangle. In the fiducial, massless-neutrino cosmology, this apex is located at a 
birth density of approximately \(\rho/\rho_{0} \simeq 3 \times 10^{-4}\) and a death density of \(\sim 1.5 \times 10^{-3}\). We refine this observation and assess the 
impact of massive neutrinos when discussing Figure~\ref{fig_persistence_diagrams_3b} later in this section.
In the \(P2\) persistence diagram (bottom row), we identify an additional prominent feature: a dense cluster of high-persistence pairs centered around birth 
densities \(\rho/\rho_{0} \simeq 3 \times 10^{-2}\) and death densities of order \(10^{2}\). This region corresponds to common and persistent halo-to-filament structures, 
which we further explore in Figure~\ref{fig_persistence_diagrams_3c}. More generally, the presence and location of apexes differ across the \(P0\), \(P1\), and \(P2\) diagrams, 
reflecting the distinct density regimes and topological roles of voids, walls, and filaments.

A key result of this work is that massive neutrinos induce a systematic shift in the position and evolution of these apex points as a function of neutrino mass and cosmic time, 
relative to the massless-neutrino case. This behavior directly reflects the impact of neutrinos on structure formation: by suppressing the growth of density perturbations, 
neutrinos modify the hierarchical buildup of the cosmic web. At higher redshifts, peaks and islands initially form on smaller scales and merge into larger structures more slowly 
than in a neutrinoless cosmology. At the same time, voids expand and evacuate matter at a reduced pace, resulting in a cosmic mass distribution characterized by progressively larger and emptier voids. 
These combined effects make filaments particularly sensitive probes of neutrino mass, as they exhibit strong and coherent neutrino-induced modifications.
Importantly, the sharp transitions marked by the apexes represent the principal processes governing cosmic web formation. 
In fact, the upper edge of each persistence diagram demarcates 
the emergence of a percolating network, signaling the global connectivity of the cosmic web. Differences in the shape, extent, and density range of the \(P0\), \(P1\), and \(P2\) diagrams reflect
fundamental differences in the multiscale nature, connectivity, and hierarchical evolution of peaks, islands, tunnels, filaments, and voids \citep[e.g.,][]{Wilding2021}. In particular, void-related features are confined to 
a narrower density range, while filamentary structures extend to higher densities and higher persistence.
In this regard, persistence diagrams are unique in their ability to capture these global connectivity transitions, which cannot be described by traditional statistics such as power spectra or correlation functions. 
Diagonal points in the diagrams correspond to topological noise, while the high-persistence regions contain a wealth of information about the morphology and topology of cosmic structures. 
The movement of the apexes---especially at high persistence---is therefore a key signature of neutrino mass effects and a central feature highlighted in Figure~\ref{fig_persistence_diagrams_3a}.
a detailed analysis of these apex points and their evolution in massive-neutrino cosmologies is presented in a companion study (Yu et al., in preparation).

In Figure~\ref{fig_persistence_diagrams_3b} we focus on the \(P0\) persistence diagrams (i.e., minima to saddle-1). These diagrams correspond to the top row of 
Figure~\ref{fig_persistence_diagrams_3a}, based on the same \texttt{MassiveNuS} simulations at \(z = 0\) with \(1\,\%\) subsampling and a simplification threshold of \(\sigma = 3\). 
The two panels of Figure~\ref{fig_persistence_diagrams_3b} show the \(P0\) persistence diagram for the fiducial, massless-neutrino cosmology in shades of blue. 
The left panel overlays the massive-neutrino model with \(M_{\nu} = 0.1\,\mathrm{eV}\) in shades of red, while the right panel shows the \(M_{\nu} = 0.6\,\mathrm{eV}\) model. 
Each persistence diagram is represented as a filled contour plot, where higher opacity indicates a larger number of persistence pairs.

While the overall shape of the diagrams remains similar across different neutrino models, we observe a shift along the baseline in the massive-neutrino models toward higher density thresholds. 
This shift increases with neutrino mass, from \(\sim 10^{-5}\) for \(M_{\nu}=0.1\,\mathrm{eV}\) to \(\sim 10^{-4}\) for \(M_{\nu}=0.6\,\mathrm{eV}\), in units of \(\rho/\rho_{0}\). 
Since the \(P0\) pairs correspond to cosmic voids, the left edge of the diagram represents a characteristic void density threshold. In Figure~\ref{fig_persistence_diagrams_3b}, 
these values are \(\rho/\rho_{0} \approx 3\times 10^{-4}\) for the massless-neutrino model and \(\approx 4\times 10^{-4}\) for the massive \(M_{\nu}=0.6\,\mathrm{eV}\) model. 
Under gravity, lower-density regions lose matter as it flows toward higher-density regions. The observed shift in the persistence diagrams therefore 
indicates that massive neutrinos delay the evacuation of cosmic voids.

Figure~\ref{fig_persistence_diagrams_3c} shows the number of persistence pairs as a function of the significance threshold 
(number of \(\sigma\)) for the \texttt{MassiveNuS} simulations at \(z = 0\), using \(1\,\%\) subsampling and a simplification threshold of \(\sigma = 3\). 
From left to right, the panels correspond to \(P2\) pairs (saddle-2 to maxima), \(P1\) pairs (saddle-1 to saddle-2), and \(P0\) pairs (minima to saddle-1). 
In each panel, the massless-neutrino model is shown as a solid line (labelled ``BG'' for the fiducial cosmology), the \(M_{\nu} = 0.1\,\mathrm{eV}\) 
model as a dotted line (``NU01''), and the \(M_{\nu} = 0.6\,\mathrm{eV}\) model as a dashed line (``NU06'').

The curves for the \(P2\) pairs feature a prominent bump centered around \(\sigma = 6\). This bump originates from the clustering of \(P2\) pairs clearly observed in the 
persistence diagrams of Figure~\ref{fig_persistence_diagrams_3a}, around birth densities of \(\rho/\rho_{0} \approx 3 \times 10^{-2}\) and death densities of \(\sim 10^{2}\). 
This cluster corresponds to typical high-persistence halos and filamentary structures. Massive neutrinos impact these structures, as evidenced by a decrease in the 
number of \(P2\) pairs near the \(\sigma = 6\) bump with increasing neutrino mass. Conversely, the number of pairs at the lower edge of the bump, around \(\sigma = 4.5\), increases with neutrino mass.
For \(P1\) pairs, a similar but less pronounced bump is present, starting around \(\sigma = 3.5\) and peaking near \(\sigma = 4.5\). Here, massive neutrinos 
increase the number of persistence pairs for \(\sigma \lesssim 5\) while decreasing it at higher persistence. 
Finally, the number of \(P0\) pairs decreases monotonically with \(\sigma\). Massive neutrinos slightly increase the number of \(P0\) pairs for \(\sigma \lesssim 4\), 
but no significant differences are observed at higher persistence thresholds.

In summary, the most striking features revealed by persistence diagrams are the shifts in density ranges and the movement of apex points, both of which are sensitive to massive neutrinos. 
These shifts reflect how neutrinos modify the hierarchical formation of cosmic structures, delaying void evacuation and altering filament connectivity. At \(z=0\), we observe notable differences in filamentary 
structures (\(P2\) pairs), while the redshift evolution highlights clear neutrino effects in \(P0\) pairs, corresponding to voids---as discussed next. 

\subsection{Redshift Evolution and Systematics}   \label{sec_z_evo_systematics} 

Building on the sensitivity of persistence diagrams to massive neutrinos identified in the previous section, we now examine how these 
signatures evolve with redshift and how robust they are to differences in simulation methodology. Specifically, we compare filament statistics 
and selected aspects of persistence diagrams from the \texttt{MassiveNuS} and \texttt{QUIJOTE} simulations to assess redshift evolution and
 systematics associated with neutrino implementations and normalization choices. This comparison clarifies how neutrino mass and simulation setup jointly 
influence the hierarchical development of voids, walls, and filaments across cosmic time.

\begin{figure*}
\centering
\includegraphics[angle=0,width=0.90\textwidth]{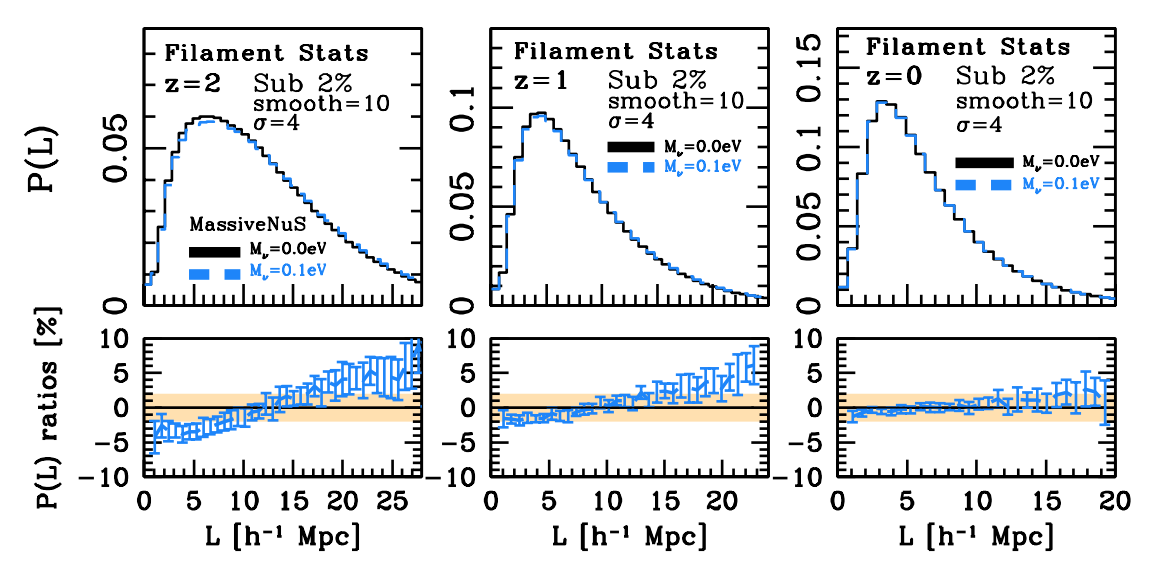}
\includegraphics[angle=0,width=0.90\textwidth]{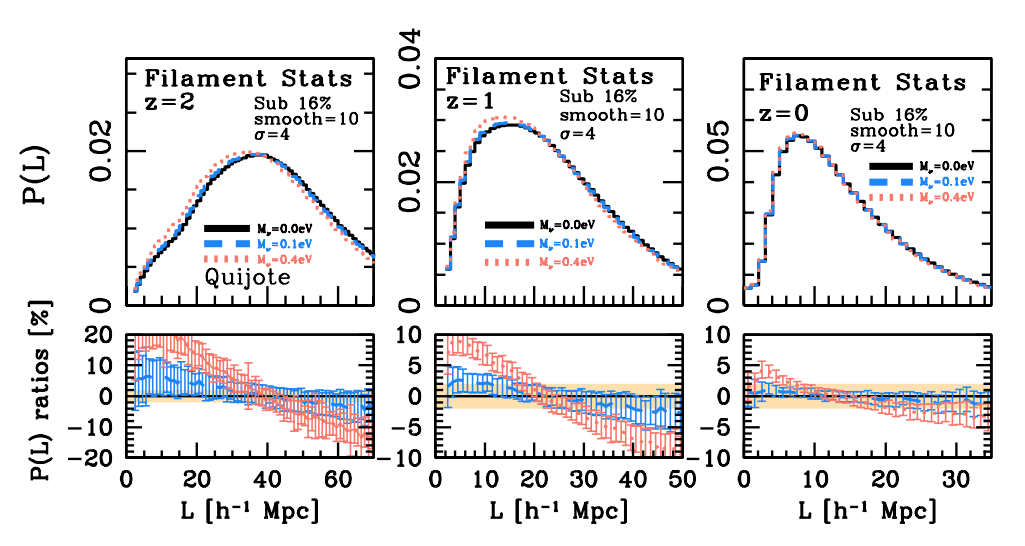}
\caption{Redshift evolution of filament lengths from the \texttt{MassiveNuS} (top) and \texttt{QUIJOTE} (bottom) simulations at \(z = 2, 1, 0\) (left to right). 
Top subpanels show the probability distributions of filament lengths, while bottom subpanels display ratios relative to the fiducial massless-neutrino model (black line). 
The \(M_{\nu}=0.1\,\mathrm{eV}\) model is shown as a dashed blue line, and the \(M_{\nu}=0.4\,\mathrm{eV}\) model (for \texttt{QUIJOTE} only) as an orange dotted line. 
Error bars on the ratio panels indicate 1-\(\sigma\) deviations across different subsamplings for \texttt{MassiveNuS}, 
and across 100 realizations for the \texttt{QUIJOTE} simulations.
The yellow shaded region marks a \(\pm 2\%\) relative difference. 
Variations between the two simulation suites arise from differences in normalization conventions, resolution, and neutrino implementations.}
\label{fig_key_plots_errorbars_columbia_2}
\end{figure*}

\begin{figure*}
\centering
\includegraphics[angle=0,width=0.90\textwidth]{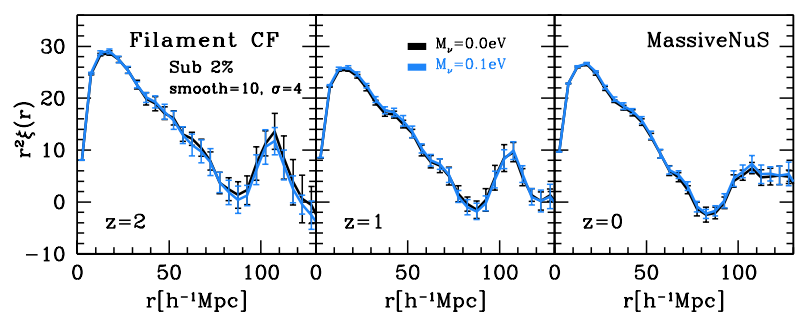}
\includegraphics[angle=0,width=0.90\textwidth]{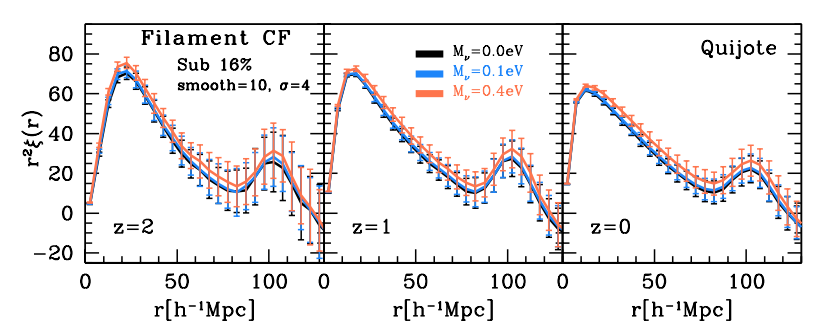}
\caption{Saddle-2 autocorrelation functions for the \texttt{MassiveNuS} (top row, 2\% subsampling) and \texttt{QUIJOTE} (bottom row, 16\% subsampling) simulations at \(z = 2, 1, 0\) (left to right). 
A simplification threshold of \(\sigma = 4\) is applied to all simulations. Line styles and error bars follow the conventions of the previous figure. 
The BAO peak around \(100\,h^{-1}\,\mathrm{Mpc}\) is clearly visible and remains consistent across redshifts, cosmologies, and simulation suites.}
\label{fig_key_plots_errorbars_columbia_3}
\end{figure*}

To this end, Figure~\ref{fig_key_plots_errorbars_columbia_2} presents the redshift evolution of filament lengths.
The top half of the figure shows results from the \texttt{MassiveNuS} simulations, while the bottom half shows the \texttt{QUIJOTE} simulations. 
From left to right, the panels correspond to \(z = 2\), 1, and 0. In all cases, a smoothing scale of 10 and a persistence threshold of \(\sigma = 4\) are 
adopted,\footnote{We note that at $4 \sigma$ persistence level the probability of
a spurious pair is $0.006\%$: hence, any arc in the DMC can be safely considered a feature of the underlying distribution.}
while subsampling of 2\% was applied for \texttt{MassiveNuS} and 16\% for \texttt{QUIJOTE}.
For each redshift and simulation suite, the top subpanel displays the probability distribution of filament lengths, 
while the bottom subpanel shows the ratios relative to the fiducial, massless-neutrino model. The fiducial model is plotted as a black line, the \(M_{\nu} = 0.1\,\mathrm{eV}\) model 
as a dashed blue line, and, for \texttt{QUIJOTE} only, the \(M_{\nu} = 0.4\,\mathrm{eV}\) model as an orange dotted line. Error bars on the ratio panels indicate 1-\(\sigma\) deviations across 
different subsamplings for \texttt{MassiveNuS}, and across multiple realizations for the \texttt{QUIJOTE} simulations (over 100 realizations). 
The yellow shaded region highlights a \(\pm 2\%\) relative difference.

As observed in Figure~\ref{fig_filaments_stats} for the \texttt{MassiveNuS} simulations, massive neutrinos tend to increase the length of longer filaments while 
decreasing the number of shorter ones, with a transition scale at \(z = 2\) around \(10\) to \(15\,h^{-1}\,\mathrm{Mpc}\). In contrast, the \texttt{QUIJOTE} simulations show the 
opposite behavior: massive neutrinos reduce the number of long filaments and increase the number of short ones. This difference arises from the distinct choices of 
power spectrum normalization in the two simulation suites. The \texttt{MassiveNuS} simulations fix the primordial amplitude \(A_{\mathrm{s}}\) and allow \(\sigma_8\) 
to vary (``UNNORM'' convention), whereas the \texttt{QUIJOTE} simulations vary \(A_{\mathrm{s}}\) to keep \(\sigma_8\) fixed (``NORM'' convention).

Another distinction between the simulation suites is the transition scale where the ratio curves cross zero. In \texttt{MassiveNuS}, this occurs 
between \(10\) and \(15\,h^{-1}\,\mathrm{Mpc}\) at \(z = 2\), while in \texttt{QUIJOTE} it occurs closer to \(40\,h^{-1}\,\mathrm{Mpc}\). This disparity likely reflects 
the different resolutions: \texttt{MassiveNuS} has roughly four times higher resolution than \texttt{QUIJOTE}. Correspondingly, the peaks of the probability 
distributions at \(z = 2\) are around \(6\,h^{-1}\,\mathrm{Mpc}\) for \texttt{MassiveNuS} and \(40\,h^{-1}\,\mathrm{Mpc}\) for \texttt{QUIJOTE}, 
as lower-resolution simulations cannot resolve the smaller filaments.
Moreover,  as redshift decreases, the relative differences between massless and massive neutrino models also diminish. For \(M_{\nu} = 0.1\,\mathrm{eV}\), the differences 
decrease from \(\pm 5\%\) at \(z = 2\) to \(\pm 2\%\) at \(z = 0\). For \(M_{\nu} = 0.4\,\mathrm{eV}\) (only in \texttt{QUIJOTE}), the differences decrease 
from \(\pm 20\%\) at \(z = 2\) to \(\pm 5\%\) at \(z = 0\). These trends indicate that the impact of massive neutrinos on filament lengths is strongest at 
high redshift and gradually weakens over cosmic time.

In Figure~\ref{fig_key_plots_errorbars_columbia_3}, we show the saddle-2 autocorrelation functions, 
using a layout similar to the right side of Figure~\ref{fig_filaments_stats}. 
The top row corresponds to the \texttt{MassiveNuS} simulations (\(2\%\) subsampling), and the 
bottom row to the \texttt{QUIJOTE} simulations (\(16\%\) subsampling). 
A simplification threshold of \(\sigma = 4\) is applied to all simulations. From left to right, each column corresponds to \(z = 2\), 1, and 0. 
Black lines represent the fiducial massless-neutrino cosmology, 
blue lines the massive \(M_{\nu} = 0.1\,\mathrm{eV}\) case, and orange lines (only for \texttt{Quijote}) the \(M_{\nu} = 0.4\,\mathrm{eV}\) case. 
Error bars indicate 1-\(\sigma\) variation across subsamplings (\texttt{MassiveNuS}) or multiple realizations (\texttt{QUIJOTE}). 
The main highlight of this figure is the clear 
BAO peak around \(100\,h^{-1}\,\mathrm{Mpc}\), which is consistent across all redshifts, cosmologies, and simulation suites. 
These saddle-2 autocorrelation functions can be naturally 
connected to the statistical properties of critical points studied in \citet{MoonRossiYu2023} in the context of massive-neutrino cosmologies, 
emphasizing that the clustering of filamentary saddle
points traces faithfully the large-scale connectivity of the cosmic web and is highly 
sensitive to massive neutrino effects.
  
\begin{figure*}
\centering     
\includegraphics[angle=0,width=1.03\textwidth]{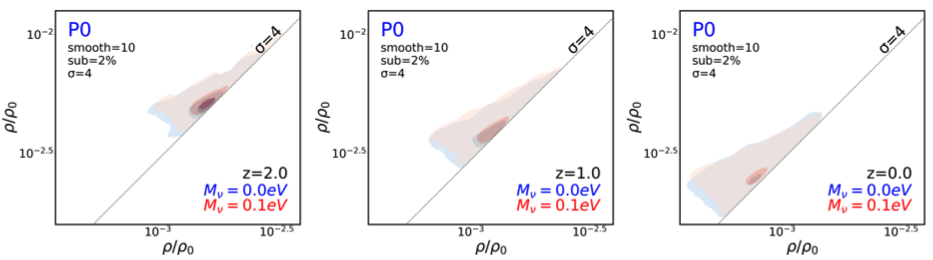}
\caption{\(P0\) persistence diagrams for the \texttt{MassiveNuS} simulations at \(z = 2, 1, 0\) (left to right), using a simplification threshold of \(\sigma = 4\) and a \(2\%\) subsampling. 
Blue shading represents the massless-neutrino model, and red shading the \(M_{\nu} = 0.1\,\mathrm{eV}\) massive-neutrino model. The diagrams shift toward lower densities with 
decreasing redshift, reflecting the evacuation of cosmic voids. The relative shift between the two cosmologies remains approximately constant, highlighting the sensitivity of voids to neutrino mass effects.}
\label{fig_key_plots_errorbars_columbia_4}
\end{figure*}
 
\begin{figure*}
\centering                                                                                                                                                              
\includegraphics[angle=0,width=0.90\textwidth]{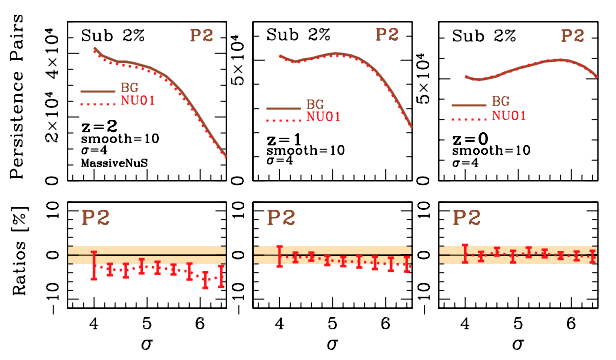}    
\caption{Number of \(P2\) persistence pairs--tracing primarily filaments--as a function of significance threshold for the \texttt{MassiveNuS} simulations at \(z = 2, 1, 0\) (left to right), using a \(2\%\) 
subsampling and a simplification threshold \(\sigma = 4\). Solid brown lines represent the massless-neutrino model, and dotted red lines the \(M_{\nu} = 0.1\,\mathrm{eV}\) massive-neutrino framework. 
Bottom panels show the relative differences between the two cosmologies with error bars from subsamplings. 
The prominence of the bump at high \(\sigma\) increases with decreasing redshift, reflecting the growth of filamentary structures, while neutrino-induced 
differences are largest at higher redshift and diminish by \(z = 0\).}
\label{fig_key_plots_errorbars_columbia_5}
\end{figure*}

Next, we briefly address selected aspects of persistence diagrams and persistence pairs across redshift. 
Specifically, Figure~\ref{fig_key_plots_errorbars_columbia_4} shows a variation of Figure~\ref{fig_persistence_diagrams_3b}, focusing on the \texttt{MassiveNuS} simulation suite. 
The three panels present filled contour plots of the \(P0\) persistence diagrams, using a simplification threshold of \(\sigma = 4\) and a \(2\%\) subsampling. 
From left to right, the panels correspond to \(z = 2\), 1, and 0. The blue shading represents the massless-neutrino model, while the red shading 
corresponds to the \(M_{\nu} = 0.1\,\mathrm{eV}\) massive-neutrino model. As time evolves from \(z = 2\) to \(z = 0\), the persistence diagrams 
shift along the \(\sigma = 4\) line toward lower densities, reflecting the evolution of cosmic voids, which progressively lose matter to surrounding higher-density regions. 
Consequently, the density of local minima in voids decreases with time, and the birth and death of persistence pairs occur at lower densities. 
The relative shift between the massless and massive neutrino models remains approximately constant during this evolution, 
highlighting that also voids are an ideal environment to detect subtle neutrino mass effects. 

Finally, Figure~\ref{fig_key_plots_errorbars_columbia_5} shows the number of \(P2\) persistence pairs (tracing primarily filaments) 
as a function of the significance threshold, 
using a layout similar to Figure~\ref{fig_persistence_diagrams_3c}. 
For this plot, we focus on the \texttt{MassiveNuS} simulations with a \(2\%\) subsampling and a simplification threshold of \(\sigma = 4\). 
From left to right, the panels correspond to \(z = 2\), 1, and 0, respectively. 
The solid brown line represents the massless-neutrino cosmology, while the dotted red line 
corresponds to the \(M_{\nu} = 0.1\,\mathrm{eV}\) massive-neutrino framework. In the bottom row, we also display the relative differences 
between the massive and massless neutrino models, including error bars from subsampling. As redshift increases, the prominent bump observed in 
Figure~\ref{fig_persistence_diagrams_3c} becomes less pronounced, since the structures responsible for the bump are only in the early stages of formation at \(z = 2\) 
and have already grown significantly by \(z = 1\). The differences between neutrino masses are larger at higher redshift (around \(\sim 2\%\) at \(z = 2\)) 
and decrease to near \(0\%\) at \(z = 0\) in the considered range (\(4 < \sigma < 7\)).

In closing this section, we note that, along with resolution and normalization choices, differences in neutrino implementation between 
the \texttt{MassiveNuS} and \texttt{QUIJOTE} simulations can affect some of the quantitative results presented above. \texttt{MassiveNuS} uses a linear-response approach for 
modeling neutrinos, 
efficient but less accurate on small scales, while \texttt{QUIJOTE} employs a particle-based method that fully captures nonlinear neutrino clustering 
at the cost of increased shot noise and computational resources. These choices may partially explain the quantitative variations in filament lengths, 
persistence diagrams, and high-significance \(P2\) pair statistics across the two suites, especially at high-$z$ and small scales. 
Nonetheless, the overall qualitative trends--such as void evacuation and the evolution of filamentary structures with neutrino mass--remain consistent, 
highlighting the robustness of the main cosmological conclusions.


\section{Summary and Outlook}       \label{sec_conclusion}

Traditional approaches to constraining neutrino mass typically rely on individual tracers (e.g., galaxies, clusters, voids) at specific scales, 
using two-point summary statistics such as the total matter power spectrum or the correlation function to quantify neutrino-induced modifications and indirectly infer their mass. 
However, recent results from DESI DR1 \citep{DESI-DR1-Cosmo} and DR2 \citep{DESI-DR2-Cosmo}  
obtained with these methods have revealed intriguing deviations from the standard $\Lambda$CDM model, 
including hints of dynamical DE and surprisingly stringent constraints on neutrino masses. Some analyses even yield 
unphysical negative $M_\nu$ values \citep[e.g.,][]{Naredo-Tuero2024,GreenMeyers2025},
highlighting the limitations of traditional two-point statistics and motivating the need to explore alternative, higher-order, 
physically grounded probes capable of capturing the full complexity of the cosmic web.

Our study addresses precisely this challenge. 
Building on the first application of persistent homology to massive neutrino cosmologies presented in \citet{Rossi2022}, we 
employed discrete Morse theory, global topology, and persistent homology to characterize 
how neutrinos simultaneously affect multiple cosmic environments (i.e., halos, walls, filaments, and voids) across the \textit{multiscale} cosmic web. 
Using two independent $N$-body simulation suites (\texttt{MassiveNuS} and \texttt{QUIJOTE}) with different neutrino implementations and normalization conventions, 
we assessed these effects within a parameter-free, scale-adaptive framework that is robust to Poisson shot noise, operates directly on particle distributions, and requires no smoothing or pre-processing of the density field.
In particular, filament-based statistics and persistence diagrams revealed that massive neutrinos imprint coherent, mass-dependent signatures on cosmic structures, 
modifying the \textit{multiscale} connectivity and skeleton of the cosmic web.
These signatures are especially pronounced at high redshift ($z \sim 2$), where filaments emerge as sensitive tracers of neutrino mass effects.
Comparing results from the two simulation suites also allowed us to address potential systematics, including the impact of resolution, neutrino implementation, 
and initial power-spectrum normalization on the measurements. 

Key highlights of our analysis include:
\begin{enumerate}
\item \textbf{Filament statistics:} Under fixed $A_{\rm s}$ normalization, massive neutrinos increase the number of long filaments and reduce shorter ones, while fixing $\sigma_8$ reverses this trend. 
The transition scale between these regimes is largely independent of neutrino mass but shows amplitudes that vary with $M_\nu$, reflecting the impact of neutrino free-streaming on the filamentary network. 
These effects reach the $\sim 10$-$15\%$ level for larger neutrino 
masses ($M_\nu \sim 0.6$ eV) and remain detectable at the few-percent level for $M_\nu \sim 0.1$ eV (Section \ref{sec_filament_stats}).
\item \textbf{Persistence diagrams:} \(P0\) pairs (minima to wall-like saddle-1) shift toward higher densities with increasing neutrino mass, reflecting delayed void evacuation.   
\(P2\) pairs (filament-like saddle-2 to maxima) show modified filament connectivity and a reduction of high-persistence clusters for higher $M_\nu$, illustrating suppression of small-scale perturbation growth. 
High-persistence apex points trace sharp connectivity transitions and percolation processes in the cosmic web, encoding dominant neutrino effects (Section \ref{sec_homtop_z0_persistence}).
\item \textbf{Redshift dependence and systematics:} Neutrino-induced effects are enhanced at higher redshift  on most statistics, in particular for filament length distributions, saddle-2 autocorrelations, and persistence pair counts, 
emphasizing the importance of upcoming high-redshift surveys in probing early cosmic structures.
The various trends are consistent across different simulation suites and neutrino implementations (Section \ref{sec_z_evo_systematics}).
\end{enumerate}

Notably, our method remains reliable even for sparse samples and preserves the native network resolution after persistence-based simplification. In particular, it enables:
\begin{itemize}
\item \textbf{Simultaneous multiscale identification:} analyzing halos, walls, filaments, and voids jointly, thereby providing a coherent global view of neutrino effects.
\item \textbf{Robustness to noise:} quantifying feature significance relative to Poisson shot noise, thereby ensuring reliability in particle-based neutrino simulations.
\item \textbf{Scale-free and parameter-free operation:} extracting structure directly from the Delaunay tessellation of particle distributions, requiring no smoothing, resampling, or predefined scales.
\end{itemize}

Moreover, our topological framework---particularly through persistence diagrams, which encode substantially more information than commonly used topological summary statistics---provides \textit{simultaneous} 
access to \textit{multiscale} information across tracers and epochs, allowing for a more comprehensive view of how massive neutrinos shape cosmic structures
by exploiting their \textit{combined} imprints rather than isolated effects. 
Taken together, these results demonstrate that cosmic web topology, persistent homology, and environment-based statistics provide a powerful, 
physically grounded and rich complementary framework for studying massive neutrino signatures beyond conventional two-point clustering analyses. 
More broadly, this work lays the foundation for a comprehensive theoretical and observational program in which higher-order statistics 
and global descriptors of structure formation play a central role in precision neutrino cosmology, 
offering a powerful pathway toward robust constraints on neutrino properties with current (e.g., DES, DESI, Euclid) and next-generation survey data (e.g., Rubin-LSST, Roman, PFS). 

While the present study establishes the physical relevance of these methods, fully exploiting persistent homology as a precision 
probe of massive neutrinos---and more broadly as a tool for cosmological parameter inference---requires further development, to be presented in forthcoming companion papers. 
In particular, in a follow-up study (Yu et al. 2026), we extend the analysis to Betti curves using both particle and halo catalogs and include a detailed investigation of high-persistence apex points in persistence diagrams. 
Their systematic shift with neutrino mass and redshift (highlighted in Section \ref{sec_homtop_z0_persistence}) emerges as a robust, physically interpretable signature of neutrino-induced modifications 
to cosmic-web connectivity, whose origin and cosmological dependence are examined there.
Complementary efforts are in progress, including systematic studies of observational effects---such as redshift-space distortions, survey incompleteness, and tracer bias---as well as the influence of baryonic physics, 
expected to be particularly relevant in filamentary environments where a large fraction of baryons resides in diffuse hot gas.   
Additional work explores alternative neutrino implementations and 
extended cosmological scenarios with next-generation simulations. Parallel developments target optimized summary statistics, likelihood frameworks, and machine-learning strategies to translate persistence-based 
observables into quantitative constraints on neutrino mass (potentially with sensitivity to the mass hierarchy) and other cosmological parameters, with applications to ongoing state-of-the-art survey data such as DESI.
A further promising avenue is the direct connection to forthcoming observations of the high-redshift cosmic web. Probing large-scale structure at $z \gtrsim 2$ is rapidly becoming feasible, and this regime
is particularly favorable for neutrino studies because the associated signatures are cleaner and less affected by late-time nonlinearities. Analyses targeting the Lyman-$\alpha$ forest and other 
high-redhsift tracers are underway. In this regards, our framework also establishes a foundation for next-generation surveys such as DESI-II and Stage-V to 
constrain neutrino mass and fundamental cosmological parameters beyond standard approaches.
  

\begin{acknowledgements}

We warmly thank Jeongin Moon for assistance with critical-point statistics. This work was supported by the National Research Foundation of Korea (NRF) through 
Grants No. 2020R1A2C1005655 and No. NRF-RS-2026-25472461, funded by the Korean Ministry of Education, Science, and Technology (MoEST). 
We also acknowledge the use of our computing resources at Sejong University (Xeon Silver 4114 master node and Xeon Gold 6126 compute node architecture).

\end{acknowledgements}


\bibliographystyle{aasjournal}
\bibliography{references}{

@ARTICLE{ACT2025,
       author = {{Louis}, Thibaut and {La Posta}, Adrien and {Atkins}, Zachary and {Jense}, Hidde T. and {Abril-Cabezas}, Irene and {Addison}, Graeme E. and {Ade}, Peter A.~R. and {Aiola}, Simone and {Alford}, Tommy and {Alonso}, David and {Amiri}, Mandana and {An}, Rui and {Austermann}, Jason E. and {Barbavara}, Eleonora and {Battaglia}, Nicholas and {Battistelli}, Elia Stefano and {Beall}, James A. and {Bean}, Rachel and {Beheshti}, Ali and {Beringue}, Benjamin and {Bhandarkar}, Tanay and {Biermann}, Emily and {Bolliet}, Boris and {Bond}, J. Richard and {Calabrese}, Erminia and {Capalbo}, Valentina and {Carrero}, Felipe and {Chen}, Shi-Fan and {Chesmore}, Grace and {Cho}, Hsiao-mei and {Choi}, Steve K. and {Clark}, Susan E. and {Cothard}, Nicholas F. and {Coughlin}, Kevin and {Coulton}, William and {Crichton}, Devin and {Crowley}, Kevin T. and {Darwish}, Omar and {Devlin}, Mark J. and {Dicker}, Simon and {Duell}, Cody J. and {Duff}, Shannon M. and {Duivenvoorden}, Adriaan J. and {Dunkley}, Jo and {Dunner}, Rolando and {Embil Villagra}, Carmen and {Fankhanel}, Max and {Farren}, Gerrit S. and {Ferraro}, Simone and {Foster}, Allen and {Freundt}, Rodrigo and {Fuzia}, Brittany and {Gallardo}, Patricio A. and {Garrido}, Xavier and {Gerbino}, Martina and {Giardiello}, Serena and {Gill}, Ajay and {Givans}, Jahmour and {Gluscevic}, Vera and {Goldstein}, Samuel and {Golec}, Joseph E. and {Gong}, Yulin and {Guan}, Yilun and {Halpern}, Mark and {Harrison}, Ian and {Hasselfield}, Matthew and {Healy}, Erin and {Henderson}, Shawn and {Hensley}, Brandon and {Herv{\'\i}as-Caimapo}, Carlos and {Hill}, J. Colin and {Hilton}, Gene C. and {Hilton}, Matt and {Hincks}, Adam D. and {Hlo{\v{z}}ek}, Ren{\'e}e and {Ho}, Shuay-Pwu Patty and {Hood}, John and {Hornecker}, Erika and {Huber}, Zachary B. and {Hubmayr}, Johannes and {Huffenberger}, Kevin M. and {Hughes}, John P. and {Ikape}, Margaret and {Irwin}, Kent and {Isopi}, Giovanni and {Joshi}, Neha and {Keller}, Ben and {Kim}, Joshua and {Knowles}, Kenda and {Koopman}, Brian J. and {Kosowsky}, Arthur and {Kramer}, Darby and {Kusiak}, Aleksandra and {Lagu{\"e}}, Alex and {Lakey}, Victoria and {Lee}, Eunseong and {Li}, Yaqiong and {Li}, Zack and {Limon}, Michele and {Lokken}, Martine and {Lungu}, Marius and {MacCrann}, Niall and {MacInnis}, Amanda and {Madhavacheril}, Mathew S. and {Maldonado}, Diego and {Maldonado}, Felipe and {Mallaby-Kay}, Maya and {Marques}, Gabriela A. and {van Marrewijk}, Joshiwa and {McCarthy}, Fiona and {McMahon}, Jeff and {Mehta}, Yogesh and {Menanteau}, Felipe and {Moodley}, Kavilan and {Morris}, Thomas W. and {Mroczkowski}, Tony and {Naess}, Sigurd and {Namikawa}, Toshiya and {Nati}, Federico and {Nerval}, Simran K. and {Newburgh}, Laura and {Nicola}, Andrina and {Niemack}, Michael D. and {Nolta}, Michael R. and {Orlowski-Scherer}, John and {Pagano}, Luca and {Page}, Lyman A. and {Pandey}, Shivam and {Partridge}, Bruce and {Perez Sarmiento}, Karen and {Prince}, Heather and {Puddu}, Roberto and {Qu}, Frank J. and {Ragavan}, Damien C. and {Ried Guachalla}, Bernardita and {Rogers}, Keir K. and {Rojas}, Felipe and {Sakuma}, Tai and {Schaan}, Emmanuel and {Schmitt}, Benjamin L. and {Sehgal}, Neelima and {Shaikh}, Shabbir and {Sherwin}, Blake D. and {Sierra}, Carlos and {Sievers}, Jon and {Sif{\'o}n}, Crist{\'o}bal and {Simon}, Sara and {Sonka}, Rita and {Spergel}, David N. and {Staggs}, Suzanne T. and {Storer}, Emilie and {Surrao}, Kristen and {Switzer}, Eric R. and {Tampier}, Niklas and {Thornton}, Robert and {Trac}, Hy and {Tucker}, Carole and {Ullom}, Joel and {Vale}, Leila R. and {Van Engelen}, Alexander and {Van Lanen}, Jeff and {Vargas}, Cristian and {Vavagiakis}, Eve M. and {Wagoner}, Kasey and {Wang}, Yuhan and {Wenzl}, Lukas and {Wollack}, Edward J. and {Zheng}, Kaiwen and {The Atacama Cosmology Telescope collaboration}},
        title = "{The Atacama Cosmology Telescope: DR6 power spectra, likelihoods and {\ensuremath{\Lambda}}CDM parameters}",
      journal = {\jcap},
         year = 2025,
        month = nov,
       volume = {2025},
       number = {11},
          eid = {062},
        pages = {062},
          doi = {10.1088/1475-7516/2025/11/062},
archivePrefix = {arXiv},
       eprint = {2503.14452},
 primaryClass = {astro-ph.CO},
       adsurl = {https://ui.adsabs.harvard.edu/abs/2025JCAP...11..062L}
}

@ARTICLE{Ahlen2025,
       author = {{Ahlen}, S.~P. and {Aviles}, A. and {Cartwright}, B. and {Croker}, K.~S. and {Elbers}, W. and {Farrah}, D. and {Fernandez}, N. and {Niz}, G. and {Rohlf}, J.~W. and {Tarl{\'e}}, G. and {Windhorst}, R.~A. and {Aguilar}, J. and {Andrade}, U. and {Bianchi}, D. and {Brooks}, D. and {Claybaugh}, T. and {de la Macorra}, A. and {de Mattia}, A. and {Dey}, B. and {Doel}, P. and {Forero-Romero}, J.~E. and {Gazta{\~n}aga}, E. and {Gontcho}, S. Gontcho A. and {Gutierrez}, G. and {Huterer}, D. and {Ishak}, M. and {Kehoe}, R. and {Kirkby}, D. and {Kremin}, A. and {Lahav}, O. and {Lamman}, C. and {Landriau}, M. and {Le Guillou}, L. and {Levi}, M.~E. and {Manera}, M. and {Miquel}, R. and {Moustakas}, J. and {P{\'e}rez-R{\`a}fols}, I. and {Prada}, F. and {Rossi}, G. and {Sanchez}, E. and {Schubnell}, M. and {Seo}, H. and {Silber}, J. and {Sprayberry}, D. and {Walther}, M. and {Weaver}, B.~A. and {Wechsler}, R.~H. and {Zou}, H. and {DESI Collaboration}},
        title = "{Positive Neutrino Masses with DESI DR2 via Matter Conversion to Dark Energy}",
      journal = {\prl},
         year = 2025,
        month = aug,
       volume = {135},
       number = {8},
          eid = {081003},
        pages = {081003},
          doi = {10.1103/yb2k-kn7h},
archivePrefix = {arXiv},
       eprint = {2504.20338},
 primaryClass = {astro-ph.CO},
       adsurl = {https://ui.adsabs.harvard.edu/abs/2025PhRvL.135h1003A}
}

@ARTICLE{Ali-Haimoud2013,
       author = {{Ali-Ha{\"\i}moud}, Yacine and {Bird}, Simeon},
        title = "{An efficient implementation of massive neutrinos in non-linear structure formation simulations}",
      journal = {\mnras},
         year = 2013,
        month = feb,
       volume = {428},
       number = {4},
        pages = {3375-3389},
          doi = {10.1093/mnras/sts286},
archivePrefix = {arXiv},
       eprint = {1209.0461},
 primaryClass = {astro-ph.CO},
       adsurl = {https://ui.adsabs.harvard.edu/abs/2013MNRAS.428.3375A}
}

@ARTICLE{Ajani2020,
       author = {{Ajani}, Virginia and {Peel}, Austin and {Pettorino}, Valeria and {Starck}, Jean-Luc and {Li}, Zack and {Liu}, Jia},
        title = "{Constraining neutrino masses with weak-lensing multiscale peak counts}",
      journal = {\prd},
         year = 2020,
        month = nov,
       volume = {102},
       number = {10},
          eid = {103531},
        pages = {103531},
          doi = {10.1103/PhysRevD.102.103531},
archivePrefix = {arXiv},
       eprint = {2001.10993},
 primaryClass = {astro-ph.CO},
       adsurl = {https://ui.adsabs.harvard.edu/abs/2020PhRvD.102j3531A}
}

@ARTICLE{Asai2024,
       author = {{Asai}, Shoji and {Ballarino}, Amalia and {Bose}, Tulika and {Cranmer}, Kyle and {Cyr-Racine}, Francis-Yan and {Demers}, Sarah and {Geddes}, Cameron and {Gershtein}, Yuri and {Heeger}, Karsten and {Heinemann}, Beate and {Hewett}, JoAnne and {Huber}, Patrick and {Mahn}, Kendall and {Mandelbaum}, Rachel and {Maricic}, Jelena and {Merkel}, Petra and {Monahan}, Christopher and {Murayama}, Hitoshi and {Onyisi}, Peter and {Palmer}, Mark and {Raubenheimer}, Tor and {Sanchez}, Mayly and {Schnee}, Richard and {Seidel}, Sally and {Seo}, Seon-Hee and {Thaler}, Jesse and {Touramanis}, Christos and {Vieregg}, Abigail and {Weinstein}, Amanda and {Winslow}, Lindley and {Yu}, Tien-Tien and {Zwaska}, Robert},
        title = "{Exploring the Quantum Universe: Pathways to Innovation and Discovery in Particle Physics}",
      journal = {arXiv e-prints},
         year = 2024,
        month = jul,
          eid = {arXiv:2407.19176},
        pages = {arXiv:2407.19176},
          doi = {10.48550/arXiv.2407.19176},
archivePrefix = {arXiv},
       eprint = {2407.19176},
 primaryClass = {hep-ex},
       adsurl = {https://ui.adsabs.harvard.edu/abs/2024arXiv240719176A}
}

@ARTICLE{Blanton2017,
       author = {{Blanton}, Michael R. and {Bershady}, Matthew A. and {Abolfathi}, Bela and
         {Albareti}, Franco D. and {Allende Prieto}, Carlos and
         {Almeida}, Andres and {Alonso-Garc{\'\i}a}, Javier and
         {Anders}, Friedrich and {Anderson}, Scott F. and {Andrews}, Brett and
         {Aquino-Ort{\'\i}z}, Erik and {Arag{\'o}n-Salamanca}, Alfonso and
         {Argudo-Fern{\'a}ndez}, Maria and {Armengaud}, Eric and
         {Aubourg}, Eric and {Avila-Reese}, Vladimir and {Badenes}, Carles and
         {Bailey}, Stephen and {Barger}, Kathleen A. and
         {Barrera-Ballesteros}, Jorge and {Bartosz}, Curtis and
         {Bates}, Dominic and {Baumgarten}, Falk and {Bautista}, Julian and
         {Beaton}, Rachael and {Beers}, Timothy C. and {Belfiore}, Francesco and
         {Bender}, Chad F. and {Berlind}, Andreas A. and {Bernardi}, Mariangela and
         {Beutler}, Florian and {Bird}, Jonathan C. and {Bizyaev}, Dmitry and
         {Blanc}, Guillermo A. and {Blomqvist}, Michael and {Bolton}, Adam S. and
         {Boquien}, M{\'e}d{\'e}ric and {Borissova}, Jura and
         {van den Bosch}, Remco and {Bovy}, Jo and {Brandt}, William N. and
         {Brinkmann}, Jonathan and {Brownstein}, Joel R. and {Bundy}, Kevin and
         {Burgasser}, Adam J. and {Burtin}, Etienne and {Busca}, Nicol{\'a}s G. and
         {Cappellari}, Michele and {Delgado Carigi}, Maria Leticia and
         {Carlberg}, Joleen K. and {Carnero Rosell}, Aurelio and
         {Carrera}, Ricardo and {Chanover}, Nancy J. and {Cherinka}, Brian and
         {Cheung}, Edmond and {G{\'o}mez Maqueo Chew}, Yilen and
         {Chiappini}, Cristina and {Choi}, Peter Doohyun and {Chojnowski}, Drew and
         {Chuang}, Chia-Hsun and {Chung}, Haeun and {Cirolini}, Rafael Fernando and
         {Clerc}, Nicolas and {Cohen}, Roger E. and {Comparat}, Johan and
         {da Costa}, Luiz and {Cousinou}, Marie-Claude and {Covey}, Kevin and
         {Crane}, Jeffrey D. and {Croft}, Rupert A.~C. and
         {Cruz-Gonzalez}, Irene and {Garrido Cuadra}, Daniel and {Cunha}, Katia and
         {Damke}, Guillermo J. and {Darling}, Jeremy and {Davies}, Roger and
         {Dawson}, Kyle and {de la Macorra}, Axel and {Dell'Agli}, Flavia and
         {De Lee}, Nathan and {Delubac}, Timoth{\'e}e and {Di Mille}, Francesco and
         {Diamond-Stanic}, Aleks and {Cano-D{\'\i}az}, Mariana and
         {Donor}, John and {Downes}, Juan Jos{\'e} and {Drory}, Niv and
         {du Mas des Bourboux}, H{\'e}lion and {Duckworth}, Christopher J. and
         {Dwelly}, Tom and {Dyer}, Jamie and {Ebelke}, Garrett and
         {Eigenbrot}, Arthur D. and {Eisenstein}, Daniel J. and
         {Emsellem}, Eric and {Eracleous}, Mike and {Escoffier}, Stephanie and
         {Evans}, Michael L. and {Fan}, Xiaohui and {Fern{\'a}ndez-Alvar}, Emma and
         {Fernandez-Trincado}, J.~G. and {Feuillet}, Diane K. and
         {Finoguenov}, Alexis and {Fleming}, Scott W. and {Font-Ribera}, Andreu and
         {Fredrickson}, Alexander and {Freischlad}, Gordon and
         {Frinchaboy}, Peter M. and {Fuentes}, Carla E. and
         {Galbany}, Llu{\'\i}s and {Garcia-Dias}, R. and
         {Garc{\'\i}a-Hern{\'a}ndez}, D.~A. and {Gaulme}, Patrick and
         {Geisler}, Doug and {Gelfand}, Joseph D. and
         {Gil-Mar{\'\i}n}, H{\'e}ctor and {Gillespie}, Bruce A. and
         {Goddard}, Daniel and {Gonzalez-Perez}, Violeta and
         {Grabowski}, Kathleen and {Green}, Paul J. and {Grier}, Catherine J. and
         {Gunn}, James E. and {Guo}, Hong and {Guy}, Julien and {Hagen}, Alex and
         {Hahn}, ChangHoon and {Hall}, Matthew and {Harding}, Paul and
         {Hasselquist}, Sten and {Hawley}, Suzanne L. and {Hearty}, Fred and
         {Gonzalez Hern{\'a}ndez}, Jonay I. and {Ho}, Shirley and
         {Hogg}, David W. and {Holley-Bockelmann}, Kelly and {Holtzman}, Jon A. and
         {Holzer}, Parker H. and {Huehnerhoff}, Joseph and
         {Hutchinson}, Timothy A. and {Hwang}, Ho Seong and
         {Ibarra-Medel}, H{\'e}ctor J. and {da Silva Ilha}, Gabriele and
         {Ivans}, Inese I. and {Ivory}, KeShawn and {Jackson}, Kelly and
         {Jensen}, Trey W. and {Johnson}, Jennifer A. and {Jones}, Amy and
         {J{\"o}nsson}, Henrik and {Jullo}, Eric and {Kamble}, Vikrant and
         {Kinemuchi}, Karen and {Kirkby}, David and {Kitaura}, Francisco-Shu and
         {Klaene}, Mark and {Knapp}, Gillian R. and {Kneib}, Jean-Paul and
         {Kollmeier}, Juna A. and {Lacerna}, Ivan and {Lane}, Richard R. and
         {Lang}, Dustin and {Law}, David R. and {Lazarz}, Daniel and
         {Lee}, Youngbae and {Le Goff}, Jean-Marc and {Liang}, Fu-Heng and
         {Li}, Cheng and {Li}, Hongyu and {Lian}, Jianhui and {Lima}, Marcos and
         {Lin}, Lihwai and {Lin}, Yen-Ting and {Bertran de Lis}, Sara and
         {Liu}, Chao and {de Icaza Lizaola}, Miguel Angel C. and {Long}, Dan and
         {Lucatello}, Sara and {Lundgren}, Britt and {MacDonald}, Nicholas K. and
         {Deconto Machado}, Alice and {MacLeod}, Chelsea L. and
         {Mahadevan}, Suvrath and {Geimba Maia}, Marcio Antonio and
         {Maiolino}, Roberto and {Majewski}, Steven R. and
         {Malanushenko}, Elena and {Malanushenko}, Viktor and
         {Manchado}, Arturo and {Mao}, Shude and {Maraston}, Claudia and
         {Marques-Chaves}, Rui and {Masseron}, Thomas and {Masters}, Karen L. and
         {McBride}, Cameron K. and {McDermid}, Richard M. and
         {McGrath}, Brianne and {McGreer}, Ian D. and
         {Medina Pe{\~n}a}, Nicol{\'a}s and {Melendez}, Matthew and
         {Merloni}, Andrea and {Merrifield}, Michael R. and
         {Meszaros}, Szabolcs and {Meza}, Andres and {Minchev}, Ivan and
         {Minniti}, Dante and {Miyaji}, Takamitsu and {More}, Surhud and
         {Mulchaey}, John and {M{\"u}ller-S{\'a}nchez}, Francisco and
         {Muna}, Demitri and {Munoz}, Ricardo R. and {Myers}, Adam D. and
         {Nair}, Preethi and {Nandra}, Kirpal and
         {Correa do Nascimento}, Janaina and {Negrete}, Alenka and
         {Ness}, Melissa and {Newman}, Jeffrey A. and {Nichol}, Robert C. and
         {Nidever}, David L. and {Nitschelm}, Christian and {Ntelis}, Pierros and
         {O'Connell}, Julia E. and {Oelkers}, Ryan J. and {Oravetz}, Audrey and
         {Oravetz}, Daniel and {Pace}, Zach and {Padilla}, Nelson and
         {Palanque-Delabrouille}, Nathalie and {Alonso Palicio}, Pedro and
         {Pan}, Kaike and {Parejko}, John K. and {Parikh}, Taniya and
         {P{\^a}ris}, Isabelle and {Park}, Changbom and {Patten}, Alim Y. and
         {Peirani}, Sebastien and {Pellejero-Ibanez}, Marcos and
         {Penny}, Samantha and {Percival}, Will J. and {Perez-Fournon}, Ismael and
         {Petitjean}, Patrick and {Pieri}, Matthew M. and {Pinsonneault}, Marc and
         {Pisani}, Alice and {Poleski}, Rados{\l}aw and {Prada}, Francisco and
         {Prakash}, Abhishek and {Queiroz}, Anna B{\'a}rbara de Andrade and
         {Raddick}, M. Jordan and {Raichoor}, Anand and {Barboza Rembold}, Sand
        ro and {Richstein}, Hannah and {Riffel}, Rogemar A. and
         {Riffel}, Rog{\'e}rio and {Rix}, Hans-Walter and {Robin}, Annie C. and
         {Rockosi}, Constance M. and {Rodr{\'\i}guez-Torres}, Sergio and
         {Roman-Lopes}, A. and {Rom{\'a}n-Z{\'u}{\~n}iga}, Carlos and
         {Rosado}, Margarita and {Ross}, Ashley J. and {Rossi}, Graziano and
         {Ruan}, John and {Ruggeri}, Rossana and {Rykoff}, Eli S. and
         {Salazar-Albornoz}, Salvador and {Salvato}, Mara and
         {S{\'a}nchez}, Ariel G. and {Aguado}, D.~S. and
         {S{\'a}nchez-Gallego}, Jos{\'e} R. and {Santana}, Felipe A. and
         {Santiago}, Bas{\'\i}lio Xavier and {Sayres}, Conor and
         {Schiavon}, Ricardo P. and {da Silva Schimoia}, Jaderson and
         {Schlafly}, Edward F. and {Schlegel}, David J. and
         {Schneider}, Donald P. and {Schultheis}, Mathias and
         {Schuster}, William J. and {Schwope}, Axel and {Seo}, Hee-Jong and
         {Shao}, Zhengyi and {Shen}, Shiyin and {Shetrone}, Matthew and
         {Shull}, Michael and {Simon}, Joshua D. and {Skinner}, Danielle and
         {Skrutskie}, M.~F. and {Slosar}, An{\v{z}}e and {Smith}, Verne V. and
         {Sobeck}, Jennifer S. and {Sobreira}, Flavia and {Somers}, Garrett and
         {Souto}, Diogo and {Stark}, David V. and {Stassun}, Keivan and
         {Stauffer}, Fritz and {Steinmetz}, Matthias and
         {Storchi-Bergmann}, Thaisa and {Streblyanska}, Alina and
         {Stringfellow}, Guy S. and {Su{\'a}rez}, Genaro and {Sun}, Jing and
         {Suzuki}, Nao and {Szigeti}, Laszlo and {Taghizadeh-Popp}, Manuchehr and
         {Tang}, Baitian and {Tao}, Charling and {Tayar}, Jamie and
         {Tembe}, Mita and {Teske}, Johanna and {Thakar}, Aniruddha R. and
         {Thomas}, Daniel and {Thompson}, Benjamin A. and {Tinker}, Jeremy L. and
         {Tissera}, Patricia and {Tojeiro}, Rita and {Hernandez Toledo}, Hector and
         {de la Torre}, Sylvain and {Tremonti}, Christy and
         {Troup}, Nicholas W. and {Valenzuela}, Octavio and
         {Martinez Valpuesta}, Inma and {Vargas-Gonz{\'a}lez}, Jaime and
         {Vargas-Maga{\~n}a}, Mariana and {Vazquez}, Jose Alberto and
         {Villanova}, Sandro and {Vivek}, M. and {Vogt}, Nicole and
         {Wake}, David and {Walterbos}, Rene and {Wang}, Yuting and
         {Weaver}, Benjamin Alan and {Weijmans}, Anne-Marie and
         {Weinberg}, David H. and {Westfall}, Kyle B. and {Whelan}, David G. and
         {Wild}, Vivienne and {Wilson}, John and {Wood-Vasey}, W.~M. and
         {Wylezalek}, Dominika and {Xiao}, Ting and {Yan}, Renbin and
         {Yang}, Meng and {Ybarra}, Jason E. and {Y{\`e}che}, Christophe and
         {Zakamska}, Nadia and {Zamora}, Olga and {Zarrouk}, Pauline and
         {Zasowski}, Gail and {Zhang}, Kai and {Zhao}, Gong-Bo and
         {Zheng}, Zheng and {Zheng}, Zheng and {Zhou}, Xu and {Zhou}, Zhi-Min and
         {Zhu}, Guangtun B. and {Zoccali}, Manuela and {Zou}, Hu},
        title = "{Sloan Digital Sky Survey IV: Mapping the Milky Way, Nearby Galaxies, and the Distant Universe}",
      journal = {\aj},
         year = 2017,
        month = jul,
       volume = {154},
       number = {1},
          eid = {28},
        pages = {28},
          doi = {10.3847/1538-3881/aa7567},
archivePrefix = {arXiv},
       eprint = {1703.00052},
 primaryClass = {astro-ph.GA},
       adsurl = {https://ui.adsabs.harvard.edu/abs/2017AJ....154...28B}
}

@ARTICLE{Biagetti2022,
       author = {{Biagetti}, Matteo and {Calles}, Juan and {Castiblanco}, Lina and {Cole}, Alex and {Nore{\~n}a}, Jorge},
        title = "{Fisher forecasts for primordial non-Gaussianity from persistent homology}",
      journal = {\jcap},
         year = 2022,
        month = oct,
       volume = {2022},
       number = {10},
          eid = {002},
        pages = {002},
          doi = {10.1088/1475-7516/2022/10/002},
archivePrefix = {arXiv},
       eprint = {2203.08262},
 primaryClass = {astro-ph.CO},
       adsurl = {https://ui.adsabs.harvard.edu/abs/2022JCAP...10..002B}
}

@ARTICLE{Bird2018,
       author = {{Bird}, Simeon and {Ali-Ha{\"\i}moud}, Yacine and {Feng}, Yu and {Liu}, Jia},
        title = "{An efficient and accurate hybrid method for simulating non-linear neutrino structure}",
      journal = {\mnras},
         year = 2018,
        month = dec,
       volume = {481},
       number = {2},
        pages = {1486-1500},
          doi = {10.1093/mnras/sty2376},
archivePrefix = {arXiv},
       eprint = {1803.09854},
 primaryClass = {astro-ph.CO},
       adsurl = {https://ui.adsabs.harvard.edu/abs/2018MNRAS.481.1486B}
}

@ARTICLE{Bolliet2020,
       author = {{Bolliet}, Boris and {Brinckmann}, Thejs and {Chluba}, Jens and {Lesgourgues}, Julien},
        title = "{Including massive neutrinos in thermal Sunyaev Zeldovich power spectrum and cluster counts analyses}",
      journal = {\mnras},
         year = 2020,
        month = sep,
       volume = {497},
       number = {2},
        pages = {1332-1347},
          doi = {10.1093/mnras/staa1835},
archivePrefix = {arXiv},
       eprint = {1906.10359},
 primaryClass = {astro-ph.CO},
       adsurl = {https://ui.adsabs.harvard.edu/abs/2020MNRAS.497.1332B}
}

@ARTICLE{Bose2021,
       author = {{Bose}, Benjamin and {Wright}, Bill S. and {Cataneo}, Matteo and {Pourtsidou}, Alkistis and {Giocoli}, Carlo and {Lombriser}, Lucas and {McCarthy}, Ian G. and {Baldi}, Marco and {Pfeifer}, Simon and {Xia.}, Qianli},
        title = "{On the road to per cent accuracy - V. The non-linear power spectrum beyond {\ensuremath{\Lambda}}CDM with massive neutrinos and baryonic feedback}",
      journal = {\mnras},
         year = 2021,
        month = dec,
       volume = {508},
       number = {2},
        pages = {2479-2491},
          doi = {10.1093/mnras/stab2731},
archivePrefix = {arXiv},
       eprint = {2105.12114},
 primaryClass = {astro-ph.CO},
       adsurl = {https://ui.adsabs.harvard.edu/abs/2021MNRAS.508.2479B}
}

@ARTICLE{Cadiou2020,
       author = {{Cadiou}, C. and {Pichon}, C. and {Codis}, S. and {Musso}, M. and {Pogosyan}, D. and {Dubois}, Y. and {Cardoso}, J. -F. and {Prunet}, S.},
        title = "{When do cosmic peaks, filaments, or walls merge? A theory of critical events in a multiscale landscape}",
      journal = {\mnras},
         year = 2020,
        month = aug,
       volume = {496},
       number = {4},
        pages = {4787-4821},
          doi = {10.1093/mnras/staa1853},
archivePrefix = {arXiv},
       eprint = {2003.04413},
 primaryClass = {astro-ph.CO},
       adsurl = {https://ui.adsabs.harvard.edu/abs/2020MNRAS.496.4787C}
}

@ARTICLE{Calles2025,
       author = {{Calles}, Juan and {Yip}, Jacky H.~T. and {Contardo}, Gabriella and {Nore{\~n}a}, Jorge and {Rouhiainen}, Adam and {Shiu}, Gary},
        title = "{Cosmology with persistent homology: parameter inference via machine learning}",
      journal = {\jcap},
         year = 2025,
        month = sep,
       volume = {2025},
       number = {9},
          eid = {064},
        pages = {064},
          doi = {10.1088/1475-7516/2025/09/064},
archivePrefix = {arXiv},
       eprint = {2412.15405},
 primaryClass = {astro-ph.CO},
       adsurl = {https://ui.adsabs.harvard.edu/abs/2025JCAP...09..064C}
}

@ARTICLE{Cautun2014,
       author = {{Cautun}, Marius and {van de Weygaert}, Rien and {Jones}, Bernard J.~T. and {Frenk}, Carlos S.},
        title = "{Evolution of the cosmic web}",
      journal = {\mnras},
         year = 2014,
        month = jul,
       volume = {441},
       number = {4},
        pages = {2923-2973},
          doi = {10.1093/mnras/stu768},
archivePrefix = {arXiv},
       eprint = {1401.7866},
 primaryClass = {astro-ph.CO},
       adsurl = {https://ui.adsabs.harvard.edu/abs/2014MNRAS.441.2923C}
}

@ARTICLE{Cueli2024,
       author = {{Cueli}, M.~M. and {Cabo}, S.~R. and {Gonz{\'a}lez-Nuevo}, J. and {Bonavera}, L. and {Lapi}, A. and {Viel}, M. and {Crespo}, D. and {Casas}, J.~M. and {Fern{\'a}ndez-Fern{\'a}ndez}, R.},
        title = "{Toward the measurement of neutrino masses: Performance of cosmic magnification with submillimeter galaxies}",
      journal = {\aap},
         year = 2024,
        month = jul,
       volume = {687},
          eid = {A300},
        pages = {A300},
          doi = {10.1051/0004-6361/202449315},
archivePrefix = {arXiv},
       eprint = {2406.03236},
 primaryClass = {astro-ph.CO},
       adsurl = {https://ui.adsabs.harvard.edu/abs/2024A&A...687A.300C}
}

@ARTICLE{Dawson2016,
       author = {{Dawson}, Kyle S. and {Kneib}, Jean-Paul and {Percival}, Will J. and
         {Alam}, Shadab and {Albareti}, Franco D. and {Anderson}, Scott F. and
         {Armengaud}, Eric and {Aubourg}, {\'E}ric and {Bailey}, Stephen and
         {Bautista}, Julian E. and {Berlind}, Andreas A. and
         {Bershady}, Matthew A. and {Beutler}, Florian and {Bizyaev}, Dmitry and
         {Blanton}, Michael R. and {Blomqvist}, Michael and {Bolton}, Adam S. and
         {Bovy}, Jo and {Brandt}, W.~N. and {Brinkmann}, Jon and
         {Brownstein}, Joel R. and {Burtin}, Etienne and {Busca}, N.~G. and
         {Cai}, Zheng and {Chuang}, Chia-Hsun and {Clerc}, Nicolas and
         {Comparat}, Johan and {Cope}, Frances and {Croft}, Rupert A.~C. and
         {Cruz-Gonzalez}, Irene and {da Costa}, Luiz N. and
         {Cousinou}, Marie-Claude and {Darling}, Jeremy and
         {de la Macorra}, Axel and {de la Torre}, Sylvain and
         {Delubac}, Timoth{\'e}e and {du Mas des Bourboux}, H{\'e}lion and
         {Dwelly}, Tom and {Ealet}, Anne and {Eisenstein}, Daniel J. and
         {Eracleous}, Michael and {Escoffier}, S. and {Fan}, Xiaohui and
         {Finoguenov}, Alexis and {Font-Ribera}, Andreu and {Frinchaboy}, Peter and
         {Gaulme}, Patrick and {Georgakakis}, Antonis and {Green}, Paul and
         {Guo}, Hong and {Guy}, Julien and {Ho}, Shirley and {Holder}, Diana and
         {Huehnerhoff}, Joe and {Hutchinson}, Timothy and {Jing}, Yipeng and
         {Jullo}, Eric and {Kamble}, Vikrant and {Kinemuchi}, Karen and
         {Kirkby}, David and {Kitaura}, Francisco-Shu and {Klaene}, Mark A. and
         {Laher}, Russ R. and {Lang}, Dustin and {Laurent}, Pierre and
         {Le Goff}, Jean-Marc and {Li}, Cheng and {Liang}, Yu and
         {Lima}, Marcos and {Lin}, Qiufan and {Lin}, Weipeng and
         {Lin}, Yen-Ting and {Long}, Daniel C. and {Lundgren}, Britt and
         {MacDonald}, Nicholas and {Geimba Maia}, Marcio Antonio and
         {Malanushenko}, Elena and {Malanushenko}, Viktor and
         {Mariappan}, Vivek and {McBride}, Cameron K. and {McGreer}, Ian D. and
         {M{\'e}nard}, Brice and {Merloni}, Andrea and {Meza}, Andres and
         {Montero-Dorta}, Antonio D. and {Muna}, Demitri and {Myers}, Adam D. and
         {Nandra}, Kirpal and {Naugle}, Tracy and {Newman}, Jeffrey A. and
         {Noterdaeme}, Pasquier and {Nugent}, Peter and {Ogando}, Ricardo and
         {Olmstead}, Matthew D. and {Oravetz}, Audrey and {Oravetz}, Daniel J. and
         {Padmanabhan}, Nikhil and {Palanque-Delabrouille}, Nathalie and
         {Pan}, Kaike and {Parejko}, John K. and {P{\^a}ris}, Isabelle and
         {Peacock}, John A. and {Petitjean}, Patrick and {Pieri}, Matthew M. and
         {Pisani}, Alice and {Prada}, Francisco and {Prakash}, Abhishek and
         {Raichoor}, Anand and {Reid}, Beth and {Rich}, James and
         {Ridl}, Jethro and {Rodriguez-Torres}, Sergio and
         {Carnero Rosell}, Aurelio and {Ross}, Ashley J. and {Rossi}, Graziano and
         {Ruan}, John and {Salvato}, Mara and {Sayres}, Conor and
         {Schneider}, Donald P. and {Schlegel}, David J. and {Seljak}, Uros and
         {Seo}, Hee-Jong and {Sesar}, Branimir and {Shandera}, Sarah and
         {Shu}, Yiping and {Slosar}, An{\v{z}}e and {Sobreira}, Flavia and
         {Streblyanska}, Alina and {Suzuki}, Nao and {Taylor}, Donna and
         {Tao}, Charling and {Tinker}, Jeremy L. and {Tojeiro}, Rita and
         {Vargas-Maga{\~n}a}, Mariana and {Wang}, Yuting and
         {Weaver}, Benjamin A. and {Weinberg}, David H. and {White}, Martin and
         {Wood-Vasey}, W.~M. and {Yeche}, Christophe and {Zhai}, Zhongxu and
         {Zhao}, Cheng and {Zhao}, Gong-bo and {Zheng}, Zheng and
         {Ben Zhu}, Guangtun and {Zou}, Hu},
        title = "{The SDSS-IV Extended Baryon Oscillation Spectroscopic Survey: Overview and Early Data}",
      journal = {\aj},
         year = "2016",
        month = "Feb",
       volume = {151},
       number = {2},
          eid = {44},
        pages = {44},
          doi = {10.3847/0004-6256/151/2/44},
archivePrefix = {arXiv},
       eprint = {1508.04473},
 primaryClass = {astro-ph.CO},
       adsurl = {https://ui.adsabs.harvard.edu/abs/2016AJ....151...44D}
}

@ARTICLE{DES2005,
       author = {{The Dark Energy Survey Collaboration}},
        title = "{The Dark Energy Survey}",
      journal = {arXiv e-prints},
         year = 2005,
        month = oct,
          eid = {astro-ph/0510346},
        pages = {astro-ph/0510346},
archivePrefix = {arXiv},
       eprint = {astro-ph/0510346},
 primaryClass = {astro-ph},
       adsurl = {https://ui.adsabs.harvard.edu/abs/2005astro.ph.10346T}
}

@ARTICLE{DESI-DR1-Cosmo,
       author = {{DESI Collaboration} and {Adame}, A.~G. and {Aguilar}, J. and {Ahlen}, S. and {Alam}, S. and {Alexander}, D.~M. and {Alvarez}, M. and {Alves}, O. and {Anand}, A. and {Andrade}, U. and {Armengaud}, E. and {Avila}, S. and {Aviles}, A. and {Awan}, H. and {Bahr-Kalus}, B. and {Bailey}, S. and {Baltay}, C. and {Bault}, A. and {Behera}, J. and {BenZvi}, S. and {Bera}, A. and {Beutler}, F. and {Bianchi}, D. and {Blake}, C. and {Blum}, R. and {Brieden}, S. and {Brodzeller}, A. and {Brooks}, D. and {Buckley-Geer}, E. and {Burtin}, E. and {Calderon}, R. and {Canning}, R. and {Carnero Rosell}, A. and {Cereskaite}, R. and {Cervantes-Cota}, J.~L. and {Chabanier}, S. and {Chaussidon}, E. and {Chaves-Montero}, J. and {Chen}, S. and {Chen}, X. and {Claybaugh}, T. and {Cole}, S. and {Cuceu}, A. and {Davis}, T.~M. and {Dawson}, K. and {de la Macorra}, A. and {de Mattia}, A. and {Deiosso}, N. and {Dey}, A. and {Dey}, B. and {Ding}, Z. and {Doel}, P. and {Edelstein}, J. and {Eftekharzadeh}, S. and {Eisenstein}, D.~J. and {Elliott}, A. and {Fagrelius}, P. and {Fanning}, K. and {Ferraro}, S. and {Ereza}, J. and {Findlay}, N. and {Flaugher}, B. and {Font-Ribera}, A. and {Forero-S{\'a}nchez}, D. and {Forero-Romero}, J.~E. and {Frenk}, C.~S. and {Garcia-Quintero}, C. and {Gazta{\~n}aga}, E. and {Gil-Mar{\'\i}n}, H. and {Gontcho a Gontcho}, S. and {Gonzalez-Morales}, A.~X. and {Gonzalez-Perez}, V. and {Gordon}, C. and {Green}, D. and {Gruen}, D. and {Gsponer}, R. and {Gutierrez}, G. and {Guy}, J. and {Hadzhiyska}, B. and {Hahn}, C. and {Hanif}, M.~M.~S. and {Herrera-Alcantar}, H.~K. and {Honscheid}, K. and {Howlett}, C. and {Huterer}, D. and {Ir{\v{s}}i{\v{c}}}, V. and {Ishak}, M. and {Juneau}, S. and {Kara{\c{c}}ayl{\i}}, N.~G. and {Kehoe}, R. and {Kent}, S. and {Kirkby}, D. and {Kremin}, A. and {Krolewski}, A. and {Lai}, Y. and {Lan}, T.-W. and {Landriau}, M. and {Lang}, D. and {Lasker}, J. and {Le Goff}, J.~M. and {Le Guillou}, L. and {Leauthaud}, A. and {Levi}, M.~E. and {Li}, T.~S. and {Linder}, E. and {Lodha}, K. and {Magneville}, C. and {Manera}, M. and {Margala}, D. and {Martini}, P. and {Maus}, M. and {McDonald}, P. and {Medina-Varela}, L. and {Meisner}, A. and {Mena-Fern{\'a}ndez}, J. and {Miquel}, R. and {Moon}, J. and {Moore}, S. and {Moustakas}, J. and {Mueller}, E. and {Mu{\~n}oz-Guti{\'e}rrez}, A. and {Myers}, A.~D. and {Nadathur}, S. and {Napolitano}, L. and {Neveux}, R. and {Newman}, J.~A. and {Nguyen}, N.~M. and {Nie}, J. and {Niz}, G. and {Noriega}, H.~E. and {Padmanabhan}, N. and {Paillas}, E. and {Palanque-Delabrouille}, N. and {Pan}, J. and {Penmetsa}, S. and {Percival}, W.~J. and {Pieri}, M.~M. and {Pinon}, M. and {Poppett}, C. and {Porredon}, A. and {Prada}, F. and {P{\'e}rez-Fern{\'a}ndez}, A. and {P{\'e}rez-R{\`a}fols}, I. and {Rabinowitz}, D. and {Raichoor}, A. and {Ram{\'\i}rez-P{\'e}rez}, C. and {Ramirez-Solano}, S. and {Rashkovetskyi}, M. and {Ravoux}, C. and {Rezaie}, M. and {Rich}, J. and {Rocher}, A. and {Rockosi}, C. and {Roe}, N.~A. and {Rosado-Marin}, A. and {Ross}, A.~J. and {Rossi}, G. and {Ruggeri}, R. and {Ruhlmann-Kleider}, V. and {Samushia}, L. and {Sanchez}, E. and {Saulder}, C. and {Schlafly}, E.~F. and {Schlegel}, D. and {Schubnell}, M. and {Seo}, H. and {Shafieloo}, A. and {Sharples}, R. and {Silber}, J. and {Slosar}, A. and {Smith}, A. and {Sprayberry}, D. and {Tan}, T. and {Tarl{\'e}}, G. and {Taylor}, P. and {Trusov}, S. and {Ure{\~n}a-L{\'o}pez}, L.~A. and {Vaisakh}, R. and {Valcin}, D. and {Valdes}, F. and {Vargas-Maga{\~n}a}, M. and {Verde}, L. and {Walther}, M. and {Wang}, B. and {Wang}, M.~S. and {Weaver}, B.~A. and {Weaverdyck}, N. and {Wechsler}, R.~H. and {Weinberg}, D.~H. and {White}, M. and {Yu}, J. and {Yu}, Y. and {Yuan}, S. and {Y{\`e}che}, C. and {Zaborowski}, E.~A. and {Zarrouk}, P. and {Zhang}, H. and {Zhao}, C. and {Zhao}, R. and {Zhou}, R. and {Zhuang}, T.},
        title = "{DESI 2024 VI: cosmological constraints from the measurements of baryon acoustic oscillations}",
      journal = {\jcap},
         year = 2025,
        month = feb,
       volume = {2025},
       number = {2},
          eid = {021},
        pages = {021},
          doi = {10.1088/1475-7516/2025/02/021},
archivePrefix = {arXiv},
       eprint = {2404.03002},
 primaryClass = {astro-ph.CO},
       adsurl = {https://ui.adsabs.harvard.edu/abs/2025JCAP...02..021A}
}

@ARTICLE{DESI-DR2-Cosmo,
       author = {{DESI Collaboration} and {Abdul Karim}, M. and {Aguilar}, J. and {Ahlen}, S. and {Alam}, S. and {Allen}, L. and {Allende Prieto}, C. and {Alves}, O. and {Anand}, A. and {Andrade}, U. and {Armengaud}, E. and {Aviles}, A. and {Bailey}, S. and {Baltay}, C. and {Bansal}, P. and {Bault}, A. and {Behera}, J. and {BenZvi}, S. and {Bianchi}, D. and {Blake}, C. and {Brieden}, S. and {Brodzeller}, A. and {Brooks}, D. and {Buckley-Geer}, E. and {Burtin}, E. and {Calderon}, R. and {Canning}, R. and {Rosell}, A. Carnero and {Carrilho}, P. and {Casas}, L. and {Castander}, F.~J. and {Charles}, M. and {Chaussidon}, E. and {Chaves-Montero}, J. and {Chebat}, D. and {Chen}, X. and {Claybaugh}, T. and {Cole}, S. and {Cooper}, A.~P. and {Cuceu}, A. and {Dawson}, K.~S. and {de la Macorra}, A. and {de Mattia}, A. and {Deiosso}, N. and {Della Costa}, J. and {Demina}, R. and {Dey}, A. and {Dey}, B. and {Ding}, Z. and {Doel}, P. and {Edelstein}, J. and {Eisenstein}, D.~J. and {Elbers}, W. and {Fagrelius}, P. and {Fanning}, K. and {Fern{\'a}ndez-Garc{\'\i}a}, E. and {Ferraro}, S. and {Font-Ribera}, A. and {Forero-Romero}, J.~E. and {Frenk}, C.~S. and {Garcia-Quintero}, C. and {Garrison}, L.~H. and {Gazta{\~n}aga}, E. and {Gil-Mar{\'\i}n}, H. and {Gontcho A Gontcho}, S. and {Gonzalez}, D. and {Gonzalez-Morales}, A.~X. and {Gordon}, C. and {Green}, D. and {Gutierrez}, G. and {Guy}, J. and {Hadzhiyska}, B. and {Hahn}, C. and {He}, S. and {Herbold}, M. and {Herrera-Alcantar}, H.~K. and {Ho}, M.-F. and {Honscheid}, K. and {Howlett}, C. and {Huterer}, D. and {Ishak}, M. and {Juneau}, S. and {Kamble}, N.~V. and {Kara{\c{c}}ayl{\i}}, N.~G. and {Kehoe}, R. and {Kent}, S. and {Kim}, A.~G. and {Kirkby}, D. and {Kisner}, T. and {Koposov}, S.~E. and {Kremin}, A. and {Krolewski}, A. and {Lahav}, O. and {Lamman}, C. and {Landriau}, M. and {Lang}, D. and {Lasker}, J. and {Le Goff}, J.~M. and {Le Guillou}, L. and {Leauthaud}, A. and {Levi}, M.~E. and {Li}, Q. and {Li}, T.~S. and {Lodha}, K. and {Lokken}, M. and {Lozano-Rodr{\'\i}guez}, F. and {Magneville}, C. and {Manera}, M. and {Martini}, P. and {Matthewson}, W.~L. and {Meisner}, A. and {Mena-Fern{\'a}ndez}, J. and {Menegas}, A. and {Mergulh{\~a}o}, T. and {Miquel}, R. and {Moustakas}, J. and {Mu{\~n}oz-Guti{\'e}rrez}, A. and {Mu{\~n}oz-Santos}, D. and {Myers}, A.~D. and {Nadathur}, S. and {Naidoo}, K. and {Napolitano}, L. and {Newman}, J.~A. and {Niz}, G. and {Noriega}, H.~E. and {Paillas}, E. and {Palanque-Delabrouille}, N. and {Pan}, J. and {Peacock}, J.~A. and {Pellejero Ibanez}, M. and {Percival}, W.~J. and {P{\'e}rez-Fern{\'a}ndez}, A. and {P{\'e}rez-R{\`a}fols}, I. and {Pieri}, M.~M. and {Poppett}, C. and {Prada}, F. and {Rabinowitz}, D. and {Raichoor}, A. and {Ram{\'\i}rez-P{\'e}rez}, C. and {Rashkovetskyi}, M. and {Ravoux}, C. and {Rich}, J. and {Rocher}, A. and {Rockosi}, C. and {Rohlf}, J. and {Rom{\'a}n-Herrera}, J.~O. and {Ross}, A.~J. and {Rossi}, G. and {Ruggeri}, R. and {Ruhlmann-Kleider}, V. and {Samushia}, L. and {Sanchez}, E. and {Sanders}, N. and {Schlegel}, D. and {Schubnell}, M. and {Seo}, H. and {Shafieloo}, A. and {Sharples}, R. and {Silber}, J. and {Sinigaglia}, F. and {Sprayberry}, D. and {Tan}, T. and {Tarl{\'e}}, G. and {Taylor}, P. and {Turner}, W. and {Ure{\~n}a-L{\'o}pez}, L.~A. and {Vaisakh}, R. and {Valdes}, F. and {Valogiannis}, G. and {Vargas-Maga{\~n}a}, M. and {Verde}, L. and {Walther}, M. and {Weaver}, B.~A. and {Weinberg}, D.~H. and {White}, M. and {Wolfson}, M. and {Y{\`e}che}, C. and {Yu}, J. and {Zaborowski}, E.~A. and {Zarrouk}, P. and {Zhai}, Z. and {Zhang}, H. and {Zhao}, C. and {Zhao}, G.~B. and {Zhou}, R. and {Zou}, H. and {DESI Collaboration}},
        title = "{DESI DR2 results. II. Measurements of baryon acoustic oscillations and cosmological constraints}",
      journal = {\prd},
         year = 2025,
        month = oct,
       volume = {112},
       number = {8},
          eid = {083515},
        pages = {083515},
          doi = {10.1103/tr6y-kpc6},
archivePrefix = {arXiv},
       eprint = {2503.14738},
 primaryClass = {astro-ph.CO},
       adsurl = {https://ui.adsabs.harvard.edu/abs/2025PhRvD.112h3515A}
}

@ARTICLE{DESICollaboration2016a,
       author = {{DESI Collaboration} and {Aghamousa}, Amir and {Aguilar}, Jessica and
         {Ahlen}, Steve and {Alam}, Shadab and {Allen}, Lori E. and
         {Allende Prieto}, Carlos and {Annis}, James and {Bailey}, Stephen and
         {Balland}, Christophe and {Ballester}, Otger and {Baltay}, Charles and
         {Beaufore}, Lucas and {Bebek}, Chris and {Beers}, Timothy C. and
         {Bell}, Eric F. and {Bernal}, Jos{\'e} Luis and {Besuner}, Robert and
         {Beutler}, Florian and {Blake}, Chris and {Bleuler}, Hannes and
         {Blomqvist}, Michael and {Blum}, Robert and {Bolton}, Adam S. and
         {Briceno}, Cesar and {Brooks}, David and {Brownstein}, Joel R. and
         {Buckley-Geer}, Elizabeth and {Burden}, Angela and {Burtin}, Etienne and
         {Busca}, Nicolas G. and {Cahn}, Robert N. and {Cai}, Yan-Chuan and
         {Cardiel-Sas}, Laia and {Carlberg}, Raymond G. and
         {Carton}, Pierre-Henri and {Casas}, Ricard and {Castand
        er}, Francisco J. and {Cervantes-Cota}, Jorge L. and
         {Claybaugh}, Todd M. and {Close}, Madeline and {Coker}, Carl T. and
         {Cole}, Shaun and {Comparat}, Johan and {Cooper}, Andrew P. and
         {Cousinou}, M. -C. and {Crocce}, Martin and {Cuby}, Jean-Gabriel and
         {Cunningham}, Daniel P. and {Davis}, Tamara M. and {Dawson}, Kyle S. and
         {de la Macorra}, Axel and {De Vicente}, Juan and
         {Delubac}, Timoth{\'e}e and {Derwent}, Mark and {Dey}, Arjun and
         {Dhungana}, Govinda and {Ding}, Zhejie and {Doel}, Peter and
         {Duan}, Yutong T. and {Ealet}, Anne and {Edelstein}, Jerry and
         {Eftekharzadeh}, Sarah and {Eisenstein}, Daniel J. and {Elliott}, Ann and
         {Escoffier}, St{\'e}phanie and {Evatt}, Matthew and
         {Fagrelius}, Parker and {Fan}, Xiaohui and {Fanning}, Kevin and
         {Farahi}, Arya and {Farihi}, Jay and {Favole}, Ginevra and {Feng}, Yu and
         {Fernandez}, Enrique and {Findlay}, Joseph R. and
         {Finkbeiner}, Douglas P. and {Fitzpatrick}, Michael J. and
         {Flaugher}, Brenna and {Flender}, Samuel and {Font-Ribera}, Andreu and
         {Forero-Romero}, Jaime E. and {Fosalba}, Pablo and {Frenk}, Carlos S. and
         {Fumagalli}, Michele and {Gaensicke}, Boris T. and {Gallo}, Giuseppe and
         {Garcia-Bellido}, Juan and {Gaztanaga}, Enrique and
         {Pietro Gentile Fusillo}, Nicola and {Gerard}, Terry and
         {Gershkovich}, Irena and {Giannantonio}, Tommaso and {Gillet}, Denis and
         {Gonzalez-de-Rivera}, Guillermo and {Gonzalez-Perez}, Violeta and
         {Gott}, Shelby and {Graur}, Or and {Gutierrez}, Gaston and
         {Guy}, Julien and {Habib}, Salman and {Heetderks}, Henry and
         {Heetderks}, Ian and {Heitmann}, Katrin and {Hellwing}, Wojciech A. and
         {Herrera}, David A. and {Ho}, Shirley and {Holland}, Stephen and
         {Honscheid}, Klaus and {Huff}, Eric and {Hutchinson}, Timothy A. and
         {Huterer}, Dragan and {Hwang}, Ho Seong and
         {Illa Laguna}, Joseph Maria and {Ishikawa}, Yuzo and {Jacobs}, Dianna and
         {Jeffrey}, Niall and {Jelinsky}, Patrick and {Jennings}, Elise and
         {Jiang}, Linhua and {Jimenez}, Jorge and {Johnson}, Jennifer and
         {Joyce}, Richard and {Jullo}, Eric and {Juneau}, St{\'e}phanie and
         {Kama}, Sami and {Karcher}, Armin and {Karkar}, Sonia and
         {Kehoe}, Robert and {Kennamer}, Noble and {Kent}, Stephen and
         {Kilbinger}, Martin and {Kim}, Alex G. and {Kirkby}, David and
         {Kisner}, Theodore and {Kitanidis}, Ellie and {Kneib}, Jean-Paul and
         {Koposov}, Sergey and {Kovacs}, Eve and {Koyama}, Kazuya and
         {Kremin}, Anthony and {Kron}, Richard and {Kronig}, Luzius and
         {Kueter-Young}, Andrea and {Lacey}, Cedric G. and {Lafever}, Robin and
         {Lahav}, Ofer and {Lambert}, Andrew and {Lampton}, Michael and {Land
        riau}, Martin and {Lang}, Dustin and {Lauer}, Tod R. and
         {Le Goff}, Jean-Marc and {Le Guillou}, Laurent and
         {Le Van Suu}, Auguste and {Lee}, Jae Hyeon and {Lee}, Su-Jeong and
         {Leitner}, Daniela and {Lesser}, Michael and {Levi}, Michael E. and
         {L'Huillier}, Benjamin and {Li}, Baojiu and {Liang}, Ming and
         {Lin}, Huan and {Linder}, Eric and {Loebman}, Sarah R. and
         {Luki{\'c}}, Zarija and {Ma}, Jun and {MacCrann}, Niall and
         {Magneville}, Christophe and {Makarem}, Laleh and {Manera}, Marc and
         {Manser}, Christopher J. and {Marshall}, Robert and {Martini}, Paul and
         {Massey}, Richard and {Matheson}, Thomas and {McCauley}, Jeremy and
         {McDonald}, Patrick and {McGreer}, Ian D. and {Meisner}, Aaron and
         {Metcalfe}, Nigel and {Miller}, Timothy N. and {Miquel}, Ramon and
         {Moustakas}, John and {Myers}, Adam and {Naik}, Milind and
         {Newman}, Jeffrey A. and {Nichol}, Robert C. and {Nicola}, Andrina and
         {Nicolati da Costa}, Luiz and {Nie}, Jundan and {Niz}, Gustavo and
         {Norberg}, Peder and {Nord}, Brian and {Norman}, Dara and
         {Nugent}, Peter and {O'Brien}, Thomas and {Oh}, Minji and
         {Olsen}, Knut A.~G. and {Padilla}, Cristobal and {Padmanabhan}, Hamsa and
         {Padmanabhan}, Nikhil and {Palanque-Delabrouille}, Nathalie and
         {Palmese}, Antonella and {Pappalardo}, Daniel and
         {P{\^a}ris}, Isabelle and {Park}, Changbom and {Patej}, Anna and
         {Peacock}, John A. and {Peiris}, Hiranya V. and {Peng}, Xiyan and
         {Percival}, Will J. and {Perruchot}, Sandrine and {Pieri}, Matthew M. and
         {Pogge}, Richard and {Pollack}, Jennifer E. and {Poppett}, Claire and
         {Prada}, Francisco and {Prakash}, Abhishek and {Probst}, Ronald G. and
         {Rabinowitz}, David and {Raichoor}, Anand and {Ree}, Chang Hee and
         {Refregier}, Alexandre and {Regal}, Xavier and {Reid}, Beth and
         {Reil}, Kevin and {Rezaie}, Mehdi and {Rockosi}, Constance M. and
         {Roe}, Natalie and {Ronayette}, Samuel and {Roodman}, Aaron and
         {Ross}, Ashley J. and {Ross}, Nicholas P. and {Rossi}, Graziano and
         {Rozo}, Eduardo and {Ruhlmann-Kleider}, Vanina and {Rykoff}, Eli S. and
         {Sabiu}, Cristiano and {Samushia}, Lado and {Sanchez}, Eusebio and
         {Sanchez}, Javier and {Schlegel}, David J. and {Schneider}, Michael and
         {Schubnell}, Michael and {Secroun}, Aur{\'e}lia and {Seljak}, Uros and
         {Seo}, Hee-Jong and {Serrano}, Santiago and {Shafieloo}, Arman and
         {Shan}, Huanyuan and {Sharples}, Ray and {Sholl}, Michael J. and
         {Shourt}, William V. and {Silber}, Joseph H. and {Silva}, David R. and
         {Sirk}, Martin M. and {Slosar}, Anze and {Smith}, Alex and
         {Smoot}, George F. and {Som}, Debopam and {Song}, Yong-Seon and
         {Sprayberry}, David and {Staten}, Ryan and {Stefanik}, Andy and
         {Tarle}, Gregory and {Sien Tie}, Suk and {Tinker}, Jeremy L. and
         {Tojeiro}, Rita and {Valdes}, Francisco and {Valenzuela}, Octavio and
         {Valluri}, Monica and {Vargas-Magana}, Mariana and {Verde}, Licia and
         {Walker}, Alistair R. and {Wang}, Jiali and {Wang}, Yuting and
         {Weaver}, Benjamin A. and {Weaverdyck}, Curtis and {Wechsler}, Risa H. and
         {Weinberg}, David H. and {White}, Martin and {Yang}, Qian and
         {Yeche}, Christophe and {Zhang}, Tianmeng and {Zhao}, Gong-Bo and
         {Zheng}, Yi and {Zhou}, Xu and {Zhou}, Zhimin and {Zhu}, Yaling and
         {Zou}, Hu and {Zu}, Ying},
        title = "{The DESI Experiment Part I: Science,Targeting, and Survey Design}",
      journal = {arXiv e-prints},
         year = 2016,
        month = oct,
          eid = {arXiv:1611.00036},
        pages = {arXiv:1611.00036},
archivePrefix = {arXiv},
       eprint = {1611.00036},
 primaryClass = {astro-ph.IM},
       adsurl = {https://ui.adsabs.harvard.edu/abs/2016arXiv161100036D}
}

@ARTICLE{eBOSS2021,
       author = {{eBOSS Collaboration} and {Alam}, Shadab and {Aubert}, Marie and {Avila}, Santiago and {Balland}, Christophe and {Bautista}, Julian E. and {Bershady}, Matthew A. and {Bizyaev}, Dmitry and {Blanton}, Michael R. and {Bolton}, Adam S. and {Bovy}, Jo and {Brinkmann}, Jonathan and {Brownstein}, Joel R. and {Burtin}, Etienne and {Chabanier}, Sol{\`e}ne and {Chapman}, Michael J. and {Choi}, Peter Doohyun and {Chuang}, Chia-Hsun and {Comparat}, Johan and {Cousinou}, Marie-Claude and {Cuceu}, Andrei and {Dawson}, Kyle S. and {de la Torre}, Sylvain and {de Mattia}, Arnaud and {Agathe}, Victoria de Sainte and {des Bourboux}, H{\'e}lion du Mas and {Escoffier}, Stephanie and {Etourneau}, Thomas and {Farr}, James and {Font-Ribera}, Andreu and {Frinchaboy}, Peter M. and {Fromenteau}, Sebastien and {Gil-Mar{\'\i}n}, H{\'e}ctor and {Le Goff}, Jean-Marc and {Gonzalez-Morales}, Alma X. and {Gonzalez-Perez}, Violeta and {Grabowski}, Kathleen and {Guy}, Julien and {Hawken}, Adam J. and {Hou}, Jiamin and {Kong}, Hui and {Parker}, James and {Klaene}, Mark and {Kneib}, Jean-Paul and {Lin}, Sicheng and {Long}, Daniel and {Lyke}, Brad W. and {de la Macorra}, Axel and {Martini}, Paul and {Masters}, Karen and {Mohammad}, Faizan G. and {Moon}, Jeongin and {Mueller}, Eva-Maria and {Mu{\~n}oz-Guti{\'e}rrez}, Andrea and {Myers}, Adam D. and {Nadathur}, Seshadri and {Neveux}, Richard and {Newman}, Jeffrey A. and {Noterdaeme}, Pasquier and {Oravetz}, Audrey and {Oravetz}, Daniel and {Palanque-Delabrouille}, Nathalie and {Pan}, Kaike and {Paviot}, Romain and {Percival}, Will J. and {P{\'e}rez-R{\`a}fols}, Ignasi and {Petitjean}, Patrick and {Pieri}, Matthew M. and {Prakash}, Abhishek and {Raichoor}, Anand and {Ravoux}, Corentin and {Rezaie}, Mehdi and {Rich}, James and {Ross}, Ashley J. and {Rossi}, Graziano and {Ruggeri}, Rossana and {Ruhlmann-Kleider}, Vanina and {S{\'a}nchez}, Ariel G. and {S{\'a}nchez}, F. Javier and {S{\'a}nchez-Gallego}, Jos{\'e} R. and {Sayres}, Conor and {Schneider}, Donald P. and {Seo}, Hee-Jong and {Shafieloo}, Arman and {Slosar}, An{\v{z}}e and {Smith}, Alex and {Stermer}, Julianna and {Tamone}, Amelie and {Tinker}, Jeremy L. and {Tojeiro}, Rita and {Vargas-Maga{\~n}a}, Mariana and {Variu}, Andrei and {Wang}, Yuting and {Weaver}, Benjamin A. and {Weijmans}, Anne-Marie and {Y{\`e}che}, Christophe and {Zarrouk}, Pauline and {Zhao}, Cheng and {Zhao}, Gong-Bo and {Zheng}, Zheng},
        title = "{Completed SDSS-IV extended Baryon Oscillation Spectroscopic Survey: Cosmological implications from two decades of spectroscopic surveys at the Apache Point Observatory}",
      journal = {\prd},
         year = 2021,
        month = apr,
       volume = {103},
       number = {8},
          eid = {083533},
        pages = {083533},
          doi = {10.1103/PhysRevD.103.083533},
archivePrefix = {arXiv},
       eprint = {2007.08991},
 primaryClass = {astro-ph.CO},
       adsurl = {https://ui.adsabs.harvard.edu/abs/2021PhRvD.103h3533A}
}

@ARTICLE{Edelsbrunner2002,
       author = {{Edelsbrunner}, H. and {Letscher}, D. and {Zomorodian}, A.},
        title = "{Topological Persistence and Simplification}",
      journal = {Discrete Comput. Geom.},
         year = 2002,
        month =  nov,
       volume = {28},
       number = {4},
        pages = {511?533},
          doi = {10.1007/s00454-002-2885-2},
archivePrefix = {Springer-Verlag, Berlin, Heidelberg},
       eprint = { },
 primaryClass = {},
       adsurl = {https://doi.org/10.1007/s00454-002-2885-2}
}

@ARTICLE{Edelsbrunner2000,
       author = {{Edelsbrunner}, H. and {Harer}, J.},
        title = "{Computational Topology: An Introduction}",
      journal = {Miscellaneous Books},
         year =  2000,
        month =  jan,
       volume = {69},
       number = { },
        pages = {XII, 241},
          doi = {10.1090/mbk/069},
archivePrefix = {American Mathematical Society},
       eprint = { },
 primaryClass = {},
       adsurl = { }
}

@ARTICLE{Esteban2024,
       author = {{Esteban}, Ivan and {Gonzalez-Garcia}, M.~C. and {Maltoni}, Michele and {Martinez-Soler}, Ivan and {Pinheiro}, Jo{\~a}o Paulo and {Schwetz}, Thomas},
        title = "{NuFit-6.0: updated global analysis of three-flavor neutrino oscillations}",
      journal = {Journal of High Energy Physics},
         year = 2024,
        month = dec,
       volume = {2024},
       number = {12},
          eid = {216},
        pages = {216},
          doi = {10.1007/JHEP12(2024)216},
archivePrefix = {arXiv},
       eprint = {2410.05380},
 primaryClass = {hep-ph},
       adsurl = {https://ui.adsabs.harvard.edu/abs/2024JHEP...12..216E}
}

@ARTICLE{Esteban2020,
       author = {{Esteban}, Ivan and {Gonzalez-Garcia}, M.~C. and {Maltoni}, Michele and {Schwetz}, Thomas and {Zhou}, Albert},
        title = "{The fate of hints: updated global analysis of three-flavor neutrino oscillations}",
      journal = {Journal of High Energy Physics},
         year = 2020,
        month = sep,
       volume = {2020},
       number = {9},
          eid = {178},
        pages = {178},
          doi = {10.1007/JHEP09(2020)178},
archivePrefix = {arXiv},
       eprint = {2007.14792},
 primaryClass = {hep-ph},
       adsurl = {https://ui.adsabs.harvard.edu/abs/2020JHEP...09..178E}
}

@ARTICLE{Feldbrugge2019,
       author = {{Feldbrugge}, Job and {van Engelen}, Matti and {van de Weygaert}, Rien and {Pranav}, Pratyush and {Vegter}, Gert},
        title = "{Stochastic homology of Gaussian vs. non-Gaussian random fields: graphs towards Betti numbers and persistence diagrams}",
      journal = {\jcap},
         year = 2019,
        month = sep,
       volume = {2019},
       number = {9},
          eid = {052},
        pages = {052},
          doi = {10.1088/1475-7516/2019/09/052},
archivePrefix = {arXiv},
       eprint = {1908.01619},
 primaryClass = {astro-ph.CO},
       adsurl = {https://ui.adsabs.harvard.edu/abs/2019JCAP...09..052F}
}

@ARTICLE{Forman2002,
       author = {{Forman},  R.},
        title = "{A User's Guide To Discrete Morse Theory}",
      journal = {Seminaire Lotharingien de Combinatoire},
         year = 2002,
        month = sep,
       volume = {48},
       number = {},
        pages = {35},
          doi = { },
archivePrefix = {arXiv},
       eprint = {math/0212354},
 primaryClass = {math.CO},
       adsurl = { }
}

@ARTICLE{Forman1998,
       author = {{Forman},  R.},
           title = {Morse Theory for Cell Complexes},
      journal = {Advances in Mathematics},
     volume = {134},
    number = {1},
      pages = {90-145},
        year = {1998},
        issn = {0001-8708},
         doi = {https://doi.org/10.1006/aima.1997.1650},
         url = {https://www.sciencedirect.com/science/article/pii/S0001870897916509}
}

@ARTICLE{Gay2012,
       author = {{Gay}, Christophe and {Pichon}, Christophe and {Pogosyan}, Dmitry},
        title = "{Non-Gaussian statistics of critical sets in 2D and 3D: Peaks, voids, saddles, genus, and skeleton}",
      journal = {\prd},
         year = 2012,
        month = jan,
       volume = {85},
       number = {2},
          eid = {023011},
        pages = {023011},
          doi = {10.1103/PhysRevD.85.023011},
archivePrefix = {arXiv},
       eprint = {1110.0261},
 primaryClass = {astro-ph.CO},
       adsurl = {https://ui.adsabs.harvard.edu/abs/2012PhRvD..85b3011G}
}

@ARTICLE{GreenMeyers2025,
       author = {{Green}, Daniel and {Meyers}, Joel},
        title = "{Cosmological preference for a negative neutrino mass}",
      journal = {\prd},
         year = 2025,
        month = apr,
       volume = {111},
       number = {8},
          eid = {083507},
        pages = {083507},
          doi = {10.1103/PhysRevD.111.083507},
archivePrefix = {arXiv},
       eprint = {2407.07878},
 primaryClass = {astro-ph.CO},
       adsurl = {https://ui.adsabs.harvard.edu/abs/2025PhRvD.111h3507G}
}

@ARTICLE{Gyulassy2008,
       author = {{Gyulassy}, A. },
        title = "{Combinatorial Construction of Morse-Smale Complexes for Data Analysis and Visualization}",
      journal = {Phd Thesis},
         year = 2008,
        month =  {},
       volume = {},
       number = {},
        pages = {},
          doi = {10.5555/1626652},
archivePrefix = {University of California at Davis},
       eprint = {},
 primaryClass = {9781109061529},
       adsurl = {}
}

@ARTICLE{Hahn2020,
       author = {{Hahn}, ChangHoon and {Villaescusa-Navarro}, Francisco and {Castorina}, Emanuele and {Scoccimarro}, Roman},
        title = "{Constraining M$_{{\ensuremath{\nu}}}$ with the bispectrum. Part I. Breaking parameter degeneracies}",
      journal = {\jcap},
         year = 2020,
        month = mar,
       volume = {2020},
       number = {3},
          eid = {040},
        pages = {040},
          doi = {10.1088/1475-7516/2020/03/040},
archivePrefix = {arXiv},
       eprint = {1909.11107},
 primaryClass = {astro-ph.CO},
       adsurl = {https://ui.adsabs.harvard.edu/abs/2020JCAP...03..040H}
}

@ARTICLE{Heydenreich2022,
       author = {{Heydenreich}, Sven and {Br{\"u}ck}, Benjamin and {Burger}, Pierre and {Harnois-D{\'e}raps}, Joachim and {Unruh}, Sandra and {Castro}, Tiago and {Dolag}, Klaus and {Martinet}, Nicolas},
        title = "{Persistent homology in cosmic shear. II. A tomographic analysis of DES-Y1}",
      journal = {\aap},
         year = 2022,
        month = nov,
       volume = {667},
          eid = {A125},
        pages = {A125},
          doi = {10.1051/0004-6361/202243868},
archivePrefix = {arXiv},
       eprint = {2204.11831},
 primaryClass = {astro-ph.CO},
       adsurl = {https://ui.adsabs.harvard.edu/abs/2022A&A...667A.125H}
}

@ARTICLE{Jalali2024,
       author = {{Jalali Kanafi}, M.~H. and {Ansarifard}, S. and {Movahed}, S.~M.~S.},
        title = "{Imprint of massive neutrinos on Persistent Homology of large-scale structure}",
      journal = {\mnras},
         year = 2024,
        month = nov,
       volume = {535},
       number = {1},
        pages = {657-674},
          doi = {10.1093/mnras/stae2044},
archivePrefix = {arXiv},
       eprint = {2311.13520},
 primaryClass = {astro-ph.CO},
       adsurl = {https://ui.adsabs.harvard.edu/abs/2024MNRAS.535..657J}
}

@ARTICLE{Kraljic2022,
       author = {{Kraljic}, Katarina and {Laigle}, Clotilde and {Pichon}, Christophe and {Peirani}, Sebastien and {Codis}, Sandrine and {Shim}, Junsup and {Cadiou}, Corentin and {Pogosyan}, Dmitri and {Arnouts}, St{\'e}phane and {Pieri}, Matthiew and {Ir{\v{s}}i{\v{c}}}, Vid and {Morrison}, Sean S. and {O{\~n}orbe}, Jose and {P{\'e}rez-R{\`a}fols}, Ignasi and {Dalton}, Gavin},
        title = "{Forecasts for WEAVE-QSO: 3D clustering and connectivity of critical points with Lyman-$\alpha$ tomography}",
      journal = {arXiv e-prints},
         year = 2022,
        month = jan,
          eid = {arXiv:2201.02606},
        pages = {arXiv:2201.02606},
archivePrefix = {arXiv},
       eprint = {2201.02606},
 primaryClass = {astro-ph.CO},
       adsurl = {https://ui.adsabs.harvard.edu/abs/2022arXiv220102606K}
}

@ARTICLE{Kuruvilla2020,
       author = {{Kuruvilla}, Joseph and {Aghanim}, Nabila and {McCarthy}, Ian G.},
        title = "{Imprint of baryons and massive neutrinos on velocity statistics}",
      journal = {\aap},
         year = 2020,
        month = dec,
       volume = {644},
          eid = {A170},
        pages = {A170},
          doi = {10.1051/0004-6361/202039115},
archivePrefix = {arXiv},
       eprint = {2010.05911},
 primaryClass = {astro-ph.CO},
       adsurl = {https://ui.adsabs.harvard.edu/abs/2020A&A...644A.170K}
}

@ARTICLE{Labate2025,
       author = {{Labate}, Andrea and {Guidi}, Massimo and {Moresco}, Michele and {Veropalumbo}, Alfonso},
        title = "{The imprints of massive neutrinos on the 3-point correlation function of large-scale structures}",
      journal = {arXiv e-prints},
         year = 2025,
        month = dec,
          eid = {arXiv:2512.16992},
        pages = {arXiv:2512.16992},
          doi = {10.48550/arXiv.2512.16992},
archivePrefix = {arXiv},
       eprint = {2512.16992},
 primaryClass = {astro-ph.CO},
       adsurl = {https://ui.adsabs.harvard.edu/abs/2025arXiv251216992L}
}

@ARTICLE{Laureijs2011,
       author = {{Laureijs}, R. and {Amiaux}, J. and {Arduini}, S. and {Augu{\`e}res}, J. -L. and {Brinchmann}, J. and {Cole}, R. and {Cropper}, M. and {Dabin}, C. and {Duvet}, L. and {Ealet}, A. and {Garilli}, B. and {Gondoin}, P. and {Guzzo}, L. and {Hoar}, J. and {Hoekstra}, H. and {Holmes}, R. and {Kitching}, T. and {Maciaszek}, T. and {Mellier}, Y. and {Pasian}, F. and {Percival}, W. and {Rhodes}, J. and {Saavedra Criado}, G. and {Sauvage}, M. and {Scaramella}, R. and {Valenziano}, L. and {Warren}, S. and {Bender}, R. and {Castander}, F. and {Cimatti}, A. and {Le F{\`e}vre}, O. and {Kurki-Suonio}, H. and {Levi}, M. and {Lilje}, P. and {Meylan}, G. and {Nichol}, R. and {Pedersen}, K. and {Popa}, V. and {Rebolo Lopez}, R. and {Rix}, H. -W. and {Rottgering}, H. and {Zeilinger}, W. and {Grupp}, F. and {Hudelot}, P. and {Massey}, R. and {Meneghetti}, M. and {Miller}, L. and {Paltani}, S. and {Paulin-Henriksson}, S. and {Pires}, S. and {Saxton}, C. and {Schrabback}, T. and {Seidel}, G. and {Walsh}, J. and {Aghanim}, N. and {Amendola}, L. and {Bartlett}, J. and {Baccigalupi}, C. and {Beaulieu}, J. -P. and {Benabed}, K. and {Cuby}, J. -G. and {Elbaz}, D. and {Fosalba}, P. and {Gavazzi}, G. and {Helmi}, A. and {Hook}, I. and {Irwin}, M. and {Kneib}, J. -P. and {Kunz}, M. and {Mannucci}, F. and {Moscardini}, L. and {Tao}, C. and {Teyssier}, R. and {Weller}, J. and {Zamorani}, G. and {Zapatero Osorio}, M.~R. and {Boulade}, O. and {Foumond}, J.~J. and {Di Giorgio}, A. and {Guttridge}, P. and {James}, A. and {Kemp}, M. and {Martignac}, J. and {Spencer}, A. and {Walton}, D. and {Bl{\"u}mchen}, T. and {Bonoli}, C. and {Bortoletto}, F. and {Cerna}, C. and {Corcione}, L. and {Fabron}, C. and {Jahnke}, K. and {Ligori}, S. and {Madrid}, F. and {Martin}, L. and {Morgante}, G. and {Pamplona}, T. and {Prieto}, E. and {Riva}, M. and {Toledo}, R. and {Trifoglio}, M. and {Zerbi}, F. and {Abdalla}, F. and {Douspis}, M. and {Grenet}, C. and {Borgani}, S. and {Bouwens}, R. and {Courbin}, F. and {Delouis}, J. -M. and {Dubath}, P. and {Fontana}, A. and {Frailis}, M. and {Grazian}, A. and {Koppenh{\"o}fer}, J. and {Mansutti}, O. and {Melchior}, M. and {Mignoli}, M. and {Mohr}, J. and {Neissner}, C. and {Noddle}, K. and {Poncet}, M. and {Scodeggio}, M. and {Serrano}, S. and {Shane}, N. and {Starck}, J. -L. and {Surace}, C. and {Taylor}, A. and {Verdoes-Kleijn}, G. and {Vuerli}, C. and {Williams}, O.~R. and {Zacchei}, A. and {Altieri}, B. and {Escudero Sanz}, I. and {Kohley}, R. and {Oosterbroek}, T. and {Astier}, P. and {Bacon}, D. and {Bardelli}, S. and {Baugh}, C. and {Bellagamba}, F. and {Benoist}, C. and {Bianchi}, D. and {Biviano}, A. and {Branchini}, E. and {Carbone}, C. and {Cardone}, V. and {Clements}, D. and {Colombi}, S. and {Conselice}, C. and {Cresci}, G. and {Deacon}, N. and {Dunlop}, J. and {Fedeli}, C. and {Fontanot}, F. and {Franzetti}, P. and {Giocoli}, C. and {Garcia-Bellido}, J. and {Gow}, J. and {Heavens}, A. and {Hewett}, P. and {Heymans}, C. and {Holland}, A. and {Huang}, Z. and {Ilbert}, O. and {Joachimi}, B. and {Jennins}, E. and {Kerins}, E. and {Kiessling}, A. and {Kirk}, D. and {Kotak}, R. and {Krause}, O. and {Lahav}, O. and {van Leeuwen}, F. and {Lesgourgues}, J. and {Lombardi}, M. and {Magliocchetti}, M. and {Maguire}, K. and {Majerotto}, E. and {Maoli}, R. and {Marulli}, F. and {Maurogordato}, S. and {McCracken}, H. and {McLure}, R. and {Melchiorri}, A. and {Merson}, A. and {Moresco}, M. and {Nonino}, M. and {Norberg}, P. and {Peacock}, J. and {Pello}, R. and {Penny}, M. and {Pettorino}, V. and {Di Porto}, C. and {Pozzetti}, L. and {Quercellini}, C. and {Radovich}, M. and {Rassat}, A. and {Roche}, N. and {Ronayette}, S. and {Rossetti}, E. and {Sartoris}, B. and {Schneider}, P. and {Semboloni}, E. and {Serjeant}, S. and {Simpson}, F. and {Skordis}, C. and {Smadja}, G. and {Smartt}, S. and {Spano}, P. and {Spiro}, S. and {Sullivan}, M. and {Tilquin}, A. and {Trotta}, R. and {Verde}, L. and {Wang}, Y. and {Williger}, G. and {Zhao}, G. and {Zoubian}, J. and {Zucca}, E.},
        title = "{Euclid Definition Study Report}",
      journal = {arXiv e-prints},
         year = 2011,
        month = oct,
          eid = {arXiv:1110.3193},
        pages = {arXiv:1110.3193},
archivePrefix = {arXiv},
       eprint = {1110.3193},
 primaryClass = {astro-ph.CO},
       adsurl = {https://ui.adsabs.harvard.edu/abs/2011arXiv1110.3193L}
}

@ARTICLE{LandySzalay1993,
       author = {{Landy}, Stephen D. and {Szalay}, Alexander S.},
        title = "{Bias and Variance of Angular Correlation Functions}",
      journal = {\apj},
         year = 1993,
        month = jul,
       volume = {412},
        pages = {64},
          doi = {10.1086/172900},
       adsurl = {https://ui.adsabs.harvard.edu/abs/1993ApJ...412...64L}
}

@ARTICLE{LesgourguesPastor2006,
       author = {{Lesgourgues}, Julien and {Pastor}, Sergio},
        title = "{Massive neutrinos and cosmology}",
      journal = {\physrep},
         year = 2006,
        month = jul,
       volume = {429},
       number = {6},
        pages = {307-379},
          doi = {10.1016/j.physrep.2006.04.001},
archivePrefix = {arXiv},
       eprint = {astro-ph/0603494},
 primaryClass = {astro-ph},
       adsurl = {https://ui.adsabs.harvard.edu/abs/2006PhR...429..307L}
}

@ARTICLE{Lewis2000,
       author = {{Lewis}, Antony and {Challinor}, Anthony and {Lasenby}, Anthony},
        title = "{Efficient Computation of Cosmic Microwave Background Anisotropies in Closed Friedmann-Robertson-Walker Models}",
      journal = {\apj},
         year = 2000,
        month = aug,
       volume = {538},
       number = {2},
        pages = {473-476},
          doi = {10.1086/309179},
archivePrefix = {arXiv},
       eprint = {astro-ph/9911177},
 primaryClass = {astro-ph},
       adsurl = {https://ui.adsabs.harvard.edu/abs/2000ApJ...538..473L}
}

@ARTICLE{Liu2018,
       author = {{Liu}, Jia and {Bird}, Simeon and {Zorrilla Matilla}, Jos{\'e} Manuel and {Hill}, J. Colin and {Haiman}, Zolt{\'a}n and {Madhavacheril}, Mathew S. and {Petri}, Andrea and {Spergel}, David N.},
        title = "{MassiveNuS: cosmological massive neutrino simulations}",
      journal = {\jcap},
         year = 2018,
        month = mar,
       volume = {2018},
       number = {3},
          eid = {049},
        pages = {049},
          doi = {10.1088/1475-7516/2018/03/049},
archivePrefix = {arXiv},
       eprint = {1711.10524},
 primaryClass = {astro-ph.CO},
       adsurl = {https://ui.adsabs.harvard.edu/abs/2018JCAP...03..049L}
}

@ARTICLE{LSST2019,
       author = {{LSST Collaboration} and {Ivezi{\'c}}, {\v{Z}}eljko and {Kahn}, Steven M. and
         {Tyson}, J. Anthony and {Abel}, Bob and {Acosta}, Emily and
         {Allsman}, Robyn and {Alonso}, David and {AlSayyad}, Yusra and
         {Anderson}, Scott F. and {Andrew}, John and {Angel}, James Roger P. and
         {Angeli}, George Z. and {Ansari}, Reza and {Antilogus}, Pierre and
         {Araujo}, Constanza and {Armstrong}, Robert and {Arndt}, Kirk T. and
         {Astier}, Pierre and {Aubourg}, {\'E}ric and {Auza}, Nicole and
         {Axelrod}, Tim S. and {Bard}, Deborah J. and {Barr}, Jeff D. and
         {Barrau}, Aurelian and {Bartlett}, James G. and {Bauer}, Amanda E. and
         {Bauman}, Brian J. and {Baumont}, Sylvain and {Bechtol}, Ellen and
         {Bechtol}, Keith and {Becker}, Andrew C. and {Becla}, Jacek and
         {Beldica}, Cristina and {Bellavia}, Steve and {Bianco}, Federica B. and
         {Biswas}, Rahul and {Blanc}, Guillaume and {Blazek}, Jonathan and {Bland
        ford}, Roger D. and {Bloom}, Josh S. and {Bogart}, Joanne and
         {Bond}, Tim W. and {Booth}, Michael T. and {Borgland}, Anders W. and
         {Borne}, Kirk and {Bosch}, James F. and {Boutigny}, Dominique and
         {Brackett}, Craig A. and {Bradshaw}, Andrew and {Brand
        t}, William Nielsen and {Brown}, Michael E. and {Bullock}, James S. and
         {Burchat}, Patricia and {Burke}, David L. and {Cagnoli}, Gianpietro and
         {Calabrese}, Daniel and {Callahan}, Shawn and {Callen}, Alice L. and
         {Carlin}, Jeffrey L. and {Carlson}, Erin L. and {Chand
        rasekharan}, Srinivasan and {Charles-Emerson}, Glenaver and
         {Chesley}, Steve and {Cheu}, Elliott C. and {Chiang}, Hsin-Fang and
         {Chiang}, James and {Chirino}, Carol and {Chow}, Derek and
         {Ciardi}, David R. and {Claver}, Charles F. and {Cohen-Tanugi}, Johann and
         {Cockrum}, Joseph J. and {Coles}, Rebecca and {Connolly}, Andrew J. and
         {Cook}, Kem H. and {Cooray}, Asantha and {Covey}, Kevin R. and
         {Cribbs}, Chris and {Cui}, Wei and {Cutri}, Roc and {Daly}, Philip N. and
         {Daniel}, Scott F. and {Daruich}, Felipe and {Daubard}, Guillaume and
         {Daues}, Greg and {Dawson}, William and {Delgado}, Francisco and
         {Dellapenna}, Alfred and {de Peyster}, Robert and
         {de Val-Borro}, Miguel and {Digel}, Seth W. and {Doherty}, Peter and
         {Dubois}, Richard and {Dubois-Felsmann}, Gregory P. and
         {Durech}, Josef and {Economou}, Frossie and {Eifler}, Tim and
         {Eracleous}, Michael and {Emmons}, Benjamin L. and
         {Fausti Neto}, Angelo and {Ferguson}, Henry and {Figueroa}, Enrique and
         {Fisher-Levine}, Merlin and {Focke}, Warren and {Foss}, Michael D. and
         {Frank}, James and {Freemon}, Michael D. and {Gangler}, Emmanuel and
         {Gawiser}, Eric and {Geary}, John C. and {Gee}, Perry and
         {Geha}, Marla and {Gessner}, Charles J.~B. and {Gibson}, Robert R. and
         {Gilmore}, D. Kirk and {Glanzman}, Thomas and {Glick}, William and
         {Goldina}, Tatiana and {Goldstein}, Daniel A. and {Goodenow}, Iain and
         {Graham}, Melissa L. and {Gressler}, William J. and {Gris}, Philippe and
         {Guy}, Leanne P. and {Guyonnet}, Augustin and {Haller}, Gunther and
         {Harris}, Ron and {Hascall}, Patrick A. and {Haupt}, Justine and {Hernand
        ez}, Fabio and {Herrmann}, Sven and {Hileman}, Edward and
         {Hoblitt}, Joshua and {Hodgson}, John A. and {Hogan}, Craig and
         {Howard}, James D. and {Huang}, Dajun and {Huffer}, Michael E. and
         {Ingraham}, Patrick and {Innes}, Walter R. and {Jacoby}, Suzanne H. and
         {Jain}, Bhuvnesh and {Jammes}, Fabrice and {Jee}, M. James and
         {Jenness}, Tim and {Jernigan}, Garrett and {Jevremovi{\'c}}, Darko and
         {Johns}, Kenneth and {Johnson}, Anthony S. and
         {Johnson}, Margaret W.~G. and {Jones}, R. Lynne and
         {Juramy-Gilles}, Claire and {Juri{\'c}}, Mario and {Kalirai}, Jason S. and
         {Kallivayalil}, Nitya J. and {Kalmbach}, Bryce and
         {Kantor}, Jeffrey P. and {Karst}, Pierre and {Kasliwal}, Mansi M. and
         {Kelly}, Heather and {Kessler}, Richard and {Kinnison}, Veronica and
         {Kirkby}, David and {Knox}, Lloyd and {Kotov}, Ivan V. and
         {Krabbendam}, Victor L. and {Krughoff}, K. Simon and
         {Kub{\'a}nek}, Petr and {Kuczewski}, John and {Kulkarni}, Shri and
         {Ku}, John and {Kurita}, Nadine R. and {Lage}, Craig S. and
         {Lambert}, Ron and {Lange}, Travis and {Langton}, J. Brian and
         {Le Guillou}, Laurent and {Levine}, Deborah and {Liang}, Ming and
         {Lim}, Kian-Tat and {Lintott}, Chris J. and {Long}, Kevin E. and
         {Lopez}, Margaux and {Lotz}, Paul J. and {Lupton}, Robert H. and
         {Lust}, Nate B. and {MacArthur}, Lauren A. and {Mahabal}, Ashish and {Mand
        elbaum}, Rachel and {Markiewicz}, Thomas W. and {Marsh}, Darren S. and
         {Marshall}, Philip J. and {Marshall}, Stuart and {May}, Morgan and
         {McKercher}, Robert and {McQueen}, Michelle and {Meyers}, Joshua and
         {Migliore}, Myriam and {Miller}, Michelle and {Mills}, David J. and
         {Miraval}, Connor and {Moeyens}, Joachim and {Moolekamp}, Fred E. and
         {Monet}, David G. and {Moniez}, Marc and {Monkewitz}, Serge and
         {Montgomery}, Christopher and {Morrison}, Christopher B. and
         {Mueller}, Fritz and {Muller}, Gary P. and
         {Mu{\~n}oz Arancibia}, Freddy and {Neill}, Douglas R. and
         {Newbry}, Scott P. and {Nief}, Jean-Yves and {Nomerotski}, Andrei and
         {Nordby}, Martin and {O'Connor}, Paul and {Oliver}, John and
         {Olivier}, Scot S. and {Olsen}, Knut and {O'Mullane}, William and
         {Ortiz}, Sandra and {Osier}, Shawn and {Owen}, Russell E. and
         {Pain}, Reynald and {Palecek}, Paul E. and {Parejko}, John K. and
         {Parsons}, James B. and {Pease}, Nathan M. and {Peterson}, J. Matt and
         {Peterson}, John R. and {Petravick}, Donald L. and
         {Libby Petrick}, M.~E. and {Petry}, Cathy E. and
         {Pierfederici}, Francesco and {Pietrowicz}, Stephen and {Pike}, Rob and
         {Pinto}, Philip A. and {Plante}, Raymond and {Plate}, Stephen and
         {Plutchak}, Joel P. and {Price}, Paul A. and {Prouza}, Michael and
         {Radeka}, Veljko and {Rajagopal}, Jayadev and {Rasmussen}, Andrew P. and
         {Regnault}, Nicolas and {Reil}, Kevin A. and {Reiss}, David J. and
         {Reuter}, Michael A. and {Ridgway}, Stephen T. and {Riot}, Vincent J. and
         {Ritz}, Steve and {Robinson}, Sean and {Roby}, William and
         {Roodman}, Aaron and {Rosing}, Wayne and {Roucelle}, Cecille and
         {Rumore}, Matthew R. and {Russo}, Stefano and {Saha}, Abhijit and
         {Sassolas}, Benoit and {Schalk}, Terry L. and {Schellart}, Pim and
         {Schindler}, Rafe H. and {Schmidt}, Samuel and {Schneider}, Donald P. and
         {Schneider}, Michael D. and {Schoening}, William and
         {Schumacher}, German and {Schwamb}, Megan E. and {Sebag}, Jacques and
         {Selvy}, Brian and {Sembroski}, Glenn H. and {Seppala}, Lynn G. and
         {Serio}, Andrew and {Serrano}, Eduardo and {Shaw}, Richard A. and
         {Shipsey}, Ian and {Sick}, Jonathan and {Silvestri}, Nicole and
         {Slater}, Colin T. and {Smith}, J. Allyn and {Smith}, R. Chris and
         {Sobhani}, Shahram and {Soldahl}, Christine and
         {Storrie-Lombardi}, Lisa and {Stover}, Edward and
         {Strauss}, Michael A. and {Street}, Rachel A. and
         {Stubbs}, Christopher W. and {Sullivan}, Ian S. and {Sweeney}, Donald and
         {Swinbank}, John D. and {Szalay}, Alexander and {Takacs}, Peter and
         {Tether}, Stephen A. and {Thaler}, Jon J. and {Thayer}, John Gregg and
         {Thomas}, Sandrine and {Thornton}, Adam J. and {Thukral}, Vaikunth and
         {Tice}, Jeffrey and {Trilling}, David E. and {Turri}, Max and
         {Van Berg}, Richard and {Vanden Berk}, Daniel and {Vetter}, Kurt and
         {Virieux}, Francoise and {Vucina}, Tomislav and {Wahl}, William and
         {Walkowicz}, Lucianne and {Walsh}, Brian and {Walter}, Christopher W. and
         {Wang}, Daniel L. and {Wang}, Shin-Yawn and {Warner}, Michael and
         {Wiecha}, Oliver and {Willman}, Beth and {Winters}, Scott E. and
         {Wittman}, David and {Wolff}, Sidney C. and {Wood-Vasey}, W. Michael and
         {Wu}, Xiuqin and {Xin}, Bo and {Yoachim}, Peter and {Zhan}, Hu},
        title = "{LSST: From Science Drivers to Reference Design and Anticipated Data Products}",
      journal = {\apj},
         year = 2019,
        month = mar,
       volume = {873},
       number = {2},
          eid = {111},
        pages = {111},
          doi = {10.3847/1538-4357/ab042c},
archivePrefix = {arXiv},
       eprint = {0805.2366},
 primaryClass = {astro-ph},
       adsurl = {https://ui.adsabs.harvard.edu/abs/2019ApJ...873..111I}
}

@ARTICLE{Luchina2025,
       author = {{Luchina}, Davide and {Roncarelli}, Mauro and {Calabrese}, Matteo and {Fabbian}, Giulio and {Carbone}, Carmelita},
        title = "{DEMNUni: the Sunyaev-Zel'dovich effect in the presence of massive neutrinos and dynamical dark energy}",
      journal = {arXiv e-prints},
         year = 2025,
        month = mar,
          eid = {arXiv:2503.16355},
        pages = {arXiv:2503.16355},
          doi = {10.48550/arXiv.2503.16355},
archivePrefix = {arXiv},
       eprint = {2503.16355},
 primaryClass = {astro-ph.CO},
       adsurl = {https://ui.adsabs.harvard.edu/abs/2025arXiv250316355L}
}

@ARTICLE{Maggiore2025,
       author = {{Maggiore}, Leonardo and {Contarini}, Sofia and {Giocoli}, Carlo and {Moscardini}, Lauro},
        title = "{Weak-lensing tunnel voids in simulated light cones: A new pipeline to investigate modified gravity and massive neutrinos signatures}",
      journal = {\aap},
         year = 2025,
        month = sep,
       volume = {701},
          eid = {A55},
        pages = {A55},
          doi = {10.1051/0004-6361/202554968},
archivePrefix = {arXiv},
       eprint = {2504.02041},
 primaryClass = {astro-ph.CO},
       adsurl = {https://ui.adsabs.harvard.edu/abs/2025A&A...701A..55M}
}

@ARTICLE{Massara2021,
       author = {{Massara}, Elena and {Villaescusa-Navarro}, Francisco and {Ho}, Shirley and {Dalal}, Neal and {Spergel}, David N.},
        title = "{Using the Marked Power Spectrum to Detect the Signature of Neutrinos in Large-Scale Structure}",
      journal = {\prl},
         year = 2021,
        month = jan,
       volume = {126},
       number = {1},
          eid = {011301},
        pages = {011301},
          doi = {10.1103/PhysRevLett.126.011301},
archivePrefix = {arXiv},
       eprint = {2001.11024},
 primaryClass = {astro-ph.CO},
       adsurl = {https://ui.adsabs.harvard.edu/abs/2021PhRvL.126a1301M}
}

@ARTICLE{Milnor1963,
       author = {{Milnor}, J.},
        title = "{Morse Theory}",
      journal = {Annals of Mathematics Studies},
         year = 1963,
        month =  may,
       volume = {51},
       number = {},
        pages = {160},
          doi = {9780691080086},
archivePrefix = {Princeton University Press},
       eprint = {},
 primaryClass = { },
       adsurl = {https://press.princeton.edu/books/paperback/9780691080086/morse-theory-am-51-volume-51}
}

@ARTICLE{MoonRossiYu2023,
       author = {{Moon}, Jeongin and {Rossi}, Graziano and {Yu}, Hogyun},
        title = "{Signature of Massive Neutrinos from the Clustering of Critical Points. I. Density-threshold-based Analysis in Configuration Space}",
      journal = {\apjs},
         year = 2023,
        month = jan,
       volume = {264},
       number = {1},
          eid = {26},
        pages = {26},
          doi = {10.3847/1538-4365/aca32a},
archivePrefix = {arXiv},
       eprint = {2302.03171},
 primaryClass = {astro-ph.CO},
       adsurl = {https://ui.adsabs.harvard.edu/abs/2023ApJS..264...26M}
}

@ARTICLE{Naredo-Tuero2024,
       author = {{Naredo-Tuero}, Daniel and {Escudero}, Miguel and {Enrique Fernandez-Martinez} and {Marcano}, Xabier and {Poulin}, Vivian},
        title = "{Critical look at the cosmological neutrino mass bound}",
      journal = {\prd},
         year = 2024,
        month = dec,
       volume = {110},
       number = {12},
          eid = {123537},
        pages = {123537},
          doi = {10.1103/PhysRevD.110.123537},
archivePrefix = {arXiv},
       eprint = {2407.13831},
 primaryClass = {astro-ph.CO},
       adsurl = {https://ui.adsabs.harvard.edu/abs/2024PhRvD.110l3537N}
}

@ARTICLE{Navas2024,
       author = {{Navas}, S. and {Amsler}, C. and {Gutsche}, T. and {Hanhart}, C. and {Hern{\'a}ndez-Rey}, J.~J. and {Louren{\c{c}}o}, C. and {Masoni}, A. and {Mikhasenko}, M. and {Mitchell}, R.~E. and {Patrignani}, C. and {Schwanda}, C. and {Spanier}, S. and {Venanzoni}, G. and {Yuan}, C.~Z. and {Agashe}, K. and {Aielli}, G. and {Allanach}, B.~C. and {Alvarez-Mu{\~n}iz}, J. and {Antonelli}, M. and {Aschenauer}, E.~C. and {Asner}, D.~M. and {Assamagan}, K. and {Baer}, H. and {Banerjee}, Sw. and {Barnett}, R.~M. and {Baudis}, L. and {Bauer}, C.~W. and {Beatty}, J.~J. and {Beringer}, J. and {Bettini}, A. and {Biebel}, O. and {Black}, K.~M. and {Blucher}, E. and {Bonventre}, R. and {Briere}, R.~A. and {Buckley}, A. and {Burkert}, V.~D. and {Bychkov}, M.~A. and {Cahn}, R.~N. and {Cao}, Z. and {Carena}, M. and {Casarosa}, G. and {Ceccucci}, A. and {Cerri}, A. and {Chivukula}, R.~S. and {Cowan}, G. and {Cranmer}, K. and {Crede}, V. and {Cremonesi}, O. and {D'Ambrosio}, G. and {Damour}, T. and {de Florian}, D. and {de Gouv{\^e}a}, A. and {DeGrand}, T. and {Demers}, S. and {Demiragli}, Z. and {Dobrescu}, B.~A. and {D'Onofrio}, M. and {Doser}, M. and {Dreiner}, H.~K. and {Eerola}, P. and {Egede}, U. and {Eidelman}, S. and {El-Khadra}, A.~X. and {Ellis}, J. and {Eno}, S.~C. and {Erler}, J. and {Ezhela}, V.~V. and {Fava}, A. and {Fetscher}, W. and {Fields}, B.~D. and {Freitas}, A. and {Gallagher}, H. and {Gershon}, T. and {Gershtein}, Y. and {Gherghetta}, T. and {Gonzalez-Garcia}, M.~C. and {Goodman}, M. and {Grab}, C. and {Gritsan}, A.~V. and {Grojean}, C. and {Groom}, D.~E. and {Gr{\"u}newald}, M. and {Gurtu}, A. and {Haber}, H.~E. and {Hamel}, M. and {Hashimoto}, S. and {Hayato}, Y. and {Hebecker}, A. and {Heinemeyer}, S. and {Hikasa}, K. and {Hisano}, J. and {H{\"o}cker}, A. and {Holder}, J. and {Hsu}, L. and {Huston}, J. and {Hyodo}, T. and {Ianni}, Al. and {Kado}, M. and {Karliner}, M. and {Katz}, U.~F. and {Kenzie}, M. and {Khoze}, V.~A. and {Klein}, S.~R. and {Krauss}, F. and {Kreps}, M. and {Kri{\v{z}}an}, P. and {Krusche}, B. and {Kwon}, Y. and {Lahav}, O. and {Lellouch}, L.~P. and {Lesgourgues}, J. and {Liddle}, A.~R. and {Ligeti}, Z. and {Lin}, C.-J. and {Lippmann}, C. and {Liss}, T.~M. and {Lister}, A. and {Littenberg}, L. and {Lugovsky}, K.~S. and {Lugovsky}, S.~B. and {Lusiani}, A. and {Makida}, Y. and {Maltoni}, F. and {Manohar}, A.~V. and {Marciano}, W.~J. and {Matthews}, J. and {Mei{\ss}ner}, U.-G. and {Melzer-Pellmann}, I.-A. and {Mertsch}, P. and {Miller}, D.~J. and {Milstead}, D. and {M{\"o}nig}, K. and {Molaro}, P. and {Moortgat}, F. and {Moskovic}, M. and {Nagata}, N. and {Nakamura}, K. and {Narain}, M. and {Nason}, P. and {Nelles}, A. and {Neubert}, M. and {Nir}, Y. and {O'Connell}, H.~B. and {O'Hare}, C.~A.~J. and {Olive}, K.~A. and {Peacock}, J.~A. and {Pianori}, E. and {Pich}, A. and {Piepke}, A. and {Pietropaolo}, F. and {Pomarol}, A. and {Pordes}, S. and {Profumo}, S. and {Quadt}, A. and {Rabbertz}, K. and {Rademacker}, J. and {Raffelt}, G. and {Ramsey-Musolf}, M. and {Richardson}, P. and {Ringwald}, A. and {Robinson}, D.~J. and {Roesler}, S. and {Rolli}, S. and {Romaniouk}, A. and {Rosenberg}, L.~J. and {Rosner}, J.~L. and {Rybka}, G. and {Ryskin}, M.~G. and {Ryutin}, R.~A. and {Safdi}, B. and {Sakai}, Y. and {Sarkar}, S. and {Sauli}, F. and {Schneider}, O. and {Sch{\"o}nert}, S. and {Scholberg}, K. and {Schwartz}, A.~J. and {Schwiening}, J. and {Scott}, D. and {Sefkow}, F. and {Seljak}, U. and {Sharma}, V. and {Sharpe}, S.~R. and {Shiltsev}, V. and {Signorelli}, G. and {Silari}, M. and {Simon}, F. and {Sj{\"o}strand}, T. and {Skands}, P. and {Skwarnicki}, T. and {Smoot}, G.~F. and {Soffer}, A. and {Sozzi}, M.~S. and {Spiering}, C. and {Stahl}, A. and {Sumino}, Y. and {Takahashi}, F. and {Tanabashi}, M. and {Tanaka}, J.},
        title = "{Review of particle physics$^{*}$}",
      journal = {\prd},
         year = 2024,
        month = aug,
       volume = {110},
       number = {3},
          eid = {030001},
        pages = {030001},
          doi = {10.1103/PhysRevD.110.030001},
       adsurl = {https://ui.adsabs.harvard.edu/abs/2024PhRvD.110c0001N}
}

@ARTICLE{Planck2016,
       author = {{Planck Collaboration} and {Ade}, P.~A.~R. and {Aghanim}, N. and {Arnaud}, M. and {Ashdown}, M. and {Aumont}, J. and {Baccigalupi}, C. and {Banday}, A.~J. and {Barreiro}, R.~B. and {Bartlett}, J.~G. and {Bartolo}, N. and {Battaner}, E. and {Battye}, R. and {Benabed}, K. and {Beno{\^\i}t}, A. and {Benoit-L{\'e}vy}, A. and {Bernard}, J. -P. and {Bersanelli}, M. and {Bielewicz}, P. and {Bock}, J.~J. and {Bonaldi}, A. and {Bonavera}, L. and {Bond}, J.~R. and {Borrill}, J. and {Bouchet}, F.~R. and {Boulanger}, F. and {Bucher}, M. and {Burigana}, C. and {Butler}, R.~C. and {Calabrese}, E. and {Cardoso}, J. -F. and {Catalano}, A. and {Challinor}, A. and {Chamballu}, A. and {Chary}, R. -R. and {Chiang}, H.~C. and {Chluba}, J. and {Christensen}, P.~R. and {Church}, S. and {Clements}, D.~L. and {Colombi}, S. and {Colombo}, L.~P.~L. and {Combet}, C. and {Coulais}, A. and {Crill}, B.~P. and {Curto}, A. and {Cuttaia}, F. and {Danese}, L. and {Davies}, R.~D. and {Davis}, R.~J. and {de Bernardis}, P. and {de Rosa}, A. and {de Zotti}, G. and {Delabrouille}, J. and {D{\'e}sert}, F. -X. and {Di Valentino}, E. and {Dickinson}, C. and {Diego}, J.~M. and {Dolag}, K. and {Dole}, H. and {Donzelli}, S. and {Dor{\'e}}, O. and {Douspis}, M. and {Ducout}, A. and {Dunkley}, J. and {Dupac}, X. and {Efstathiou}, G. and {Elsner}, F. and {En{\ss}lin}, T.~A. and {Eriksen}, H.~K. and {Farhang}, M. and {Fergusson}, J. and {Finelli}, F. and {Forni}, O. and {Frailis}, M. and {Fraisse}, A.~A. and {Franceschi}, E. and {Frejsel}, A. and {Galeotta}, S. and {Galli}, S. and {Ganga}, K. and {Gauthier}, C. and {Gerbino}, M. and {Ghosh}, T. and {Giard}, M. and {Giraud-H{\'e}raud}, Y. and {Giusarma}, E. and {Gjerl{\o}w}, E. and {Gonz{\'a}lez-Nuevo}, J. and {G{\'o}rski}, K.~M. and {Gratton}, S. and {Gregorio}, A. and {Gruppuso}, A. and {Gudmundsson}, J.~E. and {Hamann}, J. and {Hansen}, F.~K. and {Hanson}, D. and {Harrison}, D.~L. and {Helou}, G. and {Henrot-Versill{\'e}}, S. and {Hern{\'a}ndez-Monteagudo}, C. and {Herranz}, D. and {Hildebrandt}, S.~R. and {Hivon}, E. and {Hobson}, M. and {Holmes}, W.~A. and {Hornstrup}, A. and {Hovest}, W. and {Huang}, Z. and {Huffenberger}, K.~M. and {Hurier}, G. and {Jaffe}, A.~H. and {Jaffe}, T.~R. and {Jones}, W.~C. and {Juvela}, M. and {Keih{\"a}nen}, E. and {Keskitalo}, R. and {Kisner}, T.~S. and {Kneissl}, R. and {Knoche}, J. and {Knox}, L. and {Kunz}, M. and {Kurki-Suonio}, H. and {Lagache}, G. and {L{\"a}hteenm{\"a}ki}, A. and {Lamarre}, J. -M. and {Lasenby}, A. and {Lattanzi}, M. and {Lawrence}, C.~R. and {Leahy}, J.~P. and {Leonardi}, R. and {Lesgourgues}, J. and {Levrier}, F. and {Lewis}, A. and {Liguori}, M. and {Lilje}, P.~B. and {Linden-V{\o}rnle}, M. and {L{\'o}pez-Caniego}, M. and {Lubin}, P.~M. and {Mac{\'\i}as-P{\'e}rez}, J.~F. and {Maggio}, G. and {Maino}, D. and {Mandolesi}, N. and {Mangilli}, A. and {Marchini}, A. and {Maris}, M. and {Martin}, P.~G. and {Martinelli}, M. and {Mart{\'\i}nez-Gonz{\'a}lez}, E. and {Masi}, S. and {Matarrese}, S. and {McGehee}, P. and {Meinhold}, P.~R. and {Melchiorri}, A. and {Melin}, J. -B. and {Mendes}, L. and {Mennella}, A. and {Migliaccio}, M. and {Millea}, M. and {Mitra}, S. and {Miville-Desch{\^e}nes}, M. -A. and {Moneti}, A. and {Montier}, L. and {Morgante}, G. and {Mortlock}, D. and {Moss}, A. and {Munshi}, D. and {Murphy}, J.~A. and {Naselsky}, P. and {Nati}, F. and {Natoli}, P. and {Netterfield}, C.~B. and {N{\o}rgaard-Nielsen}, H.~U. and {Noviello}, F. and {Novikov}, D. and {Novikov}, I. and {Oxborrow}, C.~A. and {Paci}, F. and {Pagano}, L. and {Pajot}, F. and {Paladini}, R. and {Paoletti}, D. and {Partridge}, B. and {Pasian}, F. and {Patanchon}, G. and {Pearson}, T.~J. and {Perdereau}, O. and {Perotto}, L. and {Perrotta}, F. and {Pettorino}, V. and {Piacentini}, F. and {Piat}, M. and {Pierpaoli}, E. and {Pietrobon}, D. and {Plaszczynski}, S. and {Pointecouteau}, E. and {Polenta}, G. and {Popa}, L. and {Pratt}, G.~W. and {Pr{\'e}zeau}, G. and {Prunet}, S. and {Puget}, J. -L. and {Rachen}, J.~P. and {Reach}, W.~T. and {Rebolo}, R. and {Reinecke}, M. and {Remazeilles}, M. and {Renault}, C. and {Renzi}, A. and {Ristorcelli}, I. and {Rocha}, G. and {Rosset}, C. and {Rossetti}, M. and {Roudier}, G. and {Rouill{\'e} d'Orfeuil}, B. and {Rowan-Robinson}, M. and {Rubi{\~n}o-Mart{\'\i}n}, J.~A. and {Rusholme}, B. and {Said}, N. and {Salvatelli}, V. and {Salvati}, L. and {Sandri}, M. and {Santos}, D. and {Savelainen}, M. and {Savini}, G. and {Scott}, D. and {Seiffert}, M.~D. and {Serra}, P. and {Shellard}, E.~P.~S. and {Spencer}, L.~D. and {Spinelli}, M. and {Stolyarov}, V. and {Stompor}, R. and {Sudiwala}, R. and {Sunyaev}, R. and {Sutton}, D. and {Suur-Uski}, A. -S. and {Sygnet}, J. -F. and {Tauber}, J.~A. and {Terenzi}, L. and {Toffolatti}, L. and {Tomasi}, M. and {Tristram}, M. and {Trombetti}, T. and {Tucci}, M. and {Tuovinen}, J. and {T{\"u}rler}, M. and {Umana}, G. and {Valenziano}, L. and {Valiviita}, J. and {Van Tent}, F. and {Vielva}, P. and {Villa}, F. and {Wade}, L.~A. and {Wandelt}, B.~D. and {Wehus}, I.~K. and {White}, M. and {White}, S.~D.~M. and {Wilkinson}, A. and {Yvon}, D. and {Zacchei}, A. and {Zonca}, A.},
        title = "{Planck 2015 results. XIII. Cosmological parameters}",
      journal = {\aap},
         year = 2016,
        month = sep,
       volume = {594},
          eid = {A13},
        pages = {A13},
          doi = {10.1051/0004-6361/201525830},
archivePrefix = {arXiv},
       eprint = {1502.01589},
 primaryClass = {astro-ph.CO},
       adsurl = {https://ui.adsabs.harvard.edu/abs/2016A&A...594A..13P}
}

@ARTICLE{Planck2020cosmo,
       author = {{Planck Collaboration} and {Aghanim}, N. and {Akrami}, Y. and {Ashdown}, M. and {Aumont}, J. and {Baccigalupi}, C. and {Ballardini}, M. and {Banday}, A.~J. and {Barreiro}, R.~B. and {Bartolo}, N. and {Basak}, S. and {Battye}, R. and {Benabed}, K. and {Bernard}, J. -P. and {Bersanelli}, M. and {Bielewicz}, P. and {Bock}, J.~J. and {Bond}, J.~R. and {Borrill}, J. and {Bouchet}, F.~R. and {Boulanger}, F. and {Bucher}, M. and {Burigana}, C. and {Butler}, R.~C. and {Calabrese}, E. and {Cardoso}, J. -F. and {Carron}, J. and {Challinor}, A. and {Chiang}, H.~C. and {Chluba}, J. and {Colombo}, L.~P.~L. and {Combet}, C. and {Contreras}, D. and {Crill}, B.~P. and {Cuttaia}, F. and {de Bernardis}, P. and {de Zotti}, G. and {Delabrouille}, J. and {Delouis}, J. -M. and {Di Valentino}, E. and {Diego}, J.~M. and {Dor{\'e}}, O. and {Douspis}, M. and {Ducout}, A. and {Dupac}, X. and {Dusini}, S. and {Efstathiou}, G. and {Elsner}, F. and {En{\ss}lin}, T.~A. and {Eriksen}, H.~K. and {Fantaye}, Y. and {Farhang}, M. and {Fergusson}, J. and {Fernandez-Cobos}, R. and {Finelli}, F. and {Forastieri}, F. and {Frailis}, M. and {Fraisse}, A.~A. and {Franceschi}, E. and {Frolov}, A. and {Galeotta}, S. and {Galli}, S. and {Ganga}, K. and {G{\'e}nova-Santos}, R.~T. and {Gerbino}, M. and {Ghosh}, T. and {Gonz{\'a}lez-Nuevo}, J. and {G{\'o}rski}, K.~M. and {Gratton}, S. and {Gruppuso}, A. and {Gudmundsson}, J.~E. and {Hamann}, J. and {Handley}, W. and {Hansen}, F.~K. and {Herranz}, D. and {Hildebrandt}, S.~R. and {Hivon}, E. and {Huang}, Z. and {Jaffe}, A.~H. and {Jones}, W.~C. and {Karakci}, A. and {Keih{\"a}nen}, E. and {Keskitalo}, R. and {Kiiveri}, K. and {Kim}, J. and {Kisner}, T.~S. and {Knox}, L. and {Krachmalnicoff}, N. and {Kunz}, M. and {Kurki-Suonio}, H. and {Lagache}, G. and {Lamarre}, J. -M. and {Lasenby}, A. and {Lattanzi}, M. and {Lawrence}, C.~R. and {Le Jeune}, M. and {Lemos}, P. and {Lesgourgues}, J. and {Levrier}, F. and {Lewis}, A. and {Liguori}, M. and {Lilje}, P.~B. and {Lilley}, M. and {Lindholm}, V. and {L{\'o}pez-Caniego}, M. and {Lubin}, P.~M. and {Ma}, Y. -Z. and {Mac{\'\i}as-P{\'e}rez}, J.~F. and {Maggio}, G. and {Maino}, D. and {Mandolesi}, N. and {Mangilli}, A. and {Marcos-Caballero}, A. and {Maris}, M. and {Martin}, P.~G. and {Martinelli}, M. and {Mart{\'\i}nez-Gonz{\'a}lez}, E. and {Matarrese}, S. and {Mauri}, N. and {McEwen}, J.~D. and {Meinhold}, P.~R. and {Melchiorri}, A. and {Mennella}, A. and {Migliaccio}, M. and {Millea}, M. and {Mitra}, S. and {Miville-Desch{\^e}nes}, M. -A. and {Molinari}, D. and {Montier}, L. and {Morgante}, G. and {Moss}, A. and {Natoli}, P. and {N{\o}rgaard-Nielsen}, H.~U. and {Pagano}, L. and {Paoletti}, D. and {Partridge}, B. and {Patanchon}, G. and {Peiris}, H.~V. and {Perrotta}, F. and {Pettorino}, V. and {Piacentini}, F. and {Polastri}, L. and {Polenta}, G. and {Puget}, J. -L. and {Rachen}, J.~P. and {Reinecke}, M. and {Remazeilles}, M. and {Renzi}, A. and {Rocha}, G. and {Rosset}, C. and {Roudier}, G. and {Rubi{\~n}o-Mart{\'\i}n}, J.~A. and {Ruiz-Granados}, B. and {Salvati}, L. and {Sandri}, M. and {Savelainen}, M. and {Scott}, D. and {Shellard}, E.~P.~S. and {Sirignano}, C. and {Sirri}, G. and {Spencer}, L.~D. and {Sunyaev}, R. and {Suur-Uski}, A. -S. and {Tauber}, J.~A. and {Tavagnacco}, D. and {Tenti}, M. and {Toffolatti}, L. and {Tomasi}, M. and {Trombetti}, T. and {Valenziano}, L. and {Valiviita}, J. and {Van Tent}, B. and {Vibert}, L. and {Vielva}, P. and {Villa}, F. and {Vittorio}, N. and {Wandelt}, B.~D. and {Wehus}, I.~K. and {White}, M. and {White}, S.~D.~M. and {Zacchei}, A. and {Zonca}, A.},
        title = "{Planck 2018 results. VI. Cosmological parameters}",
      journal = {\aap},
         year = 2020,
        month = sep,
       volume = {641},
          eid = {A6},
        pages = {A6},
          doi = {10.1051/0004-6361/201833910},
archivePrefix = {arXiv},
       eprint = {1807.06209},
 primaryClass = {astro-ph.CO},
       adsurl = {https://ui.adsabs.harvard.edu/abs/2020A&A...641A...6P}
}

@ARTICLE{Pogosyan2009,
       author = {{Pogosyan}, D. and {Pichon}, C. and {Gay}, C. and {Prunet}, S. and {Cardoso}, J.~F. and {Sousbie}, T. and {Colombi}, S.},
        title = "{The local theory of the cosmic skeleton}",
      journal = {\mnras},
         year = 2009,
        month = jun,
       volume = {396},
       number = {2},
        pages = {635-667},
          doi = {10.1111/j.1365-2966.2009.14753.x},
archivePrefix = {arXiv},
       eprint = {0811.1530},
 primaryClass = {astro-ph},
       adsurl = {https://ui.adsabs.harvard.edu/abs/2009MNRAS.396..635P}
}

@ARTICLE{Pranav2021,
       author = {{Pranav}, Pratyush},
        title = "{Topology and geometry of Gaussian random fields II: on critical points, excursion sets, and persistent homology}",
      journal = {arXiv e-prints},
         year = 2021,
        month = sep,
          eid = {arXiv:2109.08721},
        pages = {arXiv:2109.08721},
          doi = {10.48550/arXiv.2109.08721},
archivePrefix = {arXiv},
       eprint = {2109.08721},
 primaryClass = {astro-ph.CO},
       adsurl = {https://ui.adsabs.harvard.edu/abs/2021arXiv210908721P}
}

@ARTICLE{Pranav2019,
       author = {{Pranav}, Pratyush and {van de Weygaert}, Rien and {Vegter}, Gert and {Jones}, Bernard J.~T. and {Adler}, Robert J. and {Feldbrugge}, Job and {Park}, Changbom and {Buchert}, Thomas and {Kerber}, Michael},
        title = "{Topology and geometry of Gaussian random fields I: on Betti numbers, Euler characteristic, and Minkowski functionals}",
      journal = {\mnras},
         year = 2019,
        month = may,
       volume = {485},
       number = {3},
        pages = {4167-4208},
          doi = {10.1093/mnras/stz541},
archivePrefix = {arXiv},
       eprint = {1812.07310},
 primaryClass = {astro-ph.CO},
       adsurl = {https://ui.adsabs.harvard.edu/abs/2019MNRAS.485.4167P}
}

@ARTICLE{Prat2026,
       author = {{Prat}, J. and {Gatti}, M. and {Doux}, C. and {Pranav}, P. and {Chang}, C. and {Jeffrey}, N. and {Whiteway}, L. and {Anbajagane}, D. and {Sugiyama}, S. and {Thomsen}, A. and {Alarcon}, A. and {Amon}, A. and {Bechtol}, K. and {Bernstein}, G.~M. and {Campos}, A. and {Chen}, R. and {Choi}, A. and {Davis}, C. and {DeRose}, J. and {Dodelson}, S. and {Eckert}, K. and {Elvin-Poole}, J. and {Everett}, S. and {Fert{\'e}}, A. and {Gruen}, D. and {Huff}, E.~M. and {Harrison}, I. and {Herner}, K. and {Jarvis}, M. and {Kuropatkin}, N. and {Leget}, P.-F. and {MacCrann}, N. and {McCullough}, J. and {Myles}, J. and {Navarro-Alsina}, A. and {Pandey}, S. and {Raveri}, M. and {Rollins}, R.~P. and {Roodman}, A. and {S{\'a}nchez}, C. and {Secco}, L.~F. and {Sheldon}, E. and {Shin}, T. and {Troxel}, M.~A. and {Tutusaus}, I. and {Varga}, T.~N. and {Yanny}, B. and {Yin}, B. and {Zhang}, Y. and {Zuntz}, J. and {Abbott}, T.~M.~C. and {Aguena}, M. and {Allam}, S. and {Andrade-Oliveira}, F. and {Blazek}, J. and {Bocquet}, S. and {Brooks}, D. and {Carretero}, J. and {Carnero Rosell}, A. and {Cawthon}, R. and {De Vicente}, J. and {Desai}, S. and {da Silva Pereira}, M.~E. and {Diehl}, H.~T. and {Flaugher}, B. and {Frieman}, J. and {Garc{\'\i}a-Bellido}, J. and {Gruendl}, R.~A. and {Gutierrez}, G. and {Hinton}, S.~R. and {Hollowood}, D.~L. and {Honscheid}, K. and {James}, D.~J. and {Kuehn}, K. and {da Costa}, L.~N. and {Lahav}, O. and {Lee}, S. and {Marshall}, J.~L. and {Mena-Fern{\'a}ndez}, J. and {Miquel}, R. and {Mohr}, J.~J. and {Ogando}, R.~L.~C. and {Plazas Malag{\'o}n}, A.~A. and {Porredon}, A. and {Samuroff}, S. and {Sanchez}, E. and {Santiago}, B. and {Sevilla-Noarbe}, I. and {Smith}, M. and {Suchyta}, E. and {Swanson}, M.~E.~C. and {Thomas}, D. and {To}, C. and {Vikram}, V. and {Walker}, A.~R. and {Weaverdyck}, N. and {Weller}, J.},
        title = "{Dark Energy Survey Year 3 results: wCDM cosmology from simulation-based inference with persistent homology on the sphere}",
      journal = {\mnras},
         year = 2026,
        month = jan,
       volume = {545},
       number = {3},
          eid = {staf2152},
        pages = {staf2152},
          doi = {10.1093/mnras/staf2152},
archivePrefix = {arXiv},
       eprint = {2506.13439},
 primaryClass = {astro-ph.CO},
       adsurl = {https://ui.adsabs.harvard.edu/abs/2026MNRAS.545f2152P}
}

@ARTICLE{Robins2000,
       author = {{Robins}, V. and {Meiss}, J.~D. and {Bradley}, E.},
        title = "{Computing connectedness: disconnectedness and discreteness}",
      journal = {Physica D Nonlinear Phenomena},
         year = 2000,
        month = may,
       volume = {139},
       number = {3-4},
        pages = {276-300},
          doi = {10.1016/S0167-2789(99)00228-6},
       adsurl = {https://ui.adsabs.harvard.edu/abs/2000PhyD..139..276R}
}

@ARTICLE{Robins1998,
       author = {{Robins}, V. and {Meiss}, J.~D. and {Bradley}, E.},
        title = "{Computing connectedness: An exercise in computational topology}",
      journal = {Nonlinearity},
         year = 1998,
        month = jul,
       volume = {11},
       number = {4},
        pages = {913-922},
          doi = {10.1088/0951-7715/11/4/009},
       adsurl = {https://ui.adsabs.harvard.edu/abs/1998Nonli..11..913R}
}

@INPROCEEDINGS{Rossi2022,
       author = {{Rossi}, Graziano},
        title = "{Cosmic Topology, Persistent Homology, and Massive Neutrinos}",
    booktitle = {American Astronomical Society Meeting \#240},
         year = 2022,
       series = {American Astronomical Society Meeting Abstracts},
       volume = {240},
        month = jun,
          eid = {312.08},
        pages = {312.08},
       adsurl = {https://ui.adsabs.harvard.edu/abs/2022AAS...24031208R}
}

@ARTICLE{Rossi2020,
       author = {{Rossi}, Graziano},
        title = "{The Sejong Suite: Cosmological Hydrodynamical Simulations with Massive Neutrinos, Dark Radiation, and Warm Dark Matter}",
      journal = {\apjs},
         year = 2020,
        month = aug,
       volume = {249},
       number = {2},
          eid = {19},
        pages = {19},
          doi = {10.3847/1538-4365/ab9d1e},
archivePrefix = {arXiv},
       eprint = {2007.15279},
 primaryClass = {astro-ph.CO},
       adsurl = {https://ui.adsabs.harvard.edu/abs/2020ApJS..249...19R}
}

@ARTICLE{Rossi2017,
       author = {{Rossi}, Graziano},
        title = "{Impact of Massive Neutrinos and Dark Radiation on the High-redshift Cosmic Web. I. Ly{\ensuremath{\alpha}} Forest Observables}",
      journal = {\apjs},
         year = 2017,
        month = nov,
       volume = {233},
       number = {1},
          eid = {12},
        pages = {12},
          doi = {10.3847/1538-4365/aa93d6},
archivePrefix = {arXiv},
       eprint = {1712.00230},
 primaryClass = {astro-ph.CO},
       adsurl = {https://ui.adsabs.harvard.edu/abs/2017ApJS..233...12R}
}

@ARTICLE{Shim2021,
       author = {{Shim}, J. and {Codis}, S. and {Pichon}, C. and {Pogosyan}, D. and {Cadiou}, C.},
        title = "{The clustering of critical points in the evolving cosmic web}",
      journal = {\mnras},
         year = 2021,
        month = apr,
       volume = {502},
       number = {3},
        pages = {3885-3910},
          doi = {10.1093/mnras/stab263},
archivePrefix = {arXiv},
       eprint = {2011.04321},
 primaryClass = {astro-ph.CO},
       adsurl = {https://ui.adsabs.harvard.edu/abs/2021MNRAS.502.3885S}
}

@ARTICLE{Spergel2015,
       author = {{Spergel}, D. and {Gehrels}, N. and {Baltay}, C. and {Bennett}, D. and {Breckinridge}, J. and {Donahue}, M. and {Dressler}, A. and {Gaudi}, B.~S. and {Greene}, T. and {Guyon}, O. and {Hirata}, C. and {Kalirai}, J. and {Kasdin}, N.~J. and {Macintosh}, B. and {Moos}, W. and {Perlmutter}, S. and {Postman}, M. and {Rauscher}, B. and {Rhodes}, J. and {Wang}, Y. and {Weinberg}, D. and {Benford}, D. and {Hudson}, M. and {Jeong}, W. -S. and {Mellier}, Y. and {Traub}, W. and {Yamada}, T. and {Capak}, P. and {Colbert}, J. and {Masters}, D. and {Penny}, M. and {Savransky}, D. and {Stern}, D. and {Zimmerman}, N. and {Barry}, R. and {Bartusek}, L. and {Carpenter}, K. and {Cheng}, E. and {Content}, D. and {Dekens}, F. and {Demers}, R. and {Grady}, K. and {Jackson}, C. and {Kuan}, G. and {Kruk}, J. and {Melton}, M. and {Nemati}, B. and {Parvin}, B. and {Poberezhskiy}, I. and {Peddie}, C. and {Ruffa}, J. and {Wallace}, J.~K. and {Whipple}, A. and {Wollack}, E. and {Zhao}, F.},
        title = "{Wide-Field InfrarRed Survey Telescope-Astrophysics Focused Telescope Assets WFIRST-AFTA 2015 Report}",
      journal = {arXiv e-prints},
         year = 2015,
        month = mar,
          eid = {arXiv:1503.03757},
        pages = {arXiv:1503.03757},
archivePrefix = {arXiv},
       eprint = {1503.03757},
 primaryClass = {astro-ph.IM},
       adsurl = {https://ui.adsabs.harvard.edu/abs/2015arXiv150303757S}
}

@ARTICLE{SPT2025,
       author = {{Camphuis}, E. and {Quan}, W. and {Balkenhol}, L. and {Khalife}, A.~R. and {Ge}, F. and {Guidi}, F. and {Huang}, N. and {Lynch}, G.~P. and {Omori}, Y. and {Trendafilova}, C. and {Anderson}, A.~J. and {Ansarinejad}, B. and {Archipley}, M. and {Barry}, P.~S. and {Benabed}, K. and {Bender}, A.~N. and {Benson}, B.~A. and {Bianchini}, F. and {Bleem}, L.~E. and {Bouchet}, F.~R. and {Bryant}, L. and {Campitiello}, M.~G. and {Carlstrom}, J.~E. and {Chang}, C.~L. and {Chaubal}, P. and {Chichura}, P.~M. and {Chokshi}, A. and {Chou}, T. -L. and {Coerver}, A. and {Crawford}, T.~M. and {Daley}, C. and {de Haan}, T. and {Dibert}, K.~R. and {Dobbs}, M.~A. and {Doohan}, M. and {Doussot}, A. and {Dutcher}, D. and {Everett}, W. and {Feng}, C. and {Ferguson}, K.~R. and {Fichman}, K. and {Foster}, A. and {Galli}, S. and {Gambrel}, A.~E. and {Gardner}, R.~W. and {Goeckner-Wald}, N. and {Gualtieri}, R. and {Guns}, S. and {Halverson}, N.~W. and {Hivon}, E. and {Holder}, G.~P. and {Holzapfel}, W.~L. and {Hood}, J.~C. and {Hryciuk}, A. and {K{\'e}ruzor{\'e}}, F. and {Knox}, L. and {Korman}, M. and {Kornoelje}, K. and {Kuo}, C. -L. and {Levy}, K. and {Lowitz}, A.~E. and {Lu}, C. and {Maniyar}, A. and {Martsen}, E.~S. and {Menanteau}, F. and {Millea}, M. and {Montgomery}, J. and {Nakato}, Y. and {Natoli}, T. and {Noble}, G.~I. and {Ouellette}, A. and {Pan}, Z. and {Paschos}, P. and {Phadke}, K.~A. and {Pollak}, A.~W. and {Prabhu}, K. and {Raghunathan}, S. and {Rahimi}, M. and {Rahlin}, A. and {Reichardt}, C.~L. and {Rouble}, M. and {Ruhl}, J.~E. and {Schiappucci}, E. and {Simpson}, A. and {Sobrin}, J.~A. and {Stark}, A.~A. and {Stephen}, J. and {Tandoi}, C. and {Thorne}, B. and {Umilta}, C. and {Vieira}, J.~D. and {Vitrier}, A. and {Wan}, Y. and {Whitehorn}, N. and {Wu}, W.~L.~K. and {Young}, M.~R. and {Zebrowski}, J.~A.},
        title = "{SPT-3G D1: CMB temperature and polarization power spectra and cosmology from 2019 and 2020 observations of the SPT-3G Main field}",
      journal = {arXiv e-prints},
         year = 2025,
        month = jun,
          eid = {arXiv:2506.20707},
        pages = {arXiv:2506.20707},
          doi = {10.48550/arXiv.2506.20707},
archivePrefix = {arXiv},
       eprint = {2506.20707},
 primaryClass = {astro-ph.CO},
       adsurl = {https://ui.adsabs.harvard.edu/abs/2025arXiv250620707C}
}

@ARTICLE{Sousbie2011a,
       author = {{Sousbie}, T.},
        title = "{The persistent cosmic web and its filamentary structure - I. Theory and implementation}",
      journal = {\mnras},
         year = 2011,
        month = jun,
       volume = {414},
       number = {1},
        pages = {350-383},
          doi = {10.1111/j.1365-2966.2011.18394.x},
archivePrefix = {arXiv},
       eprint = {1009.4015},
 primaryClass = {astro-ph.CO},
       adsurl = {https://ui.adsabs.harvard.edu/abs/2011MNRAS.414..350S}
}

@ARTICLE{Sousbie2011b,
       author = {{Sousbie}, T. and {Pichon}, C. and {Kawahara}, H.},
        title = "{The persistent cosmic web and its filamentary structure - II. Illustrations}",
      journal = {\mnras},
         year = 2011,
        month = jun,
       volume = {414},
       number = {1},
        pages = {384-403},
          doi = {10.1111/j.1365-2966.2011.18395.x},
archivePrefix = {arXiv},
       eprint = {1009.4014},
 primaryClass = {astro-ph.CO},
       adsurl = {https://ui.adsabs.harvard.edu/abs/2011MNRAS.414..384S}
}

@ARTICLE{Sousbie2008,
       author = {{Sousbie}, T. and {Pichon}, C. and {Colombi}, S. and {Novikov}, D. and {Pogosyan}, D.},
        title = "{The 3D skeleton: tracing the filamentary structure of the Universe}",
      journal = {\mnras},
         year = 2008,
        month = feb,
       volume = {383},
       number = {4},
        pages = {1655-1670},
          doi = {10.1111/j.1365-2966.2007.12685.x},
archivePrefix = {arXiv},
       eprint = {0707.3123},
 primaryClass = {astro-ph},
       adsurl = {https://ui.adsabs.harvard.edu/abs/2008MNRAS.383.1655S}
}

@ARTICLE{Springel2005,
       author = {{Springel}, Volker},
        title = "{The cosmological simulation code GADGET-2}",
      journal = {\mnras},
         year = 2005,
        month = dec,
       volume = {364},
       number = {4},
        pages = {1105-1134},
          doi = {10.1111/j.1365-2966.2005.09655.x},
archivePrefix = {arXiv},
       eprint = {astro-ph/0505010},
 primaryClass = {astro-ph},
       adsurl = {https://ui.adsabs.harvard.edu/abs/2005MNRAS.364.1105S}
}

@ARTICLE{Takada2014,
       author = {{Takada}, Masahiro and {Ellis}, Richard S. and {Chiba}, Masashi and {Greene}, Jenny E. and {Aihara}, Hiroaki and {Arimoto}, Nobuo and {Bundy}, Kevin and {Cohen}, Judith and {Dor{\'e}}, Olivier and {Graves}, Genevieve and {Gunn}, James E. and {Heckman}, Timothy and {Hirata}, Christopher M. and {Ho}, Paul and {Kneib}, Jean-Paul and {Le F{\`e}vre}, Olivier and {Lin}, Lihwai and {More}, Surhud and {Murayama}, Hitoshi and {Nagao}, Tohru and {Ouchi}, Masami and {Seiffert}, Michael and {Silverman}, John D. and {Sodr{\'e}}, Laerte and {Spergel}, David N. and {Strauss}, Michael A. and {Sugai}, Hajime and {Suto}, Yasushi and {Takami}, Hideki and {Wyse}, Rosemary},
        title = "{Extragalactic science, cosmology, and Galactic archaeology with the Subaru Prime Focus Spectrograph}",
      journal = {\pasj},
         year = 2014,
        month = feb,
       volume = {66},
       number = {1},
          eid = {R1},
        pages = {R1},
          doi = {10.1093/pasj/pst019},
archivePrefix = {arXiv},
       eprint = {1206.0737},
 primaryClass = {astro-ph.CO},
       adsurl = {https://ui.adsabs.harvard.edu/abs/2014PASJ...66R...1T}
}

@ARTICLE{Thiele2024,
       author = {{Thiele}, Leander and {Massara}, Elena and {Pisani}, Alice and {Hahn}, ChangHoon and {Spergel}, David N. and {Ho}, Shirley and {Wandelt}, Benjamin},
        title = "{Neutrino Mass Constraint from an Implicit Likelihood Analysis of BOSS Voids}",
      journal = {\apj},
         year = 2024,
        month = jul,
       volume = {969},
       number = {2},
          eid = {89},
        pages = {89},
          doi = {10.3847/1538-4357/ad434e},
archivePrefix = {arXiv},
       eprint = {2307.07555},
 primaryClass = {astro-ph.CO},
       adsurl = {https://ui.adsabs.harvard.edu/abs/2024ApJ...969...89T}
}

@ARTICLE{Verza2023,
       author = {{Verza}, Giovanni and {Carbone}, Carmelita and {Pisani}, Alice and {Renzi}, Alessandro},
        title = "{DEMNUni: disentangling dark energy from massive neutrinos with the void size function}",
      journal = {\jcap},
         year = 2023,
        month = dec,
       volume = {2023},
       number = {12},
          eid = {044},
        pages = {044},
          doi = {10.1088/1475-7516/2023/12/044},
archivePrefix = {arXiv},
       eprint = {2212.09740},
 primaryClass = {astro-ph.CO},
       adsurl = {https://ui.adsabs.harvard.edu/abs/2023JCAP...12..044V}
}

@ARTICLE{VillaescusaNavarro2020,
       author = {{Villaescusa-Navarro}, Francisco and {Hahn}, ChangHoon and {Massara}, Elena and {Banerjee}, Arka and {Delgado}, Ana Maria and {Ramanah}, Doogesh Kodi and {Charnock}, Tom and {Giusarma}, Elena and {Li}, Yin and {Allys}, Erwan and {Brochard}, Antoine and {Uhlemann}, Cora and {Chiang}, Chi-Ting and {He}, Siyu and {Pisani}, Alice and {Obuljen}, Andrej and {Feng}, Yu and {Castorina}, Emanuele and {Contardo}, Gabriella and {Kreisch}, Christina D. and {Nicola}, Andrina and {Alsing}, Justin and {Scoccimarro}, Roman and {Verde}, Licia and {Viel}, Matteo and {Ho}, Shirley and {Mallat}, Stephane and {Wandelt}, Benjamin and {Spergel}, David N.},
        title = "{The Quijote Simulations}",
      journal = {\apjs},
         year = 2020,
        month = sep,
       volume = {250},
       number = {1},
          eid = {2},
        pages = {2},
          doi = {10.3847/1538-4365/ab9d82},
archivePrefix = {arXiv},
       eprint = {1909.05273},
 primaryClass = {astro-ph.CO},
       adsurl = {https://ui.adsabs.harvard.edu/abs/2020ApJS..250....2V}
}

@ARTICLE{Whitford2022,
       author = {{Whitford}, Abb{\'e} M. and {Howlett}, Cullan and {Davis}, Tamara M.},
        title = "{Using peculiar velocity surveys to constrain neutrino masses}",
      journal = {\mnras},
         year = 2022,
        month = jun,
       volume = {513},
       number = {1},
        pages = {345-362},
          doi = {10.1093/mnras/stac783},
archivePrefix = {arXiv},
       eprint = {2112.10302},
 primaryClass = {astro-ph.CO},
       adsurl = {https://ui.adsabs.harvard.edu/abs/2022MNRAS.513..345W}
}

@ARTICLE{Wilding2021,
       author = {{Wilding}, Georg and {Nevenzeel}, Keimpe and {van de Weygaert}, Rien and {Vegter}, Gert and {Pranav}, Pratyush and {Jones}, Bernard J.~T. and {Efstathiou}, Konstantinos and {Feldbrugge}, Job},
        title = "{Persistent homology of the cosmic web - I. Hierarchical topology in {\ensuremath{\Lambda}}CDM cosmologies}",
      journal = {\mnras},
         year = 2021,
        month = oct,
       volume = {507},
       number = {2},
        pages = {2968-2990},
          doi = {10.1093/mnras/stab2326},
archivePrefix = {arXiv},
       eprint = {2011.12851},
 primaryClass = {astro-ph.CO},
       adsurl = {https://ui.adsabs.harvard.edu/abs/2021MNRAS.507.2968W}
}

@ARTICLE{Yankelevich2023,
       author = {{Yankelevich}, Victoria and {McCarthy}, Ian G. and {Kwan}, Juliana and {Stafford}, Sam G. and {Liu}, Jia},
        title = "{The halo bispectrum as a sensitive probe of massive neutrinos and baryon physics}",
      journal = {\mnras},
         year = 2023,
        month = may,
       volume = {521},
       number = {1},
        pages = {1448-1461},
          doi = {10.1093/mnras/stad571},
archivePrefix = {arXiv},
       eprint = {2202.07680},
 primaryClass = {astro-ph.CO},
       adsurl = {https://ui.adsabs.harvard.edu/abs/2023MNRAS.521.1448Y}
}

@ARTICLE{Yip2024,
       author = {{Yip}, Jacky H.~T. and {Biagetti}, Matteo and {Cole}, Alex and {Viswanathan}, Karthik and {Shiu}, Gary},
        title = "{Cosmology with persistent homology: a Fisher forecast}",
      journal = {\jcap},
         year = 2024,
        month = sep,
       volume = {2024},
       number = {9},
          eid = {034},
        pages = {034},
          doi = {10.1088/1475-7516/2024/09/034},
archivePrefix = {arXiv},
       eprint = {2403.13985},
 primaryClass = {astro-ph.CO},
       adsurl = {https://ui.adsabs.harvard.edu/abs/2024JCAP...09..034Y}
}

@ARTICLE{York2000,
       author = {{York}, Donald G. and {Adelman}, J. and {Anderson}, John E., Jr. and
         {Anderson}, Scott F. and {Annis}, James and {Bahcall}, Neta A. and
         {Bakken}, J.~A. and {Barkhouser}, Robert and {Bastian}, Steven and
         {Berman}, Eileen and {Boroski}, William N. and {Bracker}, Steve and
         {Briegel}, Charlie and {Briggs}, John W. and {Brinkmann}, J. and
         {Brunner}, Robert and {Burles}, Scott and {Carey}, Larry and
         {Carr}, Michael A. and {Castander}, Francisco J. and {Chen}, Bing and
         {Colestock}, Patrick L. and {Connolly}, A.~J. and {Crocker}, J.~H. and
         {Csabai}, Istv{\'a}n and {Czarapata}, Paul C. and {Davis}, John Eric and
         {Doi}, Mamoru and {Dombeck}, Tom and {Eisenstein}, Daniel and
         {Ellman}, Nancy and {Elms}, Brian R. and {Evans}, Michael L. and
         {Fan}, Xiaohui and {Federwitz}, Glenn R. and {Fiscelli}, Larry and
         {Friedman}, Scott and {Frieman}, Joshua A. and {Fukugita}, Masataka and
         {Gillespie}, Bruce and {Gunn}, James E. and {Gurbani}, Vijay K. and
         {de Haas}, Ernst and {Haldeman}, Merle and {Harris}, Frederick H. and
         {Hayes}, J. and {Heckman}, Timothy M. and {Hennessy}, G.~S. and
         {Hindsley}, Robert B. and {Holm}, Scott and {Holmgren}, Donald J. and
         {Huang}, Chi-hao and {Hull}, Charles and {Husby}, Don and
         {Ichikawa}, Shin-Ichi and {Ichikawa}, Takashi and
         {Ivezi{\'c}}, {\v{Z}}eljko and {Kent}, Stephen and {Kim}, Rita S.~J. and
         {Kinney}, E. and {Klaene}, Mark and {Kleinman}, A.~N. and
         {Kleinman}, S. and {Knapp}, G.~R. and {Korienek}, John and
         {Kron}, Richard G. and {Kunszt}, Peter Z. and {Lamb}, D.~Q. and
         {Lee}, B. and {Leger}, R. French and {Limmongkol}, Siriluk and
         {Lindenmeyer}, Carl and {Long}, Daniel C. and {Loomis}, Craig and
         {Loveday}, Jon and {Lucinio}, Rich and {Lupton}, Robert H. and
         {MacKinnon}, Bryan and {Mannery}, Edward J. and {Mantsch}, P.~M. and
         {Margon}, Bruce and {McGehee}, Peregrine and {McKay}, Timothy A. and
         {Meiksin}, Avery and {Merelli}, Aronne and {Monet}, David G. and
         {Munn}, Jeffrey A. and {Narayanan}, Vijay K. and {Nash}, Thomas and
         {Neilsen}, Eric and {Neswold}, Rich and {Newberg}, Heidi Jo and
         {Nichol}, R.~C. and {Nicinski}, Tom and {Nonino}, Mario and
         {Okada}, Norio and {Okamura}, Sadanori and {Ostriker}, Jeremiah P. and
         {Owen}, Russell and {Pauls}, A. George and {Peoples}, John and
         {Peterson}, R.~L. and {Petravick}, Donald and {Pier}, Jeffrey R. and
         {Pope}, Adrian and {Pordes}, Ruth and {Prosapio}, Angela and
         {Rechenmacher}, Ron and {Quinn}, Thomas R. and {Richards}, Gordon T. and
         {Richmond}, Michael W. and {Rivetta}, Claudio H. and
         {Rockosi}, Constance M. and {Ruthmansdorfer}, Kurt and {Sand
        ford}, Dale and {Schlegel}, David J. and {Schneider}, Donald P. and
         {Sekiguchi}, Maki and {Sergey}, Gary and {Shimasaku}, Kazuhiro and
         {Siegmund}, Walter A. and {Smee}, Stephen and {Smith}, J. Allyn and
         {Snedden}, S. and {Stone}, R. and {Stoughton}, Chris and
         {Strauss}, Michael A. and {Stubbs}, Christopher and {SubbaRao}, Mark and
         {Szalay}, Alexander S. and {Szapudi}, Istvan and {Szokoly}, Gyula P. and
         {Thakar}, Anirudda R. and {Tremonti}, Christy and {Tucker}, Douglas L. and
         {Uomoto}, Alan and {Vanden Berk}, Dan and {Vogeley}, Michael S. and
         {Waddell}, Patrick and {Wang}, Shu-i. and {Watanabe}, Masaru and
         {Weinberg}, David H. and {Yanny}, Brian and {Yasuda}, Naoki and
         {SDSS Collaboration}},
        title = "{The Sloan Digital Sky Survey: Technical Summary}",
      journal = {\aj},
         year = "2000",
        month = "Sep",
       volume = {120},
       number = {3},
        pages = {1579-1587},
          doi = {10.1086/301513},
archivePrefix = {arXiv},
       eprint = {astro-ph/0006396},
 primaryClass = {astro-ph},
       adsurl = {https://ui.adsabs.harvard.edu/abs/2000AJ....120.1579Y}
}

@ARTICLE{Zeldovich1970,
       author = {{Zel'dovich}, Ya. B.},
        title = "{Gravitational instability: An approximate theory for large density perturbations.}",
      journal = {\aap},
         year = 1970,
        month = mar,
       volume = {5},
        pages = {84-89},
       adsurl = {https://ui.adsabs.harvard.edu/abs/1970A&A.....5...84Z}
}

@ARTICLE{Zennaro2017,
       author = {{Zennaro}, M. and {Bel}, J. and {Villaescusa-Navarro}, F. and {Carbone}, C. and {Sefusatti}, E. and {Guzzo}, L.},
        title = "{Initial conditions for accurate N-body simulations of massive neutrino cosmologies}",
      journal = {\mnras},
         year = 2017,
        month = apr,
       volume = {466},
       number = {3},
        pages = {3244-3258},
          doi = {10.1093/mnras/stw3340},
archivePrefix = {arXiv},
       eprint = {1605.05283},
 primaryClass = {astro-ph.CO},
       adsurl = {https://ui.adsabs.harvard.edu/abs/2017MNRAS.466.3244Z}
}

@ARTICLE{Zhang2020,
       author = {{Zhang}, Gemma and {Li}, Zack and {Liu}, Jia and {Spergel}, David N. and {Kreisch}, Christina D. and {Pisani}, Alice and {Wandelt}, Benjamin D.},
        title = "{Void halo mass function: A promising probe of neutrino mass}",
      journal = {\prd},
         year = 2020,
        month = oct,
       volume = {102},
       number = {8},
          eid = {083537},
        pages = {083537},
          doi = {10.1103/PhysRevD.102.083537},
archivePrefix = {arXiv},
       eprint = {1910.07553},
 primaryClass = {astro-ph.CO},
       adsurl = {https://ui.adsabs.harvard.edu/abs/2020PhRvD.102h3537Z}
}

@ARTICLE{Zomorodian2009,
       author = {{Zomorodian}, A.J.},
        title = "{Topology for Computing}",
      journal = {Phd Thesis},
         year = 2009,
        month =  {},
       volume = {},
       number = {},
        pages = {},
          doi = {10.5555/1629671},
archivePrefix = {Cambridge University Press, USA},
       eprint = {},
 primaryClass = {0521136091},
       adsurl = {}
}


\appendix 
 
\section{A. Methodology, Key Algorithms, Pipeline: Details}     \label{sec_appendix_A}

We provide here more detailed information on the methodology, core algorithms, and pipeline adopted in our work for the computation 
of the \textit{discrete} Morse-Smale complex, combined with topological simplification through persistence. 
For more extensive information, we refer the reader to \cite{Sousbie2011a} upon which our methods are based.

The first step in our pipeline is to compute the Delaunay tessellation of a discrete point set, represented 
in our case by a subset of DM particles from selected $N$-body simulations.
This allows us to define the simplicial complex $K$, 
namely the set of $k$-simplexes $\{\alpha_{\rm k}\}$ identified by the Delaunay tessellation.

Next, we compute the discrete density field. This is achieved using the DTFE, 
a powerful mathematical tool that reconstructs a continuous, volume-covering field from a discrete point set. 
The main advantage of DTFE over other techniques is its ability to automatically adapt to strong variations in 
density and geometry, optimally preserving the multiscale geometric and topological nature of the underlying mass distribution sampled by the 
$N$-body particles.
 
We then define the corresponding discrete Morse function over the simplicial complex, following the heuristic methodology introduced by \cite{Sousbie2011a}.
This involves assigning an appropriate density value to each simplex 
 $\alpha_{\rm k}$, ensuring it closely approximates its smooth counterpart $\rho$. 
Specifically, we set
$\rho_{\rm D} (\alpha_0) \equiv \rho(\alpha_0)$ for $k=0$,
and $\rho_{\rm D} (\alpha_{\rm k}) = {\rm max} [\rho_{\rm D} ({\rm facet} (\alpha_{\rm k}))] + \epsilon^{\rm k}  \sum \rho_{\rm D} [{\rm vertex} (\alpha_{\rm k})]$
for $k>0$, where
 $\rho_{\rm D}$ represents the discrete density assignment, $\epsilon$ is an infinitesimally small value,  
and $k \le d$ with $d$ being the maximal number of dimensions.
This formulation ensures that $\rho_{\rm D}$ is uniquely determined from the smooth function $\rho$,
independently of any arbitrary labeling of the simplexes. 
Consequently, this definition satisfies the necessary properties of a discrete Morse function.

The next step is the computation of the discrete gradient and the identification of critical simplexes.
A discrete gradient can be defined over the simplicial complex by pairing simplexes whose dimensions differ by only 1, 
while the remaining unpaired simplexes are classified as critical.
Within each pair, the simplex with the lower value is termed the tail, and the one with the higher value is the head.
Notably, each critical simplex has an associated ascending or descending manifold.
Following \cite{Sousbie2011a}, the key algorithm proceeds as follows:
We scan the set of $k$-simplexes in $K$ one by one,
in ascending order of their dimension.
Within each set, we iterate over the simplexes 
$\alpha_{\rm k}$ in ascending order of their assigned  value $\rho_{\rm D} (\alpha_{\rm k})$.
For each simplex, if it is not already part of a gradient pair, we retrieve the lowest-valued cofacet   
$\alpha_{\rm k+1}$ belonging to $\{\alpha_{\rm k} \}$
that is not yet paired and whose value is only infinitesimally higher than
$\rho_{\rm  D} (\alpha_{\rm k})$.
If such a cofacet exists, we pair them; otherwise,
$\alpha_{\rm k}$ remains unpaired. 
The algorithm terminates once all simplexes have been processed.

Once the discrete gradient has been defined over the simplicial complex, 
we compute the discrete Morse-Smale complex using the recursive method of \cite{Sousbie2011a}. 
This task is relatively straightforward, requiring only the ability to trace and compare simplexes. 
In essence, it involves a slight modification of the key algorithm used to compute manifolds. 
At this stage, we obtain an output consisting of ascending or descending 
manifolds as a function of the persistence level.

We then perform a filtration process and pair critical simplexes of the discrete Morse-Smale complex into persistence pairs. 
The filtration of $K$ determines how many simplexes must enter the filtration before a given topological feature is destroyed. 
The goal is to identify which critical simplexes create or destroy a given $k$-cycle 
and to quantify the significance of different topological features based on the function 
that defines the order in which each simplex enters the filtration. 
In this regard, the algorithm is a slight variation of the one used to compute the discrete gradient: 
critical simplexes that survive the discrete gradient computation (i.e., those belonging to the discrete Morse-Smale complex) 
are paired into persistence pairs. Each simplex is tagged as positive if it creates 
a $k$-cycle and negative if it destroys one. In 3D, critical vertices (minima) and tetrahedrons (maxima) can only create a 
$0$-cycle and destroy a $2$-cycle, respectively. 
Therefore, all critical $0$-simplexes are tagged as positive, while all critical 
$3$-simplexes are all tagged negative.
The sign of the remaining critical simplexes is determined by tracking the growth and merging of each component in the filtration, 
using a union-find type data structure to monitor the creation and destruction of $1$- and $2$-cycles.
Once each critical simplex has been assigned a sign, the filtration of the 
critical simplexes is revisited in ascending order, and persistence pairs are identified.
 
The final step  involves performing symbolic (persistence-based) topological simplification,
which assesses the relative significance of topological features to recover the true structure of the Morse-Smale complex. 
 Essentially, this process simplifies the complex by canceling 
persistence pairs that are likely noise-induced, based on a confidence level threshold expressed in units of 
$\sigma$, analogous to the Gaussian case.
Specifically, persistence levels of 1, 2, 3, and 4 $\sigma$ correspond to probabilities of 0.68, 0.95, 0.997, and 0.99937, respectively, and so on.
It is important to note that canceling a pair may lead to an increase in the total number of arcs in the complex. 
However, without simplification, filaments are detected almost everywhere in the distribution. Additionally, there are two specific cases 
where cancellation is impossible: when critical simplexes do not share an arc in the discrete Morse complex, and when critical simplexes share more than one arc.

At the conclusion of this process, we derive the discrete Morse-Smale complex as a function of the persistence level. 
This allows us to identify actual cosmic structures (i.e., voids, walls, filaments, and peak patches or halos) within a given cosmology. 


\end{document}